\documentclass[12pt,letterpaper]{article}
 
\usepackage[margin=1.2in]{geometry}
\usepackage[english]{babel}
\usepackage[utf8x]{inputenc}
\usepackage{amsmath}
\usepackage{amsthm}
\usepackage{placeins}
\usepackage{graphicx,caption,subcaption}
\usepackage{tabularx}
\usepackage{siunitx}
\usepackage{array}
\usepackage{colortbl}
\usepackage{multirow}
\usepackage[flushleft]{threeparttable}
\usepackage[usenames, dvipsnames]{xcolor}  
\usepackage[colorlinks=true, allcolors=blue]{hyperref}
\usepackage{tikz}
\usepackage{pgfplots}
\colorlet{MyGray}{gray!120}
\usepackage{eurosym}
\usepackage{bm}
\usepackage{siunitx}
\usepackage{booktabs}
\usepackage{epstopdf}
\usepackage{rotating}
\usepackage{pdflscape}
\usepackage[]{apacite}
\usepackage[title]{appendix}
\usepackage[english]{isodate}
\usepackage{float}
\usepackage{placeins}
\usepackage{subcaption}
\usepackage{multicol}
\usepackage{graphicx,caption}
\usepackage{amsmath}
\usepackage{amssymb}
\usepackage{bbm}
\usepackage{comment}

\usepackage{tikz}
\usetikzlibrary{calc}
\usetikzlibrary{positioning}
\tikzset{
solid node/.style={circle,draw,inner sep=1.5,fill=black},
hollow node/.style={circle,draw,inner sep=1.5}
}

\usepackage[round]{natbib}
\usepackage{kpfonts}    % for nice fonts

\pgfplotsset{compat=1.16}

\begin{document}

\newgeometry{top=1mm, bottom=15mm}

\title{
%\begin{center}
%\textbf{{\small Preliminary version -- do not circulate or cite without authors' approval}}
%\end{center}

% sell a lemon

% Does a market frame trigger selfish behavior? 

% An experimental study on the role of social
% preferences and Kantian moral concerns
%in lemons-like situations

%Are sellers willing to sell lemons? 
%Market Frames and Moral Dilemmas: Exploring Kantian Concerns in Decision Making
%Market Framing and Moral Decision Making: Insights from an Experimental Study
%Framing the Marketplace: Understanding Moral Motivations in Economic Transactions
%Navigating Market Morality: The Impact of Framing on Economic Behavior 

Doing the right thing (or not) in a lemons-like situation:  on the role of social preferences and Kantian moral concerns \thanks{I.A. acknowledges funding from the European Research Council (ERC) under the European Union's Horizon 2020 research and innovation programme (grant agreement No 789111 - ERC EvolvingEconomics), as well as IAST funding from the French National Research Agency (ANR) under grant ANR-17-EURE-0010 (Investissements d'Avenir program). We are grateful to Lony Bessagnet, Jean-Fran\c{c}ois Bonnefon, Eric Danan, Pierre Fleckinger, Peter Hammond, Astrid Hopfensitz, Nicolas Jacquemet, Pau Juan-Bartroli, Fabrice Le Lec, Thierry Magnac, Pauline Morault, Esteban Mu\~{n}oz-Sobrado, S\'{e}bastien Pouget, R\'{e}gis Renault, Angelo Secchi, Sigrid Suetens, Ao Wang and Boris van Leeuwen for helpful discussions, and to St\'{e}phane C\'{e}zera for invaluable assistance in the lab. We thank audiences at CY Cergy Paris University, IAST, the Paris School of Economics Summer School, Toulouse School of Economics, Catholic University of Uruguay, Max Planck Institute, Economic Science Association, Uruguayan Central Bank, Universidad de la Rep\'{u}blica and University of Gothenburg for feedback.}}

\date{\color{gray} \normalsize \today}

\author{ Ingela Alger\thanks{Toulouse School of Economics, CNRS, University of Toulouse Capitole, and Institute for Advanced Study in Toulouse, France, and CEPR. \color{blue}ingela.alger@tse-fr.eu}\\ 
Jos\'e Ignacio Rivero-Wildemauwe\thanks{Department of Business and Administration, Catholic University of Uruguay, Montevideo, Uruguay. \color{blue}joseignacio.rivero@ucu.edu.uy}}

\maketitle
\thispagestyle{empty}
{\small {\textbf{Abstract:} 
We conduct a laboratory experiment using framing to assess the willingness to ``sell a lemon'', i.e., to undertake an action that benefits self but hurts the other (the ``buyer''). We seek to disentangle the role of other-regarding preferences and (Kantian) moral concerns, and to test if it matters whether the decision is described in neutral terms or as a market situation. When evaluating an action, morally motivated individuals consider what their own payoff would be if---hypothetically---the roles were reversed and the other subject chose the same action  (universalization). We vary the salience of role uncertainty, thus varying the ease for participants to envisage the role-reversal scenario. 
%In order to mute the effects of repeated interaction considerations and social image or reciprocity concerns, 
We find that subjects are (1) more likely to ``sell a lemon'' in the market frame, and (2) less likely to do so when the role uncertainty is salient. 
%The latter is consistent with the presence of moral concerns of a Kantian nature. 
We also structurally estimate other-regarding and Kantian moral concern parameters.}}

\smallskip 

{\small \textbf{JEL codes}: C91, D01, D91.
%C49, C72, C9, C91, D03, D84.

}
\restoregeometry

\newpage

\section{Introduction}

There is by now robust evidence from experimental games played by subjects in laboratory settings in a host of countries, that a large share of individuals give up own material payoff to increase that of others (e.g., \cite{Zelmer2003,Engel2011}), and exhibit lying aversion (e.g., \citet*{Gneezy2013,Abeler2019}).
%; data collected through field experiments and surveys have led to qualitatively similar conclusions \citep<e.g.,>{Nagin2002,falk2018global}. 
%If externally valid, this evidence suggests that real individuals might differ quite drastically from the selfish materialist that still populates many economic models. 
What are the implications of this evidence for behaviour in real market settings? 
In particular, is there reason to believe that sellers are less willing to sell goods at a price above their value than many theoretical models assume \citep{Akerlof1970}?
% This view of human nature appears to be quite restrictive, however: experimental games played by subjects in laboratory 
%Inspired by theoretical results showing that markets populated by selfish materialists are sometimes unable to reach first-best outcomes  in 
%\citep{Akerlof1970}, 
%\citep{Helpman1975,Rothschild1976,Myerson1983,Akerlof1970}, 
We aim at providing some insights by conducting a laboratory experiment that we designed with two objectives in mind: (1) to examine behavior of subjects in a situation reminiscent of that of a seller of a ``lemon'', using both a neutral and a market frame; and (2) to disentangle the role of other-regarding, or distributional, preferences from Kantian moral concerns in this context. 

Subjects in our experiment face a series of anonymous binary choices, each of which affects the payoff of a randomly drawn other subject, hence avoiding that subjects' decisions be influenced by strategic or repeated interaction considerations, \citep{Roth1983,Andreoni1993}, and social image or reciprocity concerns \citep{benabou2006image,Levine1998,charness2002understanding}. In each binary choice one of the options---the \textit{selfish option}---entails a gain for self and a loss for the other, compared to the other option---the \textit{status quo} option (in other words, each decision situation is a Dictator game, with taking rather than giving \citep*{Dreber2013}. The payoff consequences of this choice thus capture a lemons-like situation, in which the selfish  option amounts to selling an object of low quality at such a price that the seller gets better off but the buyer gets worse off, while the \textit{status quo} option amounts to keeping the object. 
%All the binary choices are such that the payoff of the decision-maker exceeds that of the passive subject independent of the selected option. 

To achieve our first objective, we let half of the decisions be taken in a standard \textit{neutral frame} and the other half in a \textit{market frame}, keeping the payoff consequences identical under both frames. Only the wording used to describe the situation differs. Specifically, in the market frame subjects get the following information: ``As you can see on the decision screen above, the Buyer would be better off if you chose Not Sell, while you are better off if you Sell. Think of this as representing a situation in which the good that you sell has a defect which makes the Buyer enjoy owning the good less than you do.'' The wording was chosen to mimic a situation in which a seller can sell a ``lemon'' to an uninformed buyer (who in the experiment is passive). In the neutral frame the options are simply referred to as X and Y, and no context is given.

%\textbf{INGELA PLEASE CHECK}

Our second objective is to disentangle two distinct motivations for refraining from the selfish option. The first motivation is consequentialist: since the selfish option increases the payoff gap between the decision-maker and the passive subject, the \textit{status quo} may be selected due to an other-regarding concern in the form of altruism or inequity aversion, formalized as a weight $\beta$ on the payoff gap \citep{Becker1974,fehr1999theory}. The second motivation relies on a (possibly partial) universalization argument: it makes the subject evaluate each decision in the light of what his payoff would be, if---hypothetically---the other were to take the same decision as the subject himself with some probability $\kappa$, if the roles were reversed. This motivation, which is reminiscent of a Kantian moral concern, can be expressed with \textit{Homo moralis} preferences, where $\kappa$ is the individual's degree of morality \citep{alger2013homo}. We posit that subjects have preferences that combine other-regard with a Kantian moral concern. Such a combination is indeed predicted by recent theoretical work on the evolutionary foundations of preferences in a model where interactions yield  payoff consequences with negligible effects on overall reproductive success \citep*{alger2019evolution}, and it has also received empirical support in another experimental study \citep{vanleeuwen23}. Moreover, recent theoretical studies on the effects of Kantian moral concerns on market outcomes that feature an unequal distribution of information (including cases where product quality is the relevant information asymmetry) document important effects of such moral concerns on equilibrium outcomes \citep*{wildemauwe2023,wildemauwe2023a}. In equilibrium, Kantian motivations reduce the likelihood of ``lemons'' being exchanged, or may even eliminate it altogether. Importantly, these references also underline that other-regard may have similar effects. In light of this, it is of particular relevance to be able to distinguish between both behavioral drivers.

 %\textbf{----------------------------------------------------------------------------------------------------------}
 
Because the Kantian concern is triggered by consideration of a possible role reversal, the experimental design consists in varying the salience of role reversal. Specifically, half of the decisions we observe are taken in a \textit{VOI frame} and the other half in a \textit{non-VOI frame}, where VOI stands for \textit{veil of ignorance}. In the VOI frame, a subject is told that (s)he stands an equal chance of being the decision-maker and the passive subject. In the non-VOI frame (s)he is simply told that (s)he is the decision-maker. In reality, however, the two situations are the same, since upon entering the experiment any subject effectively stands an equal chance of being the decision-maker and the passive subject in any given match. Hence, the only difference is that we make this role uncertainty explicit in the VOI frame. 
Our theory thus predicts that if subjects are driven by a Kantian moral concern and if they are not fully aware of the arbitrariness in the role distribution in the non-VOI decisions, they are more likely to sell in non-VOI decisions than in VOI decisions. We also estimate the contribution of the other-regarding and the Kantian motivation to the subjects' decisions, and examine whether the estimated motivations differ in the market and the neutral frames.

We find that, on average, subjects are significantly more prone to selecting the selfish option in the market frame compared to the neutral frame. In line with our theoretical prediction, they also select the selfish option significantly more often in non-VOI than in VOI decision situations. Comparing the effect of the VOI frame on the propensity to select the selfish decision between the market frame and the neutral frame, the magnitude in absolute terms is larger in the market frame, but the effect of VOI relative to non-VOI decisions is the same in both frames. The effects of the market frame and of the VOI frame are larger in magnitude in the within-subjects regressions, suggesting significant heterogeneity across subjects.

We also structurally estimate the preference parameters $\beta$ and $\kappa$, in line with several earlier experimental studies (e.g., \citet*{fisman2007individual,bruhin2018many,vanleeuwen23}). A novelty of our study compared to existing ones, is that we obtain estimates of the preference parameters in two different frames: the market frame and the neutral frame. Overall, we find differences in both the degree of aheadness aversion and the Kantian morality estimates across the market and the neutral frame, some of the differences being large.

Our study contributes to three strands of the literature. First, it adds insights into an old question: do market interactions render people more selfish \citep{Bowles1998b}? Some authors have relied on lab-in-the-field experiments to compare the behavior of individuals with varying degrees of exposure to market interactions (e.g., \citet{Henrich2005,Henrich2016} and \citet*{agneman2022}), while others have used laboratory to examine how variation in the competition between subjects affects behaviors (e.g., \citet*{Cabrales2010,  Sutter2020, Engel2020,Dufwenberg2022,Bartling2023replacement}). Our experiment explores a different possibility, by adopting a design where we examine the effect of framing the task as the sale of an lemon (recall the wording above), compared to a standard neutral framing, the objective being to quantify the effect of the market frame (for a recent discussion of the use of frames in economic experiments, see \citet*{ALEKSEEV2017}). Our hypothesis is close to that put forward elsewhere (e.g., \cite{Bowles1998b,Kirman2010}): preferences may be endogenous and thus shaped differently by market interactions than by, say, interactions with friends and family. Some experiments lend support to the importance of frames, for example the famous study that showed that a Prisoner's dilemma generates more cooperation if it is labeled as ``the community game'' than if it is called ``the Wall street game'' \citep*{Liberman2004} (see also \citep{KAY2003}). Other studies, however, found small or no effects \citep*{Dreber2013,DUFWENBERG2011}. By contrast to other studies that use some kind of market-related framing (e.g., \citet*{DUFWENBERG2011,THONI2015}), our instructions to the participants explicitly describe an action as the sale of a lemon, and we compare behavior under this wording to behavior under neutral wording. Furthermore, our design allows us to compare the estimated preference parameters between the neutral and the market frames. 

Second, our study adds to a recent set of experiments that have sought to detect moral concerns as drivers of behavior, where moral concerns are distinct from distributional preferences \citep*{capraro2018right, Bursztyn2019, weibull2019horserace, Chen2022, Feess2022, benabou22a, benabou22b, vanleeuwen23}. 
The most closely related experiments are those by \citet*{weibull2019horserace} and \cite{vanleeuwen23}, who also posit a utility function with a Kantian moral concern \`a la \textit{Homo moralis} \citep{alger2013homo}, and who also seek to disentangle this concern from distributional preferences. Our study differs from both of them in two important ways. First, while they only use neutral wording, our experimental design allows us to test whether preferences differ between the neutral and the market frame. Second, they rely exclusively on what we call VOI decisions; our comparison of decisions in VOI and non-VOI decisions gives an indication about the extent to which the explicit mention of the role uncertainty helps trigger the Kantian moral concern.

This last point brings us to the third literature to which our study contributes: that which examines if decisions depend on whether they are elicited using the direct-response or the strategy method --- i.e., whether they are taken after or before subjects learn the role distribution. Several studies find that decisions differ significantly, while other studies find mixed or no effects (see the surveys by \citet*{Brandts2011} and \citet*{Grech2022}). In the studies where an effect appears, decisions taken behind the veil of ignorance tend to be more pro-social on average \citep*{Sutter2003, Iriberri2011, Huang2019, Levine2020, Garcia-Pola2020, Ortiz-Riomalo2021, Herne2022}. Our findings are in line with this empirical regularity. But while previous studies have sought to explain the greater pro-sociality in decisions taken under role uncertainty by referring mainly to aspects of the decision-making process (e.g., number of decisions taken, complexity of the information presented, likelihood that emotions are triggered by the task), our experiment relies on a preference-based theory which predicts and thus explains it. According to our theory, the explicit mention of role uncertainty awakens or reinforces a Kantian moral concern, which triggers pro-social behavior in our dictator game design. 

The next section describes the experimental design and procedures, and in Section 3 we test whether behaviors are different under the market and the neutral frames, and under the VOI and the non-VOI frames. In Section 4 we report the results from the structural estimations, and Section 5 concludes.

\section{The experiment}\label{sec:preferences}

\subsection{Game protocols and preferences}

\begin{figure}[t]
	%\begin{subfigure}[b]{0.30\textwidth}
	\begin{center}
	\begin{tikzpicture}[scale=0.7]
	% Specify spacing for each level of the tree
	\tikzstyle{level 1}=[level distance=20mm,sibling distance=128mm]
	\tikzstyle{level 2}=[level distance=20mm,sibling distance=64mm]
	%\tikzstyle{level 3}=[level distance=20mm,sibling distance=32mm]
	\scriptsize
	\node(0)[hollow node,label=above:{Nature}]{}
	child{node[solid node,label=above left:{Subject $i$ is \textit{Player 1}}]{}
		child{node[hollow node, label=below:{$(e_1,e_2)$}]{} 
			edge from parent node[left,xshift=-5]{X}}
		child{node[hollow node, label=below:{$(e_1+G,e_2-L)$}]{}
			edge from parent node[right,xshift=5]{Y}}
		edge from parent node[left,yshift=9]{$1/2$}
	}
	child{node[solid node,label=above right:{Subject $j$ is \textit{Player 1}}]{}
		child{node[hollow node, label=below:{$(e_2,e_1)$}]{} 
			edge from parent node[left,xshift=-5]{X}}
		child{node[hollow node, label=below:{$(e_2-L,e_1+G)$}]{}
			edge from parent node[right,xshift=5]{Y}}
		edge from parent node[right,yshift=9]{$1/2$}
	};
	\end{tikzpicture}
	\end{center}
	%\end{subfigure}
	\caption{The game protocol used in the experiment. The payoff vectors state subject $i$'s and $j$'s payoff as the first and second component, respectively.}
	\label{fig:subgame_tree_VOI}
\end{figure}
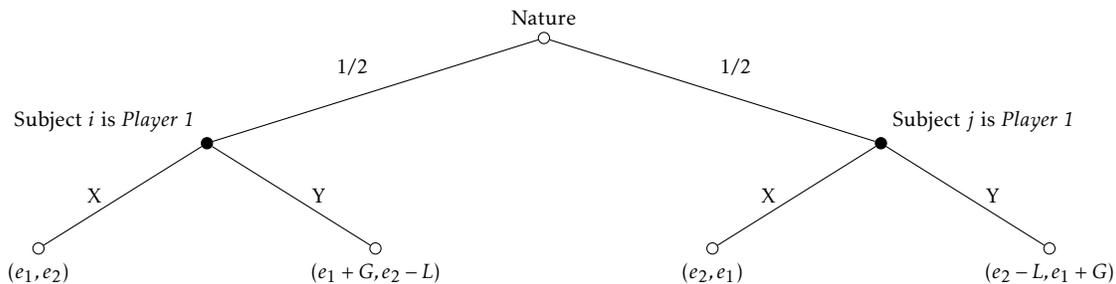

In the experiment subjects are matched into pairs to play anonymous one-shot interactions. In each interaction one subject is assigned to the \textit{Player 1} role and the other to the \textit{Player 2} role, each role assignment being equally likely. Player 1 has to choose between two actions, call them $X$ and $Y$. If she chooses $X$, both individuals obtain their initial endowment, denoted $e_1$ and $e_2$, respectively, while if she chooses $Y$, she gets $e_1 +G > e_1$ while Player 2 gets $e_2 -L < e_2$, where $e_1> e_2$, $G>0$, and $L>0$. That is, Player 1 makes a net gain $G$ from choosing $Y$ rather than $X$, while this choice entails a net loss $L$ for Player 2. Note that whether she chooses $X$ or $Y$, Player 1 receives a higher payoff than Player 2. The game tree for any matched pair of subjects---here called $i$ and $j$--- is depicted in Figure \ref{fig:subgame_tree_VOI}.

We use two frames in the experiment: the \textbf{Neutral frame} and the \textbf{Market frame}. In the latter, we replace \textit{Player 1} by \textit{Seller} and \textit{Player 2} by \textit{Buyer}, and we describe actions $X$ and $Y$ by \textit{Not Sell} and \textit{Sell}, respectively. Indeed, the payoff structure captures in a stylised manner the case where a seller must decide whether to sell a low-quality item to a buyer in exchange for the going (fixed) price in the market---a classic ``lemons'' situation, in which $Y$ amounts to selling and $X$ to keeping the good. The Buyer has no say here: we adopt this simplification to concentrate on the Seller's willingness to sell a lemon. To facilitate the exposition, we will henceforth use the term \textit{Selfish option} to refer to \textit{Sell} in the Market frame and action $Y$ in the Neutral frame, and the term \textit{Status quo option} to refer to \textit{Not Sell} in the Market frame and action $X$ in the Neutral frame. We will also refer to the two roles as the active and the passive roles.

\begin{figure}[t]
	\begin{center}
	\begin{tikzpicture}[scale=0.7]
	% Specify spacing for each level of the tree
	\tikzstyle{level 1}=[level distance=20mm,sibling distance=128mm]
	\tikzstyle{level 2}=[level distance=20mm,sibling distance=64mm]
	%\tikzstyle{level 3}=[level distance=20mm,sibling distance=32mm]
	\scriptsize
	\node(0)[hollow node,label=above:{Nature}]{}
	child{node[red][solid node,fill=red,label=above left:{\textcolor{red}{Subject $i$ is Player 1}}]{}
		child[red]{node[hollow node, label=below:{$(e_1,e_2)$}]{} 
			edge from parent node[left,xshift=-5]{Y}}
		child[red]{node[hollow node, label=below:{$(e_1+G,e_2-L)$}]{}
			edge from parent node[right,xshift=5]{X}}
		edge from parent node[left,yshift=9]{$1/2$}
	}
	child{node[solid node,label=above right:{Subject $j$ is Player 1 }]{}
		child{node[hollow node, label=below:{$(e_2,e_1)$}]{} 
			edge from parent node[left,xshift=-5]{Y}}
		child{node[hollow node, label=below:{$(e_2-L,e_1+G)$}]{}
			edge from parent node[right,xshift=5]{X}}
		edge from parent node[right,yshift=9]{$1/2$}
	};
	\end{tikzpicture}
	\end{center}
		\caption{{{The matched subjects ($i$ and $j$) are informed about the whole game tree in a VOI decision, but only about the realized role distribution in a non-VOI decision  (if the Player 1 role is assigned to subject $i$, she only receives information about the part highlighted in red). }}}
\label{fig:subgame_tree_nonVOI}
	%\end{subfigure}
	%\caption{The non-VOI game protocol}
\end{figure}
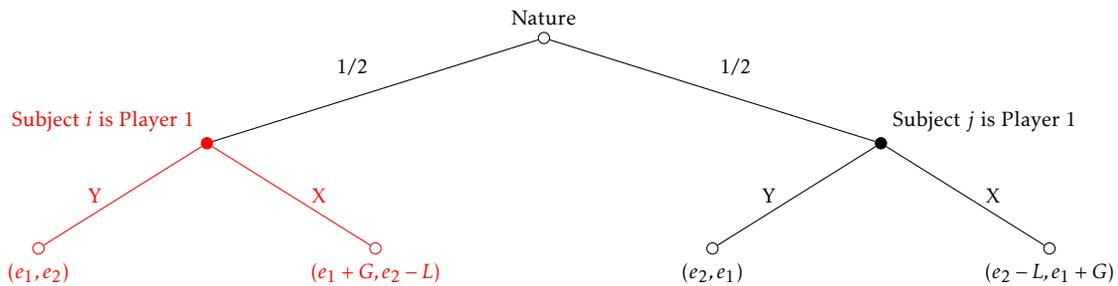

In addition to the aforementioned frames, we vary the information that subjects receive about the decision situation. In the \textbf{VOI decision situation} subjects are asked to state their choice in the active role \textit{before} learning the role distribution. Participants are told that they have an equal chance of being cast in either role. Being thus explicitly informed about the whole game tree in Figure \ref{fig:subgame_tree_nonVOI}, they are led to reason behind the ``Veil of Ignorance'' (VOI) with respect to the role distribution. In the \textbf{non-VOI decision situation}, a subject is asked to state her decision \textit{after} being informed that she is in the active  role and is therefore not behind the ``Veil of Ignorance''. Hence, even though the game actually being played between any given matched subject pair in the experiment is the one depicted in Figure \ref{fig:subgame_tree_nonVOI} (since each subject stands an equal chance of being handed the active role), in a non-VOI decision situation a subject cast in the active role is given explicit information from us only about part of the game tree (the part highlighted in red in Figure \ref{fig:subgame_tree_nonVOI}, should the active player be if $i$ is the active subject).

The decisions being anonymous and one-shot, the experimental design removes motivations such as social image concerns, repeated interaction, and reciprocity effects. Hence, a subject's decision should be driven by his or her intrinsic preferences and beliefs. We posit a utility function that combines material self-interest, other-regard, and a Kantian moral concern.
As we will now show the key point is that the Kantian moral concern is expected to be fully triggered in a VOI decision situation, but only partially so or not at all in a non-VOI decision situation.
The reason is that in the former, subjects explicitly take into account the possibility that they may end up in either role, while in the latter individuals are either fully unaware or only partially aware of the arbitrariness of the role allocation. 

%For simplicity, in the description of the posited utility function we will use the vocabulary corresponding to the Market frame.

%While the experiment also uses a \textbf{Neutral frame}, we use the vocabulary of the \textbf{Market frame} introduced above to show how the posited preferences operate as described intuitively above. The behavior strategy of the Seller consists of the probability to sell when the quality is low and the probability to sell when the quality is high. Formally, consider a pair of matched subjects, $i$ and $j$, and denote by $x=(x_L,x_H)$ the behavior strategy of $i$ and $y=(y_L,y_H)$ that of $j$. Thus, $x_Q \in [0,1]$ is the probability that $i$ sells when in the Seller role and the quality is $Q$, and $y_Q \in [0,1]$ is the probability that $j$ sells when in the Seller role and the quality is $Q$.

%Our hypothesis that the VOI treatments might, on average, provoke a more pronounced Kantian moral concern than the non-VOI treatments. We formalize this hypothesis as follows. Consider, first, a subject $i$ who, when told that (s)he is the Seller in the non-VOI treatment, does not reflect on the fact that (s)he could have been cast in the Buyer role instead. We posit that when the quality is $Q$ the utility of such a subject is:

Formally, consider subjects $i$ and $j$ and denote their decisions by $x \in \{0,1\}$ and $y \in \{0,1\}$ respectively, where $x = 1$ if $i$ chooses the selfish option and $x=0$ if she selects the \textit{status quo} option when active, while $y = 1$ if $j$ selects the selfish option and $y=0$ if she chooses the \textit{status quo} option when active.
%\footnote{In the Neutral frame, $x=0$ if $i$ selects $X$ and $x=1$ if (s)he selects $Y$, while $y=0$ if $j$ selects $X$ and $y=1$ if she selects $Y$.} 
We posit the following (expected) utility obtained by $i$ in the VOI decision situation:
    \begin{eqnarray}
    \label{eq:prefsVOI}
U_{i}\left( x,y\right) &=&
(1-\kappa_i) \cdot 
\Big[\frac{1}{2} \cdot (e_1+x G)
+\frac{1}{2} \cdot 
(e_2-yL) \Big] \\ 
&+&\kappa_i \cdot 
\Big[\frac{1}{2} \cdot (e_1+x G)+\frac{1}{2} \cdot 
(e_2-x L) \Big] \nonumber\\ 
&-& \frac{1}{2}\cdot \beta _{i}\cdot[e_1-e_2+x(G+L)]
 \nonumber\\
&-&\frac{1}{2}\cdot \alpha_{i}\cdot [e_1-e_2 +y (G+L) ] \nonumber
\end{eqnarray}

The term inside the square brackets in the first line is $i$'s expected material payoff: with probability $1/2$ $i$ gets the active role, and $i$'s decision $x$ determines his material payoff; with probability $1/2$ $j$ gets the active role, and $j$'s decision $y$ determines $i$'s material payoff. The term inside the square brackets in the second line captures $i$'s Kantian moral concern: it is the expected material payoff that $i$ (in fact, any subject in this interaction) would get if---hypothetically---the strategy $x$ was universalized. The parameter $\kappa_i \in [0,1]$ is $i$'s \textit{degree of morality}.\footnote{Mathematically, the Kantian moral concern in \textit{Homo moralis} preferences has a similar effect as the false belief that one's action affects the action of the opponent, a belief known as magical (or quasi-magical) thinking \citep{Shafir1992,Daley2017}. Moreover, under the special case of \textit{Homo moralis} preferences where the degree of morality $\kappa$ is equal to one, predictions based on Nash equilibrium play sometimes coincide with predictions based on Kantian equilibrium, an equilibrium concept introduced by \cite{Roemer2010}. The preference-based approach adopted here presents several advantages: it avoids reliance on false beliefs, it allows for intermediate degrees of Kantian concern, it allows for the co-existence of other-regarding and Kantian concerns, and it allows for the adoption of the standard Nash equilibrium concept.}

The third line measures the effects on $i$'s utility of being materially ahead of the other when in the active role, while the last line measures the effects on $i$'s utility of being materially behind the other when in the passive role. The parameter $\beta _{i}$ represents $i$'s \textit{aheadness aversion} (if $\beta _{i}>0$) or \textit{love for aheadness} (if $\beta _{i}<0$), where aheadness means that $i$'s monetary payoff is larger than $j$'s.
 
Similarly, the parameter $\alpha _{i}$ represents $i$'s \textit{behindness aversion} (if $\alpha _{i}>0$) or \textit{love for behindness}  (if $\alpha _{i}<0$), where behindness means that $i$'s monetary payoff is smaller than $j$'s. While it is natural to include attitudes towards being both ahead and behind materially, in our experimental design any active decision-maker is always ahead of the other, passive subject; hence we henceforth omit the last term of the utility function.\footnote{The utility function in \eqref{eq:prefsVOI} is the same as that posited in the main analysis of \citep{vanleeuwen23}. Moreover, such a utility function, which describes the individual's preferences over (potentially trivial) material payoffs, has been shown to be favored by evolution by natural selection (Alger, Weibull, and Lehmann, 2020).}
%\footnote{Since $w_2<e_2< e_1<w_1$, only the attitude towards aheadness (resp. behindness) matters when the individual is in the \textit{Seller} role (resp. the \textit{Buyer} role).}

% The utility parameter that appears in the last line, $\kappa _{i} \in [0,1]$, captures the Kantian moral concern. It places weight on the change in the expected monetary payoff the subject would experience if, hypothetically, the opponent were to switch from using strategy $y$ to using strategy $x$. Put differently, the last line captures the idea that a Kantian moral concern amounts to pondering what would happen if the opponent chose the same strategy as the subject him/herself.

To see why the Kantian moral concern is expected to generate different decisions in the VOI and the non-VOI decision situations, consider now a subject in a non-VOI decision situation, cast in the active role. Assuming that this subject is completely unaware of the fact that she could have been allocated to the passive role instead, her utility reduces to:
 \begin{equation}
 \label{eq:prefsnonVOI}
\begin{aligned}
V_{i}\left( x,y\right) = e_1  +x G- \beta _{i}\cdot [e_1-e_2 +x(G+L)].
\end{aligned}
\end{equation}
Comparing this with \eqref{eq:prefsVOI}, we see that as long as $L > 0$, a subject may select a different $x$ in VOI and non-VOI decisions if $\kappa_i>0$. 
Specifically, Kantian morality reduces the subject's willingness to select the selfish decision: 
%a fully unaware individual (i.e., who believes that her probability of being cast into the active  role is 1) 
from \eqref{eq:prefsVOI} and \eqref{eq:prefsnonVOI}, we see that she is willing to select the selfish decision if 
\begin{equation}\label{eq:condition_nonVOI_un}
G \geq \beta_{i} (G+L)
\end{equation}
in the non-VOI decision situation, but only if 
\begin{equation}\label{eq:condition_VOI}
G - \kappa_i  L \geq \beta_{i} (G+L)
\end{equation}
in the VOI decision situation. The VOI decision situation renders explicit the fact that the selfish decision would hurt her if she were cast in the passive rather than the active role and her strategy was universalized to the other individual. 

Importantly for our experiment, these two conditions imply that for a given payoff vector $(e_1,e_2,e_1+G,e_2-L)$ a subject \textit{makes a switch} from the selfish action in the non-VOI decision to the \textit{status quo} act in the VOI decision situation if
her aheadness aversion $\beta_i$ is not too pronounced, 
\begin{equation} \label{betabar}
	 \beta_{i}\leq  \frac{G}{G+L} \equiv z,
\end{equation}
and her Kantian moral concern $\kappa_i$ is sufficiently pronounced,
\begin{equation} \label{kappabar}
\kappa_i \geq  \frac{G -\beta_{i}(G+L)}{ L} =\frac{z-\beta_i}{1-z}.
\end{equation}
%for any given payoffs $(e_1,e_2,w_1,w_2)$, a (fully unaware) subject switches from the selfish option in the non-VOI decision situation to the \textit{status quo} option in the VOI decision situation if both conditions  \eqref{betabar} and \eqref{kappabar} are satisfied. 
Figure \ref{fig:unaw_lemon1} shows these threshold values for two payoff configurations used in the experiment: the solid vertical and oblique lines show, respectively, the threshold values $z$ and $(z-\beta_i)/(1-z)$ for  
payoff configuration $(e_1,e_2,e_1+G,e_2-L)=(150,100,165,90)$; the dashed lines for payoff configuration $(e_1,e_2,e_1+G,e_2-L)=(200,190,210,100)$. To see how variation of the payoffs enables to distinguish between different preference types, note that the two solid lines and the two dashed lines divide the $(\beta,\kappa)$-space into six regions. An individual with a preference type $(\beta_i,\kappa_i)$:
\begin{itemize}
\vspace{-0.3cm}
   \item  in region A always selects the selfish option; 
\vspace{-0.3cm}
\item  in region B selects the selfish option under payoff $(150,100,165,90)$, but makes a switch under payoff $(200,190,210,100)$;
\vspace{-0.3cm}
    \item  in region C makes a switch under both payoffs;
\vspace{-0.3cm}
    \item  in region D selects the selfish option under payoff $(150,100,165,90)$, but always the \textit{status quo} option under payoff $(200,190,210,100)$;
\vspace{-0.3cm}
    \item  in region E makes a switch under payoff $(150,100,165,90)$, but always selects the \textit{status quo} option under payoff $(200,190,210,100)$;
\vspace{-0.3cm}
    \item  in region F always selects the \textit{status quo} option.
 \end{itemize}

% three sets of preference types $(\beta,\kappa)$, depending on whether the type would lead an individual to select the selfish option under both non-VOI and VOI, select the \textit{status quo} option under both non-VOI and VOI, or switch from the selfish option under non-VOI to the \textit{status quo} option under VOI.\footnote{Note that a switch from the  \textit{status quo} option under non-VOI to the selfish option under VOI would require a negative value of $\kappa$.}

\usepgfplotslibrary{fillbetween}
\begin{figure}[ht] 	
  \captionsetup{width=0.7\linewidth}
\begin{center}
	\begin{tikzpicture}[scale=1.2]
%	\tikzstyle{every node}=[ball color=red,circle,text=white]
	\begin{axis}[
	ticklabel style = {font=\footnotesize},
	 legend style={font=\footnotesize},
	axis lines=center,
	grid = none,
	ymin=0,
	ymax=2.3,
    xtick={0.1,0.6},
    ytick={0.11,1.5},
	xmin = -0.6,
	xmax=1.0,
    %	extra y ticks={0},
	xlabel={$\beta$},
	ylabel={$\kappa$}]

	\addplot[smooth,very thick, solid,name path=A] {1.5-2.5*x}; % actual curve
%	\addlegendentry{No. 1}
  	\addplot[smooth,very thick, solid,name path=F]coordinates {(0.6,0)(0.6,2.3)}; % actual curve
% 	\addlegendentry{No. 12}
	\addplot[smooth,very thick,dashed,name path=C] {0.11-1.11*x}; % actual curve
%	\addlegendentry{No. 1}
  	\addplot[smooth,very thick,dashed,name path=F]coordinates {(0.1,0)(0.1,2.3)}; % actual curve
  	\node[ball color=red,circle,text=white] at (0.35,1.3) {E};
  	\node[ball color=red,circle,text=white] at (0.75,1.3) {F};
  	\node[ball color=red,circle,text=white] at (0.3,0.3) {D};
  	\node[ball color=red,circle,text=white] at (-0.4,0.22) {A};
  	\node[ball color=red,circle,text=white] at (-0.05,0.8) {B};
  	\node[ball color=red,circle,text=white] at (-0.015,1.95) {C};
  	
% 	\addlegendentry{No. 12}
%	\addplot[draw=none,name path=Z] {-2};     % “fictional” curve
%	\addplot[MyGray,fill opacity=0.2] fill between[of=A and Z,soft clip={domain=0:1}]; % filling
% 	\addplot[MyGray,fill opacity=0.2] fill between[of=B and Z,soft clip={domain=0:1}]; % filling
% 	\addplot[MyGray,fill opacity=0.2] fill between[of=C and Z,soft clip={domain=0:1}]; % filling
% 	\addplot[MyGray,fill opacity=0.2] fill between[of=D and Z,soft clip={domain=0:1}]; % filling
% 	\addplot[MyGray,fill opacity=0.2] fill between[of=E and Z,soft clip={domain=0:1}]; % filling
% 	\addplot[MyGray,fill opacity=0.2] fill between[of=F and Z,soft clip={domain=0:1}]; % filling
	\end{axis}
	\end{tikzpicture}
	
\vspace{0.2cm}
\begin{minipage}{0.5\textwidth} % choose width suitably
\caption{\label{fig:unaw_lemon1} Threshold values $z$ and $(z-\beta_i)/(1-z)$ for payoffs $(e_1,e_2,e_1+G,e_2-L)=(150,100,165,90)$ (solid) and $(e_1,e_2,e_1+G,e_2-L)=(200,190,210,100)$ (dashed)}
\end{minipage}
	\end{center}

    \end{figure}
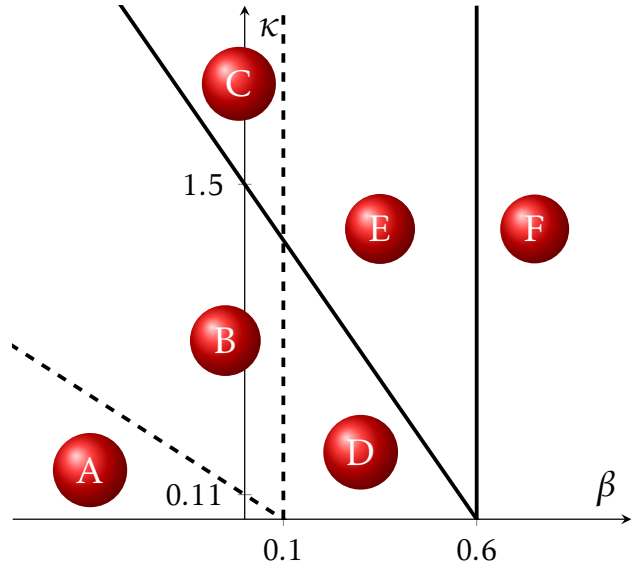

In the experiment we use 20 different payoffs $(e_1,e_2,e_1+G,e_2-L)$.  Table \ref{tab:thresholds} shows, for each of these 20 payoffs, the threshold values $z$ and $(z-\beta)/(1-z)$ (see \eqref{betabar} and \eqref{kappabar}). Note that the 20 payoff configurations used in the experiment entail 20 different values of $z$. Let $Z$ denote the set of these 20 values. Figure \ref{fig:unaw_lemon} shows the lines $\kappa=(z-\beta)/(1-z)$ for the 20 values of $z$. Except for individuals with $\beta_i \geq 0.6$ (who always select the \textit{status quo} option for all payoffs), and those with $\beta_i \leq 0.03$ and $\kappa_i \leq 0.03-1.03 \beta_i$ (who always select the selfish option for all payoffs), these payoffs are expected to generate behavioral variation across payoffs and/or across the two conditions (VOI vs. non-VOI). 

The payoffs were chosen so as to maximize statistical power in the between-subject estimation of the VOI treatment's effects (conditional on the fact that we decided to have no more than 20 decisions in each sequence), where power computations were based on the results presented by \cite{vanleeuwen23}. More precisely, we carried out 1000 simulations where in each one, the 109 subjects for which \cite{vanleeuwen23} estimate individual preference parameters were randomly assigned to either VOI (55 participants) or non-VOI (54 people). Then, we computed their decisions based on the $(\beta,\kappa)$-estimates obtained by \cite{vanleeuwen23}, assuming that subjects in the non-VOI treatment where completely unaware of the possibility of role reversal. Finally, we estimated the effect of the VOI treatment on the decision to sell, through the specifications presented in Section \ref{sec:hyptest}. 
%\textbf{[comment more on this?]} 
The results of this exercise indicate that the chosen payoffs generate behavioral variations that our main specification is able to detect a significant effect (at a 5\% confidence level) 75\% of the time.

\begin{table}
\begin{center}
\caption{Payoff combinations $(e_1,e_2,e_1+G,e_2-L)$ (first four columns) and (approximate) threshold values for $\beta$ and $\kappa$ as per equations \eqref{betabar} and \eqref{kappabar} (last two columns)}
\begin{tabular}{cccccccc} \label{tab:thresholds}
\small
Payoff number ($s$) & $e_1$  & $e_2$  & $e_1+G$  & $e_2-L$  & $z=G/(G+L)$ & $(z-\beta_i)/(1-z)$                      \\ \hline
1&  150 & 100 & 165 & 90  & 0.6         & $1.5 - \beta_i \cdot 2.5$   \\
2&  150 & 100 & 160 & 90  & 0.5         & $1 - \beta_i \cdot    2$\\
3&  150 & 100 & 165 & 80  & 0.43         & $0.75 - \beta_i  \cdot 1.75$   \\
4&  150 & 100 & 165 & 70  & 0.33         & $0.5 - \beta_i \cdot 1.5$   \\
5&  250 & 240 & 300 & 100 & 0.26         & $0.36 - \beta_i \cdot 1.36$   \\
6&  250 & 240 & 300 & 90  & 0.25         & $0.33 - \beta_i \cdot 1.33$   \\
7&  250 & 240 & 300 & 80  & 0.24         & $0.31 - \beta_i \cdot 1.31$   \\
8&  250 & 240 & 300 & 70  & 0.23         & $0.29 - \beta_i \cdot 1.29$   \\
9&  250 & 240 & 300 & 60  & 0.22         & $0.28 - \beta_i \cdot 1.28$   \\
10&  150 & 120 & 170 & 20  & 0.17         & $0.2 - \beta_i \cdot 1.2$   \\
11&  200 & 190 & 220 & 60  & 0.13         & $0.15 - \beta_i \cdot 1.15$   \\
12&  200 & 190 & 220 & 50  & 0.125         & $0.14 - \beta_i \cdot 1.14$   \\
13&  200 & 190 & 210 & 100  & 0.1         & $0.11 - \beta_i \cdot 1.11$   \\
14&  200 & 190 & 210 & 90  & 0.09         & $0.1 - \beta_i \cdot 1.1$   \\
15&  200 & 190 & 210 & 80  & 0.08         & $0.09 - \beta_i \cdot 1.09$   \\
16&  250 & 220 & 265 & 20  & 0.07         & $0.08 - \beta_i \cdot 1.08$   \\
17&  250 & 220 & 260 & 50  & 0.06         & $0.06 - \beta_i \cdot 1.06$   \\
18&  250 & 220 & 260 & 10  & 0.05         & $0.05 - \beta_i \cdot 1.05$   \\
19&  250 & 220 & 260 & 5   & 0.04         & $0.05 - \beta_i \cdot 1.05$   \\
20&  250 & 220 & 255 & 60  & 0.03         & $0.03 - \beta_i \cdot 1.03$   \\
\hline
\end{tabular}
\end{center}
\end{table}

\usepgfplotslibrary{fillbetween}
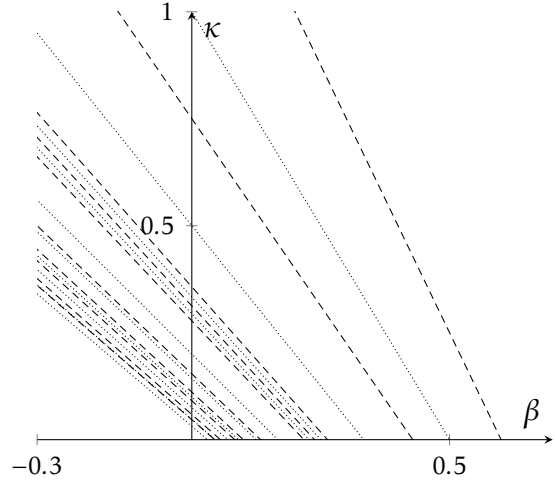
\begin{figure}[ht]
  \captionsetup{width=.9\linewidth}
\begin{center}
	\begin{tikzpicture}[scale=1]
	\begin{axis}[
	ticklabel style = {font=\footnotesize},
	 legend style={font=\footnotesize},
	axis lines=center,
	grid = none,
	ymin=0,
	ymax=1,
    xtick={-1,-0.3,0.5,1},
    ytick={0,0.5,1},
	xmin = -0.3,
	xmax=0.7,
%	extra y ticks={0},
	xlabel={$\beta$},
	ylabel={$\kappa$}]

%the lines have been put in the order in which they appear in the figure, from right to left
 	\addplot[smooth,densely dashed,name path=C] {(1.5-2.5*x}; % actual curve
% 	\addlegendentry{No. 3}
	\addplot[smooth,densely dotted,name path=A] {1-2*x}; % actual curve
%	\addlegendentry{No. 1}
 	\addplot[smooth,densely dashed,name path=B] {0.75-1.75*x}; % actual curve
% 	\addlegendentry{No. 2}
  	\addplot[smooth,densely dotted,name path=D] {0.5-1.5*x}; % actual curve
% 	\addlegendentry{No. 4}
 	\addplot[smooth,densely dashed,name path=F] {(5-19*x)/14}; % actual curve
% 	\addlegendentry{No. 16}
 	\addplot[smooth,densely dotted,name path=F] {(5-20*x)/15}; % actual curve
% 	\addlegendentry{No. 17}
 	\addplot[smooth,densely dashed,name path=F] {(5-21*x)/16}; % actual curve
% 	\addlegendentry{No. 18}
 	\addplot[smooth,densely dotted,name path=F] {(5-22*x)/17}; % actual curve
% 	\addlegendentry{No. 19}
 	\addplot[smooth,densely dashed,name path=F] {(5-23*x)/18}; % actual curve
% 	\addlegendentry{No. 20}
  	\addplot[smooth,densely dotted,name path=E] {0.2-1.2*x}; % actual curve
% 	\addlegendentry{No. 5}
  	\addplot[smooth,densely dashed,name path=F] {(2-15*x)/13}; % actual curve
% 	\addlegendentry{No. 9}
  	\addplot[smooth,densely dotted,name path=F] {(2-16*x)/14}; % actual curve
% 	\addlegendentry{No. 10}
  	\addplot[smooth,densely dashed,name path=F] {(1-10*x)/9}; % actual curve
% 	\addlegendentry{No. 8}
  	\addplot[smooth,densely dotted,name path=F] {0.1-1.1*x}; % actual curve
% 	\addlegendentry{No. 6}
  	\addplot[smooth,densely dashed,name path=F] {(1-12*x)/11}; % actual curve
% 	\addlegendentry{No. 7}
  	\addplot[smooth,densely dotted,name path=F] {(1.5-21.5*x)/20}; % actual curve
% 	\addlegendentry{No. 13}
  	\addplot[smooth,densely dashed,name path=F] {(1-18*x)/17}; % actual curve
% 	\addlegendentry{No. 11}
  	\addplot[smooth,densely dotted,name path=F] {(1-22*x)/21}; % actual curve
% 	\addlegendentry{No. 14}
  	\addplot[smooth,densely dashed,name path=F] {(1-22.5*x)/21.5}; % actual curve
% 	\addlegendentry{No. 15}
  	\addplot[smooth,densely dotted,name path=F] {(0.5-16.5*x)/16}; % actual curve
% 	\addlegendentry{No. 12}

%	\addplot[draw=none,name path=Z] {-2};     % “fictional” curve
%	\addplot[MyGray,fill opacity=0.2] fill between[of=A and Z,soft clip={domain=0:1}]; % filling
% 	\addplot[MyGray,fill opacity=0.2] fill between[of=B and Z,soft clip={domain=0:1}]; % filling
% 	\addplot[MyGray,fill opacity=0.2] fill between[of=C and Z,soft clip={domain=0:1}]; % filling
% 	\addplot[MyGray,fill opacity=0.2] fill between[of=D and Z,soft clip={domain=0:1}]; % filling
% 	\addplot[MyGray,fill opacity=0.2] fill between[of=E and Z,soft clip={domain=0:1}]; % filling
% 	\addplot[MyGray,fill opacity=0.2] fill between[of=F and Z,soft clip={domain=0:1}]; % filling
	\end{axis}
	\end{tikzpicture}

\vspace{0.2cm}
\begin{minipage}{0.5\textwidth} % choose width suitably
\caption{\label{fig:unaw_lemon} Each line represents the threshold value $\bar{\kappa}(\beta)$ for one of the 20 payoffs.}
\end{minipage}
	\end{center}

    \end{figure}

In the analysis above we assumed that in a non-VOI decision situation a subject is completely unaware of the complete game tree. However, some subjects may in fact understand that they could have been in their counterpart's place with some non-null probability. The following equation shows the utility of a subject $i$ who assigns a probability $\hat{p}_i$ that she was cast in the Seller role. 
    \begin{eqnarray}
    \label{eq:prefsnonVOIpartial}
\hat{U}_{i}\left( x,y\right) &=&
(1-\kappa_i) \cdot 
\Big[\hat{p}_i \cdot (e_1+x G)
+(1-\hat{p}_i) \cdot 
(e_2-y L) \Big] \\ 
&+&\kappa_i \cdot 
\Big[\hat{p}_i \cdot (e_1+x G)+(1-\hat{p}_i) \cdot 
(e_2-x L) \Big] \nonumber\\ 
&-& \hat{p}_i\cdot \beta _{i}\cdot[e_1-e_2+x(G+L)]
 \nonumber\\
&-&(1-\hat{p}_i)\cdot \alpha_{i}\cdot [e_1-e_2 +y (G+L) ] \nonumber
\end{eqnarray}
The expressions in \eqref{eq:prefsVOI} and \eqref{eq:prefsnonVOI} correspond to the special cases $\hat{p}_i=1/2$ and $\hat{p}_i=1$, respectively. We will say that a subject with $\hat{p}_i=1$ is \textit{fully unaware}, that one with $\hat{p}_i \in (1/2,1)$ is \textit{partially aware}, and that one with $\hat{p}_i =1/2$ is \textit{fully aware} of the full game tree. 

A partially aware subject switches from the selfish option in the non-VOI decision situation to the \textit{status quo} option in the VOI decision situation if
\begin{equation} \label{condition:partial}
	 \beta_{i}< \dfrac{G+\frac{1-\hat{p}_i}{\hat{p}_i}\kappa_i L }{G+L}
\end{equation}
and 
\begin{equation} \label{condition:partial2}
\kappa_i > \frac{G-\beta_{i} (G+L)}{L}.
\end{equation}
Our experiment is based on the premise that in the non-VOI decisions at least some subjects are partially aware or fully unaware, i.e., they hold some belief $\hat{p}>1/2$, since a fully aware subject would never switch (i.e., conditions \eqref{condition:partial} and \eqref{condition:partial2} are incompatible). We deliberately chose not to elicit the subjects' beliefs $\hat{p}$: eliciting them prior to the presentation of the decision situations might have affected the choices, and eliciting them afterwards would make no sense since the subjects would then have been exposed already to VOI decisions, in which they are told explicitly that both role distributions are equally likely. 

\subsection{Treatments}

Each subject is asked to state his/her choice in two sequences, both of which consist of the same set of 20 payoff configurations, listed in Table \ref{tab:thresholds}. Each sequence of 20 decisions is either entirely non-VOI or entirely VOI.

Clearly, to avoid raising awareness of the arbitrariness of the role distribution in the non-VOI decision situations, subjects should be presented a non-VOI sequence before being exposed to a VOI sequence. However, always letting a VOI sequence be preceded by a non-VOI one may lead to anchoring and/or fatigue effects. To deal with these issues, and to enable meaningful comparisons between the Neutral and the Market frames, we adopt the following four treatments:
\begin{enumerate}
\item \textbf{Neutral (N)}: Non-VOI + VOI --- neutral frame 

\item \textbf{Market (M)}: Non-VOI + VOI --- market frame 

\item \textbf{Mixed Frame A (A)}: VOI neutral frame + VOI market frame 

\item \textbf{Mixed Frame B (B)}: VOI market frame + VOI neutral frame  
\end{enumerate}

These treatments allow us to carry out a large number of comparisons: 
%First, we will examine the occurrence of switching from choosing $Y$ in neutrally worded non-VOI decisions to choosing $X$ in neutrally worded VOI decisions, by conducting 
(1) a within-subject comparison of decisions in the non-VOI and VOI sequences in the \textbf{Neutral} treatment; 
(2) a within-subject comparison of decisions in the non-VOI and VOI sequences in the \textbf{Market} treatment; 
(3) a between-subject comparison of decisions in the non-VOI sequence in the \textbf{Neutral} treatment and the first sequence in the \textbf{Mixed Frame A} treatment; 
(4) a between-subject comparison of decisions in the non-VOI sequence in the \textbf{Market} treatment and the first sequence in the \textbf{Mixed Frame B} treatment; 
and (5) within-subject comparisons of decisions in differently framed VOI sequences, using treatments \textbf{Mixed Frame A} and \textbf{Mixed Frame B}.

For further use below, let $D$ denote the set of eight decision sequences in our four treatments: $D=\{N1, N2, M1, M2, A1, A2, B1, B2\}$, where a 1 refers to the first sequence and a 2 to the second sequence of 20 decisions in the treatment.

\subsection{Procedures and data}

We conducted the experiment between November 2021 and March 2022 at the Toulouse School of Economics TSE Lab for Experimental Social Sciences. The software oTree was used to program the experiment, and participants were recruited via email using the Laboratory's participant pool (people who had signed up to be informed of laboratory experiments). Overall, we recruited 453 participants, all of whom  participated in only one session. 
Tables \ref{tab:subj_desc_table_age} and \ref{tab:subj_desc_table_treatment_age} in Appendix \ref{appendixA} provide information on the total number of subjects in each treatment and associated summary statistics. The experiment (the design and the empirical analysis reported below) was pre-registered on aspredicted.org on November 21st 2021, the first session taking place on November 22nd.

Each session was allocated to one treatment, and lasted between 25 and 45 minutes.
After being randomized to the lab booths, participants read the instructions on their desktop computer screens, were allowed to ask questions privately, and completed a comprehension test. Then the two sequences of 20 decisions were presented, upon which the participants filled out a post-experiment questionnaire (with questions on sex, age, nationality, and field of study). The English version of the instructions are included in Appendix \ref{app:instructions} (the experiment was conducted in French). In each decision situation the participants had to answer questions about the payoffs correctly before being allowed to state their decision (see an example of the two screens shown for one decision in Appendix \ref{app:last}).  

The participants' payoffs were determined based on two randomly drawn matches, one from each sequence. Each unit of the payoffs used in the instructions (see Table \ref{tab:thresholds}) was converted to 2.5 eurocents. For VOI decisions, for the randomly drawn match the role distribution was determined randomly, and the payoffs were calculated based on the active subject's decision. For non-VOI decisions, for the randomly drawn match each subject received the payoff corresponding to their decision, and also the payoff corresponding to the other's decision (of which they were not aware during the experiment).
Participants who answered all the questions received a show-up fee of 4 euros. Together with the payoffs obtained from the tasks, participants earned on average 15.5 euros. 
Participants privately received their payoffs in cash, upon which they left the premises. 

Table \ref{tab:total_sells_frame} below and Figures \ref{Fig:frame}-\ref{Fig:voiandmarket} in Appendix \ref{appendixA} give a first glimpse of the data, here pooled for all the $n=453$ participants' decisions across all the treatments and sessions. In the figures we define one observation as the number of selfish options selected by one participant in one sequence of 20 decisions (there are thus two observations per participant). Figure \ref{Fig:frame} shows the observations from the Neutral frame on the left and those from the Market frame on the right. The dashed (resp. solid) horizontal lines show the average (resp. median) number of selfish choices. It appears that, overall, the participants were more prone to select the selfish option in the Market than in the Neutral frame. 
Figure \ref{Fig:voi} compares the cumulative distributions of the observations in the VOI decision sequences (solid line, $n=686$) and the observations in the non-VOI decision sequences (dashed line, $n=220$). Participants were more prone to select the selfish option in the non-VOI than in the VOI decision situations.
Figure \ref{Fig:voiandmarket} confirms these two tendencies by splitting the observations into the four different frame combinations: Neutral VOI (black line), Market VOI (red line), Neutral non-VOI (blue line), and Market non-VOI (green line). 
Finally, Table \ref{tab:total_sells_frame} provides the summary statistics on the number of selfish decisions: in the first two rows for the Neutral and the Market frames, followed by two rows for the VOI and non-VOI frames, and finally the four different frame combinations. We see that the differences in the total number of selfish decisions between the Neutral and the Market frame, and between the VOI and non-VOI frame, are significantly different from zero.

\begin{table}[ht!]

	\centering

	\begin{threeparttable}

		\caption{Total number of selfish decisions per sequence of 20 decisions, by frame}
		\label{tab:total_sells_frame}

		\begin{tabular}{lccccccc}

			\hline

			Frame & N & Mean & Median & Min & Max & Q1 & Q3\\

			\hline

			Neutral &449 &7.08& 6& 0& 20& 2& 10 \\
			Market &457& 9.03& 9& 0& 20& 4 &13 \\

			\hline

			t-test and Wilcoxon test p-values &&0.00&  0.00&& \\

						\hline

						\hline

						VOI & 686 & 7.35 & 7 & 0 & 20 & 3 & 10\\
						non-VOI & 220 & 10.30 & 10 & 0 & 20 & 6 & 15\\

						\hline

						t-test and Wilcoxon test p-values & &  0.00  &0.00 & & & & \\
	
      \hline
      
      \hline
	
    VOI and Neutral & 341 & 6.52 & 6 & 0 & 20 & 2 & 9 \\
				VOI and Market & 345 & 8.17 & 8 & 0 & 20 & 4 & 11 \\
				non-VOI and Neutral & 108 & 8.86 & 8 & 0 & 20 & 4 & 13 \\
				non-VOI and Market & 112 & 11.7 & 11 & 0 & 20 & 8 & 17 \\
				\hline

		\end{tabular}
	\end{threeparttable}

\end{table}
\section{Hypothesis testing}
\label{sec:hyptest}

In this section we perform regressions to test whether the Market and VOI frames have significant effects on behavior in the experiment. 

% It is easy to verify that for an individual with a given preference type $(\beta_i,\kappa_i)$ who takes both non-VOI and VOI decisions, our theory predicts one of three scenarios. The following two are uninteresting: (A) the individual chooses the selfish option under both VOI and non-VOI for all $z \in Z$ (this requires $\beta_i \leq 0.03$ and $\kappa_i \leq 0.03-1.03 \beta_i$); (B) the individual chooses the \textit{status quo} option for all $z \in Z$ (this requires $\beta_i \geq 0.6$). Turning now to the interesting scenario, for preference types such that either $0.03< \beta_i < 0.6$, or $\beta_i \leq 0.03$ and $\kappa_i > 0.03-1.03\beta_i$, there exists at least one value $z \in Z$ such that the individual switches from the selfish option under non-VOI to the \textit{status quo} option under VOI. Furthermore, 

%In order to precisely describe our hypothesis tests, we introduce here some additional notation. 

% \subsection{Does the Market frame significantly affect behavior? (logit)}

% As per our model, the willingness to sell should increase with $z$ (both under VOI and under non-VOI). Would it thus make sense to use logit regression to evaluate impact of market treatment? 

% \begin{equation}
%     p(x) = \frac{1}{1+e^{-[\beta_0+\beta_0'M_{izt}+(\beta_1+\beta_1'M_{izt})z]}}
% \end{equation}

% And then test whether $\beta_0'=0$ and/or $\beta_1'=0$?

\subsection{Does the Market frame significantly affect behavior?}

To assess the effects of the Market frame we estimate linear models of the form: 
\begin{equation}\label{eq:modelmkt1}
    {x}_{izd}=\lambda_0 + \lambda_1 z+ \lambda_2 M_{izd} + \upsilon_{izd}
\end{equation}
% and 
% \begin{equation}\label{eq:modelmkt2}
%     {x}_{izd}=\lambda_0  + \lambda_1 c+ \lambda_2 M_{izd} + \lambda_3 M_{izd}c + \upsilon_{izd},
% \end{equation}
where ${x}_{izd} \in\{0,1\}$ is the decision of subject $i$ in payoff configuration $z=G/(G+L)$ in sequence $d \in \{N1, N2, M1, M2, A1, A2, B1, B2\}$, where ${x}_{izd}=1$ if she chooses the selfish option and ${x}_{izd}=0$ otherwise; $M_{izd}$ is a dummy variable that takes value 1 if the frame is Market and 0 if the frame is Neutral; and $\upsilon_{izd}$ is a mean-zero random variable assumed to be uncorrelated with $z$ and $M_{izd}$. Our null hypothesis is then that $\lambda_2 =0$. Recalling also condition \eqref{betabar} for a (fully unaware subject $i$ to choose the selfish option in a non-VOI decision, our theory predicts that \textit{ceteris paribus} an increase in $z=G/(G+L)$ should lead to a higher propensity to choose the selfish option, and hence we conjecture that $\lambda_1>0$. Because the values of $z$ are not uniformly spaced, we also run regressions where instead of including $z$ as an independent variable we include payoff fixed effects.

\begin{table}[H]
 \hspace*{-1.8cm}  \centering
   \begin{threeparttable}[b]
      \caption{\label{reg_frame_bet_nVOI} Effect of Frame on Selling under non-VOI conditions - between subjects (N1-M1)}
      \begin{tabular}{lcccccc}
         \tabularnewline \midrule \midrule
         Dependent Variable: & \multicolumn{6}{c}{Sell (Yes or No)}\\
         Model:              & (1)            & (2)            & (3)            & (4)            & (5)            & (6)\\  
         \midrule
         \emph{Variables}\\
         Frame               & 0.1413$^{***}$ & 0.1413$^{***}$ & 0.1413$^{***}$ & 0.1413$^{***}$ & 0.1413$^{***}$ & 0.1425$^{***}$\\   
                             & (0.0149)       & (0.0144)       & (0.0396)       & (0.0149)       & (0.0396)       & (0.0403)\\   
         z                   &                & 0.8027$^{***}$ & 0.8027$^{***}$ &                &                &   \\   
                             &                & (0.0461)       & (0.0655)       &                &                &   \\   
         Socio-dem. controls & No             & No             & No             & No             & No             & Yes\\  
         Clustering of SE    & No             & No             & Subject        & No             & Subject        & Subject and Payoff\\  
         \midrule
         \emph{Fixed-effects}\\
         Payoff              &                &                &                & Yes            & Yes            & Yes\\  
         \midrule
         \emph{Fit statistics}\\
         Observations        & 4,400          & 4,400          & 4,400          & 4,400          & 4,400          & 4,400\\  
         R$^2$               & 0.01998        & 0.08323        & 0.08323        & 0.09530        & 0.09530        & 0.10272\\  
         Within R$^2$        &                &                &                & 0.02161        & 0.02161        & 0.02964\\  
         \midrule \midrule
         \multicolumn{7}{l}{\emph{Signif. Codes: ***: 0.01, **: 0.05, *: 0.1}}\\
      \end{tabular}
      
      \begin{tablenotes}\item Note: Column (3) includes clustering of SE's at the Subject level, column (4) includes Payoff fixed effects,
                                        column (5) includes Payoff fixed effects with clustering of SE's at the Subject level, column (6) includes Payoff fixed effects with clustering of the SE's at the Subject and Payoff level. 
                                      Socio-demographic controls are: nationality, gender, couple status, previous experience in experiments and attendance to a private school.
      \end{tablenotes}
   \end{threeparttable}
\end{table}

Overall, the results reported in Tables \ref{reg_frame_bet_nVOI}-\ref{reg_frame_with_VOI} show that subjects are more prone to selecting the selfish option under the Market than under the Neutral frame, that this effect is more pronounced in non-VOI than in VOI decisions, and that within-subject effects are stronger than between-subject effects in VOI decisions.

Specifically, Table \ref{reg_frame_bet_nVOI} shows between-subject effects of the Market frame in non-VOI decisions, based on the decisions in the first decision sequences of the Neutral and the Market treatments (N1 and M1). The Market frame reduces the propensity to select the selfish option, an effect which is consistent in magnitude at 14.13 percentage points, and consistently significant across specifications.

\begin{table}[H]
 \hspace*{-1.8cm}  \centering
   \begin{threeparttable}[b]
   \caption{\label{reg_frame_bet_VOI_A1B1} Effect of Frame on Selling under VOI conditions - between subjects (A1-B1)}
   \begin{tabular}{lcccccc}
      \tabularnewline \midrule \midrule
      Dependent Variable: & \multicolumn{6}{c}{Sell (Yes or No)}\\
      Model:              & (1)           & (2)            & (3)            & (4)      & (5)      & (6)\\  
      \midrule
      \emph{Variables}\\
      Frame               & 0.0319$^{**}$ & 0.0319$^{**}$  & 0.0319         & 0.0319   & 0.0319   & 0.0488\\   
                          & (0.0142)      & (0.0138)       & (0.0367)       & (0.0185) & (0.0368) & (0.0362)\\   
      z                   &               & 0.7354$^{***}$ & 0.7354$^{***}$ &          &          &   \\   
                          &               & (0.0442)       & (0.0546)       &          &          &   \\   
      Socio-dem. controls & No            & No             & No             & No       & No       & Yes\\  
      Clustering of SE    & No            & No             & Subject        & No       & Subject  & Subject and Payoff\\  
      \midrule
      \emph{Fixed-effects}\\
      Payoff              &               &                &                & Yes      & Yes      & Yes\\  
      \midrule
      \emph{Fit statistics}\\
      Observations        & 4,660         & 4,660          & 4,660          & 4,660    & 4,660    & 4,660\\  
      R$^2$               & 0.00107       & 0.05714        & 0.05714        & 0.06439  & 0.06439  & 0.08209\\  
      Within R$^2$        &               &                &                & 0.00114  & 0.00114  & 0.02004\\  
      \midrule \midrule
      \multicolumn{7}{l}{\emph{Signif. Codes: ***: 0.01, **: 0.05, *: 0.1}}\\
   \end{tabular}
   
  \begin{tablenotes}\item  Note: Column (3) includes clustering of SE's at the Subject level, column (4) includes Payoff fixed effects,
                                  column (5) includes Payoff fixed effects with clustering of SE's at the Subject level, column (6) includes Payoff fixed 
                                    effects with clustering of the SE's at the Subject and Payoff level. Socio-demographic controls are: nationality, gender, 
                                    couple status, previous experience in experiments and attendance to a private school.
  \end{tablenotes}
   \end{threeparttable}
\end{table}

Table \ref{reg_frame_bet_VOI_A1B1} also shows between-subject effects of the Market frame, but now in VOI decisions, based on regressions using only the first decision sequences of Mixed Frame A and Mixed Frame B (A1 and B1): these subjects would thus not have been exposed to the other frame. The effect of the Market frame is much smaller in magnitude than for the non-VOI decisions, and also insignificant as soon as subject-specific and/or payoff-specific effects are accounted for.

\begin{table}[htb!]
\hspace*{-1.8cm}   \centering
   \begin{threeparttable}[b]
      \caption{\label{reg_frame_with_VOI} Effect of Frame on Selling under VOI conditions - within subjects (A1-A2-B1-B2)}
      \begin{tabular}{lcccccc}
         \tabularnewline \midrule \midrule
         Dependent Variable: & \multicolumn{6}{c}{Sell (Yes or No)}\\
         Model:              & (1)            & (2)            & (3)            & (4)            & (5)            & (6)\\  
         \midrule
         \emph{Variables}\\
         Frame               & 0.0700$^{***}$ & 0.0700$^{***}$ & 0.0700$^{***}$ & 0.0700$^{***}$ & 0.0700$^{***}$ & 0.0700$^{***}$\\   
                             & (0.0100)       & (0.0096)       & (0.0181)       & (0.0096)       & (0.0181)       & (0.0082)\\   
         z                   &                & 0.8806$^{***}$ & 0.8806$^{***}$ &                &                &   \\   
                             &                & (0.0308)       & (0.0513)       &                &                &   \\   
         Socio-dem. controls & No             & No             & No             & No             & No             & Yes\\  
         Clustering of SE    & No             & No             & Subject        & No             & Subject        & Subject and Payoff\\  
         \midrule
         \emph{Fixed-effects}\\
         Sub.                &                &                & Yes            &                &                & \\  
         Payoff              &                &                &                & Yes            & Yes            & \\  
         Sub.-Payoff         &                &                &                &                &                & Yes\\  
         \midrule
         \emph{Fit statistics}\\
         Observations        & 9,320          & 9,320          & 9,320          & 9,320          & 9,320          & 9,320\\  
         R$^2$               & 0.00517        & 0.08557        & 0.33028        & 0.09207        & 0.09207        & 0.66812\\  
         Within R$^2$        &                &                & 0.11329        & 0.00567        & 0.00567        & 0.01535\\  
         \midrule \midrule
         \multicolumn{7}{l}{\emph{Signif. Codes: ***: 0.01, **: 0.05, *: 0.1}}\\
      \end{tabular}
      
      \begin{tablenotes}\item Note: Column (3) includes clustering of SE's at the Subject level, column (4) includes Payoff fixed effects,
                                        column (5) includes Payoff fixed effects with clustering of SE's at the Subject level, column (6) includes Payoff fixed effects with clustering of the SE's at the Subject and Payoff level. Socio-demographic controls are: nationality, gender, couple status, previous experience in experiments and attendance to a private school.
      \end{tablenotes}
   \end{threeparttable}
\end{table}

Finally, Table \ref{reg_frame_with_VOI} shows within-subject effects of the Market frame in VOI decisions, based on decisions 
in the mixed treatments $A$ and $B$. Framing the decision as a market interaction on average increases the likelihood of selecting the selfish option by 7 percentage points, and the significance level is consistently high across specifications. Although large, this effect is about half as large as for the non-VOI decisions.

Note that, as expected, the estimate of $\lambda_1$ is positive and highly significant (see columns (2) and (3) in the tables), and that its inclusion substantially increases R$^2$. The specifications which instead use ``payoff fixed effects'' yield similar results.

%\clearpage

%\input{Figures/Tables/reg_frame_bet_VOI.tex}

%Effect of frame under non-VOI conditions, between subjects:   $t\in\{N1,M1\}$.

%Effect of frame under VOI conditions, within subjects: $t\in\{A1,A2\}$ and $t\in\{B1,B2\}$

%\input{Figures/Tables/reg_frame_with_VOI_A.tex}
%\input{Figures/Tables/reg_frame_with_VOI_B.tex}

\subsection{Does the VOI frame significantly affect behavior?}

To evaluate the effects of the VOI frame, we estimate linear models of the form:
\begin{equation}
    {x}_{izd}=\gamma_0 + \gamma_1 z+ \gamma_2V_{izd} + \epsilon_{izd}.
\end{equation}
%and
%\begin{equation}
%     {x}_{izd}=\gamma_0 + \gamma_1 c+ \gamma_2 V_{izd} + \gamma_3 V_{izd}c + \epsilon_{izd},
% \end{equation}
where $V_{izd}$ is a dummy variable that equals 1 if the decision situation is VOI and 0 if the decision is non-VOI, and $\epsilon_{izd}$ is a mean-zero random variable, assumed to be uncorrelated with $c$ and $V_{izd}$. Our null hypothesis is then that $\gamma_2 =0$. The results, reported in Tables \ref{reg_VOI_with_neutral}-\ref{reg_VOI_bet_mkt}, show that this hypothesis is clearly rejected, for both within- and between-subject analyses, and that the effect of VOI is large.

\begin{table}[htb!]
\hspace*{-1.8cm}   \centering
   \begin{threeparttable}[b]
      \caption{\label{reg_VOI_with_neutral} Effect of VOI on Selling under Neutral frame - within subjects (N1-N2)}
      \begin{tabular}{lcccccc}
         \tabularnewline \midrule \midrule
         Dependent \newline Variable: & \multicolumn{6}{c}{Sell (Yes or No)}\\
         Model:           & (1)             & (2)             & (3)             & (4)             & (5)             & (6)\\  
         \midrule
         \emph{Variables}\\
         VOI              & -0.1528$^{***}$ & -0.1528$^{***}$ & -0.1528$^{***}$ & -0.1528$^{***}$ & -0.1528$^{***}$ & -0.1528$^{***}$\\   
                          & (0.0145)        & (0.0140)        & (0.0277)        & (0.0140)        & (0.0278)        & (0.0122)\\   
         z                &                 & 0.7685$^{***}$  & 0.7685$^{***}$  &                 &                 &   \\   
                          &                 & (0.0447)        & (0.0828)        &                 &                 &   \\   
         Clustering of SE & No              & No              & Subject         & No              & Subject         & Subject*Payoff\\  
         \midrule
         \emph{Fixed-effects}\\
         Sub.             &                 &                 & Yes             &                 & Yes             & \\  
         Payoff           &                 &                 &                 & Yes             & Yes             & \\  
         Sub.-Payoff      &                 &                 &                 &                 &                 & Yes\\  
         \midrule
         \emph{Fit statistics}\\
         Observations     & 4,320           & 4,320           & 4,320           & 4,320           & 4,320           & 4,320\\  
         R$^2$            & 0.02513         & 0.08750         & 0.32862         & 0.09302         & 0.33415         & 0.65531\\  
         Within R$^2$     &                 &                 & 0.11530         & 0.02696         & 0.03637         & 0.06795\\  
         \midrule \midrule
         \multicolumn{7}{l}{\emph{Signif. Codes: ***: 0.01, **: 0.05, *: 0.1}}\\
      \end{tabular}
      
      \begin{tablenotes}\item Note: Column (3) includes Subject fixed effects with clustering of SE's at the Subject level, column (4) includes Subject and Payoff fixed effects with clustering of SE's at the Subject level,
                                        column (5) includes Subject and Payoff fixed effects with clustering of SE's at the Subject level, column (6) includes Subject by Payoff fixed effects with clustering of the SE's at the Subject by Payoff level.
      \end{tablenotes}
   \end{threeparttable}
\end{table}

Beginning with the neutral frame, Table \ref{reg_VOI_with_neutral} shows the results based on a within-subjects comparison  ($d\in\{N1,N2\})$, while Table \ref{reg_VOI_bet_neutral} reports results from the between-subjects comparison ($d\in\{N1,A1\}$). Independent of the specification, the effect is highly significant, and on average the VOI frame reduces the likelihood of selecting the selfish option by 15.28 percentage points.\footnote{This indicates that the within-subjects treatment does not suffer too much from subjects' desire to act consistently in the VOI and the non-VOI decision situations with the same payoff configuration.} The between-subjects effect is smaller, at 7.45 percentage points for most specifications, and also less robust in terms of significance across the specifications.

\begin{table}[htb!]
\hspace*{-1.8cm}   \centering
   \begin{threeparttable}[b]
      \caption{\label{reg_VOI_bet_neutral} Effect of VOI on Selling under Neutral frame - between subjects (N1-A1)}
      \begin{tabular}{lcccccc}
         \tabularnewline \midrule \midrule
         Dependent Variable: & \multicolumn{6}{c}{Sell (Yes or No)}\\
         Model:              & (1)             & (2)             & (3)            & (4)             & (5)           & (6)\\  
         \midrule
         \emph{Variables}\\
         VOI                 & -0.0745$^{***}$ & -0.0745$^{***}$ & -0.0745$^{*}$  & -0.0745$^{***}$ & -0.0745$^{*}$ & -0.0728$^{*}$\\   
                             & (0.0145)        & (0.0141)        & (0.0395)       & (0.0143)        & (0.0396)      & (0.0403)\\   
         z                   &                 & 0.6780$^{***}$  & 0.6780$^{***}$ &                 &               &   \\   
                             &                 & (0.0451)        & (0.0586)       &                 &               &   \\   
         Socio-dem. controls & No              & No              & No             & No              & No            & Yes\\  
         Clustering of SE    & No              & No              & Subject        & No              & Subject       & Subject and Payoff\\  
         \midrule
         \emph{Fixed-effects}\\
         Payoff              &                 &                 &                & Yes             & Yes           & Yes\\  
         \midrule
         \emph{Fit statistics}\\
         Observations        & 4,580           & 4,580           & 4,580          & 4,580           & 4,580         & 4,580\\  
         R$^2$               & 0.00574         & 0.05257         & 0.05257        & 0.05625         & 0.05625       & 0.06045\\  
         Within R$^2$        &                 &                 &                & 0.00604         & 0.00604       & 0.01047\\  
         \midrule \midrule
         \multicolumn{7}{l}{\emph{Signif. Codes: ***: 0.01, **: 0.05, *: 0.1}}\\
      \end{tabular}
      
      \begin{tablenotes}\item Note: Column (3) includes clustering of SE's at the Subject level, column (4) includes Payoff fixed effects,
                                        column (5) includes Payoff fixed effects with clustering of SE's at the Subject level, column (6) includes Payoff fixed effects with clustering of the SE's at the Subject and Payoff level. Socio-demographic controls are: nationality, gender, couple status, previous experience in experiments and attendance to a private school.
      \end{tablenotes}
   \end{threeparttable}
\end{table}

One explanation for the difference between the within- and the between-subjects effect of VOI appears to be heterogeneity across subjects, as indicated by the fact that the between-subjects  effect is only weakly significant for the specifications which cluster the standard errors at the subject level.

Turning now to the market frame, Tables \ref{reg_VOI_with_mkt} and  \ref{reg_VOI_bet_mkt} report, respectively, how VOI affects the propensity to choose Sell within subjects ($d \in\{M1,M2\}$) and between subjects ($d\in\{M1,B1\}$). Qualitatively, the within-subjects results are similar to those in the neutral frame (see Table \ref{reg_VOI_with_neutral}): the effect is highly significant and identical in size across all specifications. However, the effect is even stronger than in the neutral frame: the VOI wording reduces the propensity to sell by 18.66 percentage points. Contrary to the neutral frame, here the between-subjects effect is similar in size to the within-subjects effect, and it is also robustly significant across the specifications. Overall there is thus a large and robust effect of the VOI wording on the propensity to choose Sell in the market frame.

\begin{table}[htb!]
\hspace*{-1.8cm}   \centering
   \begin{threeparttable}[b]
      \caption{\label{reg_VOI_with_mkt} Effect of VOI on Selling under Market frame - within subjects (M1-M2)}
      \begin{tabular}{lcccccc}
         \tabularnewline \midrule \midrule
         Dependent Variable: & \multicolumn{6}{c}{Sell (Yes or No)}\\
         Model:           & (1)             & (2)             & (3)             & (4)             & (5)             & (6)\\  
         \midrule
         \emph{Variables}\\
         VOI              & -0.1866$^{***}$ & -0.1866$^{***}$ & -0.1866$^{***}$ & -0.1866$^{***}$ & -0.1866$^{***}$ & -0.1866$^{***}$\\   
                          & (0.0147)        & (0.0138)        & (0.0265)        & (0.0138)        & (0.0265)        & (0.0112)\\   
         z                &                 & 1.042$^{***}$   & 1.042$^{***}$   &                 &                 &   \\   
                          &                 & (0.0442)        & (0.0913)        &                 &                 &   \\   
         Clustering of SE & No              & No              & Subject         & No              & Subject         & Subject*Payoff\\  
         \midrule
         \emph{Fixed-effects}\\
         Sub.             &                 &                 & Yes             &                 & Yes             & \\  
         Payoff           &                 &                 &                 & Yes             & Yes             & \\  
         Sub.-Payoff      &                 &                 &                 &                 &                 & Yes\\  
         \midrule
         \emph{Fit statistics}\\
         Observations     & 4,480           & 4,480           & 4,480           & 4,480           & 4,480           & 4,480\\  
         R$^2$            & 0.03483         & 0.14142         & 0.36707         & 0.15022         & 0.37586         & 0.72045\\  
         Within R$^2$     &                 &                 & 0.18263         & 0.03938         & 0.05286         & 0.11080\\  
         \midrule \midrule
         \multicolumn{7}{l}{\emph{Signif. Codes: ***: 0.01, **: 0.05, *: 0.1}}\\
      \end{tabular}
      
      \begin{tablenotes}\item Note: Column (3) includes Subject fixed effects with clustering of SE's at the Subject level, column (4) includes Subject and Payoff fixed effects with clustering of SE's at the Subject level,
                                        column (5) includes Subject and Payoff fixed effects with clustering of SE's at the Subject level, column (6) includes Subject by Payoff fixed effects with clustering of the SE's at the Subject by Payoff level.
      \end{tablenotes}
   \end{threeparttable}
\end{table}

\begin{table}[htb!]
\hspace*{-1.8cm}   \centering
   \begin{threeparttable}[b]
      \caption{\label{reg_VOI_bet_mkt} Effect of VOI on Selling under Market frame - between subjects (M1-B1)}
      \begin{tabular}{lcccccc}
         \tabularnewline \midrule \midrule
         Dependent Variable: & \multicolumn{6}{c}{Sell (Yes or No)}\\
         Model:              & (1)             & (2)             & (3)             & (4)             & (5)             & (6)\\  
         \midrule
         \emph{Variables}\\
         VOI                 & -0.1839$^{***}$ & -0.1839$^{***}$ & -0.1839$^{***}$ & -0.1839$^{***}$ & -0.1839$^{***}$ & -0.1879$^{***}$\\   
                             & (0.0147)        & (0.0141)        & (0.0367)        & (0.0220)        & (0.0368)        & (0.0402)\\   
         z                   &                 & 0.8602$^{***}$  & 0.8602$^{***}$  &                 &                 &   \\   
                             &                 & (0.0451)        & (0.0609)        &                 &                 &   \\   
         Socio-dem. controls & No              & No              & No              & No              & No              & Yes\\  
         Clustering of SE    & No              & No              & Subject         & No              & Subject         & Subject and Payoff\\  
         \midrule
         \emph{Fixed-effects}\\
         Payoff              &                 &                 &                 & Yes             & Yes             & Yes\\  
         \midrule
         \emph{Fit statistics}\\
         Observations        & 4,480           & 4,480           & 4,480           & 4,480           & 4,480           & 4,480\\  
         R$^2$               & 0.03384         & 0.10642         & 0.10642         & 0.11804         & 0.11804         & 0.14127\\  
         Within R$^2$        &                 &                 &                 & 0.03695         & 0.03695         & 0.06231\\  
         \midrule \midrule
         \multicolumn{7}{l}{\emph{Signif. Codes: ***: 0.01, **: 0.05, *: 0.1}}\\
      \end{tabular}
      
      \begin{tablenotes}\item Note: Column (3) includes clustering of SE's at the Subject level, column (4) includes Payoff fixed effects,
                                        column (5) includes Payoff fixed effects with clustering of SE's at the Subject level, column (6) includes Payoff fixed effects with clustering of the SE's at the Subject and Payoff level. Socio-demographic controls are: nationality, gender, couple status, previous experience in experiments and attendance to a private school.
      \end{tablenotes}
   \end{threeparttable}
\end{table}

As was the case for the regressions measuring the effects of the Market frame, the estimate of the coefficient for $z$ is positive and highly significant (see columns (2) and (3) in the tables), and its inclusion substantially increases R$^2$. Moreover, the results are similar in the specifications which instead use ``payoff fixed effects''.

%In sum, the within-subjects analyses show that VOI has a robust and economically significant negative effect on the propensity to select the selfish option, in both  the neutral and the market frame, and the effect appears to be stronger in the market frame (this will be tested below). 

\clearpage

\subsection{Interaction effects}

The results reported above show that the VOI frame has a stronger absolute effect on the propensity to select the selfish option in the Market than in the Neutral frame. However, from the descriptive statistics we also know that subjects are on average more willing to choose the selfish option in the Market than in the Neutral frame. Here we test whether the relative effect of VOI is different under the two frames, by running regressions with a VOI-Market interaction, as follows: \begin{equation}
    {x}_{izd}=\gamma_0 + \gamma_1 z+ \gamma_2V_{izd} + \gamma_3V_{izd}\times M_{izd} + \epsilon_{izd}.
\end{equation}
The results are reported in Table \ref{reg_VOI_with_frame_bet} in Appendix \ref{appendixA}. 
The coefficient on the interaction VOI $\times$ Market is not significant, implying that, on average, the effect of VOI relative to non-VOI is the same in the Market as in the Neutral frame.

\section{Structural estimation of $\beta$ and $\kappa$}

Besides the reduced-form estimates of treatment effects specified above, we structurally estimate the aheadness aversion and Kantian morality parameters $\beta$ and $\kappa$. 

\subsection{Method}

We run the estimations using the data obtained from subjects exposed to both non-VOI and VOI sequences. We will first report the estimates of the preference parameters, $(\hat\beta,\hat\kappa)$, and the choice sensitivity, $\hat\sigma$, under the hypothesis that the observed data pooled over all the subjects (in a given treatment) had emanated from a representative agent. 
For non-VOI decisions we assume that any subject is fully unaware of the possible role reversal, i.e., we posit $\hat{p}=1$ (see \eqref{eq:prefsnonVOIpartial}). By contrast, for VOI decisions we assume that any subject is fully aware of the arbitrariness of the role distribution (by positing $\hat{p}=1/2$), since we explicitly inform the subjects about it. These assumptions are extreme, and we do not claim that they are in line with the subjects'  subjective beliefs about the role distribution. However, the advantage is that they imply a conservative estimate of $\hat\kappa$. Indeed, under the assumption that $\hat{p}=1$ in non-VOI decisions, only a positive $\hat\beta$ can explain why a subject would refrain from selecting the selfish option. Hence, for any subject who is in fact fully or partly aware of the role reversal in the non-VOI decisions (i.e., with $\hat{p}=1\in [1/2,1)$) and who refrains from the selfish option in such a decision due to a combination of aheadness attitude and Kantian morality (see \eqref{condition:partial}), our assumption that $\hat\kappa$ cannot be at work leads to over-estimation of $\hat\beta$ and under-estimation of $\hat\kappa$.

We use a standard random utility model, by assuming that the representative agent's true utility from using pure strategy $x$ is a random variable of the form
\begin{equation} \label{eq:RUvoi}
\tilde{U}(x;\beta,\kappa,\sigma,z) = 2U(x;\beta,\kappa,z) + \varepsilon_{x}
\end{equation}
in a VOI decision situation with payoff configuration $z \in Z$ and 
\begin{equation}
\tilde{V}(x;\beta,\sigma,z) = V(x;\beta,z) + \varepsilon_{x}
\end{equation}
in a non-VOI decision situation with payoff configuration $z \in Z$, where the function $U(x;\beta,\kappa,z)$ is that specified in \eqref{eq:prefsVOI}, the function $V(x;\beta,z)$ is that specified in \eqref{eq:prefsnonVOI},  
and $\varepsilon_{x}$ represents randomness in the utility evaluation, assumed to be independent across all the decisions. 
This random variable is taken to follow a type 1 extreme value distribution with scale parameter $1/\sigma$. The factor 2 in equation \eqref{eq:RUvoi} avoids that the magnitude of noise relative to $z$, $\beta$, and $\kappa$ be twice as large for VOI than for non-VOI decisions (recall the factor 1/2 in \eqref{eq:prefsVOI}). 

%For any subject $i$ we assume that  the aheadness attitude parameter $\beta_i$ is the same in both non-VOI and VOI sequences. Furthermore, for non-VOI decisions we assume that the subject is fully unaware of the possible role reversal, i.e., we posit $\hat{p}=1$ (see \eqref{eq:prefsnonVOIpartial}). By contrast, for VOI decisions we assume that the subject is fully aware of the arbitrariness of the role distribution (by positing $\hat{p}=1/2$), since we explicitly inform the subjects about it. These assumptions are extreme, and we do not claim that they are in line with the subjects'  subjective beliefs about the role distribution. However, the advantage is that they imply a conservative estimate of $\kappa_i$. Indeed, under the assumption that $\hat{p}=1$ in non-VOI decisions, only a positive $\beta_i$ can explain why a subject would refrain from selecting the selfish option. Hence, for any subject who is in fact fully or partly aware of the role reversal in the non-VOI decisions (i.e., with $\hat{p}=1\in [1/2,1)$) and who refrains from the selfish option in such a decision due to a combination of aheadness attitude and Kantian morality (see \eqref{condition:partial}), our assumption that $\kappa_i$ cannot be at work leads to an over-estimation of $\beta_i$.

In a VOI decision situation, the selfish option is selected if  
\begin{equation}
\tilde{U}(1;\beta,\kappa,\sigma,z)\geq \tilde{U}(0;\beta,\kappa,\sigma,z),
\end{equation}
or 
\begin{equation}
2 U\left(1;\beta,\kappa,z\right)-2 U\left(0;\beta,\kappa,z\right)\geq \varepsilon_{0}-\varepsilon_{1},
\end{equation}
which reduces to
\begin{equation}
G-\kappa L-\beta (G+L)\geq \varepsilon_{0}-\varepsilon_{1}.
\end{equation}
Likewise, in a non-VOI decision situation, the selfish option is selected if 
\begin{equation}
\tilde{V}(1;\beta,\sigma,z)\geq \tilde{V}(0;\beta,\sigma,z),
\end{equation}
or 
\begin{equation}
V\left(1;\beta,z\right)-V\left(0;\beta,z\right)\geq \varepsilon_{0}-\varepsilon_{1},
\end{equation}
which reduces to
\begin{equation}
G-\beta (G+L)\geq \varepsilon_{0}-\varepsilon_{1}.
\end{equation}
Letting $\nu$ be a dummy variable that takes the value 1 in a VOI decision situation and the value 0 in a non-VOI decision situation, it follows that the probability that the selfish option is selected in payoff configuration $z$ equals (McFadden, 1974):
\begingroup
\addtolength{\jot}{1em}
\begin{eqnarray}
\label{eq:sigmoid}
% & Pr\left(x_{i}=1;\beta_i,\kappa_i,\sigma_i,z,\nu\right)&=\nonumber \\
 \dfrac{1}{1+\exp\bigg[-\sigma\big[\nu \left[(1-\beta)G-(\beta + \kappa)L\right]+(1-\nu)\left[(1-\beta)G-\beta L\right]\big]\bigg]},
\end{eqnarray}
where $\sigma$ is the representative agent's choice sensitivity with respect to differences in deterministic utility. If $\sigma$ is close to 0, the agent chooses either option with a probability close to 1/2, independent of the difference in deterministic utilities. In contrast, arbitrarily large values of $\sigma$ indicate that the probability of choosing the option that results in the higher deterministic utility approaches 1. Noting that the expression in \eqref{eq:sigmoid} can be written as:
\begin{eqnarray}
H\big(\sigma\big[\nu \left[(1-\beta)G-(\beta + \kappa)L\right]+(1-\nu)\left[(1-\beta)G-\beta L\right]\big]\big),& \nonumber
\end{eqnarray}
\endgroup
where $H:\mathbb{R} \rightarrow (0,1)$ is the logistic function, the probability density function can be written
%We are now in a position to write each individual's contribution to the conditional density of the model:
\begingroup
\addtolength{\jot}{1em}
\begin{equation}
\label{eq:contribution}
\begin{aligned}
f\left(\bold{x},\beta,\kappa,\sigma\right)=&
\prod_{i\in I}\prod_{z_i\in Z}H\left(\sigma\left( U(x^{(z_i)}=1;\beta,\kappa,z_i)-U(x^{(z_i)}=0;\beta,\kappa,z_i)\right)\right)^{ \mathbbm{1}\{x^{(z_i)}=1\}}\\&
\times\left[1-H\left(\sigma\left( U(x^{(z_i)}=1;\beta,\kappa,z)-U(x^{(z_i)}=0;\beta,\kappa,z_i)\right)\right)\right]^{1-\mathbbm{1}\{x^{(z_i)}=1\}},
\end{aligned}
\end{equation}
\endgroup
where $I$ is the set of subjects, and $z_i$ refers to the situation in which subject $i$ faced payoff $z$: this notation addresses the fact that here all the subjects' decisions are pooled and treated as if taken by one single individual. In the expression, $\bold{x}$ is the vector of observed choices, and $\mathbbm{1}\{x^{(z_i)}=1\}$ is an indicator function that takes the value 1 if  $x^{(z_i)}=1$ and 0 if $x^{(z_i)}=0$.
Using (\ref{eq:contribution}), we obtain the log-likelihood
\begingroup
\addtolength{\jot}{1em}
\begin{equation}
\label{eq:loglikind}
\begin{aligned}
& \ln(L(\beta,\kappa,\sigma;\bold{x}))=
%\sum_{z\in Z}
\log (f\left(\bold{x},\beta,\kappa,\sigma\right)),
\end{aligned}
\end{equation}
\endgroup
and we will report the vector of estimates $(\hat\beta,\hat\kappa,\hat\sigma)$ that maximizes this log-likelihood. 

Second, we use the finite-mixture approach used by \citet*{bruhin2018many} and \citet*{vanleeuwen23} to evaluate whether the data is reasonably described by a finite set of estimated preference types, and if so, whether the Kantian moral concern appears relevant for most of them. This approach consists in estimating preference parameters for a given number $K$ of ``preference types''.
For each type $k =\{1,...,K\}$, we estimate the preference parameters $\left( \beta _{k}, \kappa _{k}\right)$ and the noise parameter $\sigma _{k}$.
The log-likelihood is then given by:%
\begin{equation}\label{eq:loglik_mixture}
    \ln(L(\boldsymbol{\beta},\boldsymbol{\kappa},\boldsymbol{\sigma};\bold{x})) = \ln \left( \sum_{k=1}^K \phi_k \cdot f( \beta_k,\kappa_k, \sigma_k; \bold{x}) \right),
\end{equation}%
where $\phi_k$ is the share of subjects of type $k$, estimated together with the preference and noise parameters. 

All structural estimations of the preference parameters $\beta$ and $\kappa$ are conducted both for the Neutral frame and for the Market frame, to examine whether any differences appear. Furthermore, we carry out all our estimation procedures for the full samples and for two subsets of the latter, which we deem ``Core 1'' and ``Core 2''. Our Core 1 samples consist of the 98 subjects in the Neutral treatment and 100 subjects in the Market treatment who made at least one decision switch between non-VOI and VOI. 
Recalling that our theoretical model predicts that a subject should be less inclined to select the selfish decision in VOI situations, we say that a subject switches in the expected direction if (s)he selects the selfish option under non-VOI and the \textit{status quo} option under VOI, and in the unexpected direction if (s)he selects the \textit{status quo} option under non-VOI and the selfish option under VOI. Thus, for our Core 2 samples we remove subjects who made at least as many unexpected as expected switches from the Core 1 samples. This leaves us with 68 subjects in the Neutral treatment and 79 subjects in the Market treatment. In the following section, all reported estimates are obtained using the Core 1 samples, while results with the full and Core 2 samples can be found in Appendix \ref{appendixA}.

\subsection{Results}

\subsubsection{Representative agent estimates}

Table \ref{tab:rep_agent_c1} shows the representative agent estimates of the preference parameters $\beta$ and $\kappa$, as well as that of the choice sensitivity parameter $\sigma$ for our Core 1 sample for the Neutral frame (first column) and the Market frame (second column).

\begin{table}[ht!]
    \begin{threeparttable}
    \centering
    \caption{\label{tab:rep_agent_c1} Estimated preferences for the representative agent in Neutral and Market frames - Core 1 sample}
    \begin{tabular}{lccc}
    \toprule
    \toprule
    & \multicolumn{1}{p{3cm}}{\centering \vspace{0.1cm} Neutral} & \multicolumn{1}{p{3cm}}{\centering  \vspace{0.1cm} Market} & \multicolumn{1}{p{3cm}}{\centering $H_0$: Neutral = Market\tnote{+}}\\
    \midrule
    \multirow{2}{*}{$\beta$: Aheadness aversion} & 0.194$^{***}$ & 0.099$^{***}$ & 0.009 \\ 
    & (0.030) - \textit{[0.012]} & (0.020) - \textit{[0.008]} &  \\ 
    \multirow{2}{*}{$\kappa$: Degree of morality} & 0.258$^{***}$ & 0.228$^{***}$ & 0.721 \\ 
    & (0.075) - \textit{[0.032]}  & (0.037) - \textit{[0.018]}&  \\ 
    \multirow{2}{*}{$\sigma$: Choice sensitivity} & 0.295$^{***}$ & 0.040$^{***}$ & 0.295 \\ 
    &  (0.009) - \textit{[0.003]}  & (0.010) - \textit{[0.003]} &  \\ 
    \midrule
    Number of Observations & 3,600 & 3,720 &  \\ 
    Number of subjects & 90 & 93 &  \\ 
    Log likelihood & -2,196.136 & -2,255.419 &  \\ 
    \hline
    \end{tabular}
    \begin{tablenotes}
    \item Notes: Standard errors clustered at the individual level in parentheses. Non-clustered standard errors in brackets. $^{***}$ Significant at 1\% using clustered standard errors.
    \item[+] $p$-value of $z$-test, using clustered standard errors.
    \end{tablenotes}
    \end{threeparttable}
\end{table}

Both the aheadness aversion and the Kantian morality estimates significantly differ from zero and are positive. They are also both higher in the Neutral than in the Market frame: $\beta=0.194$ and $\kappa=0.258$ in the Neutral frame while $\beta=0.099$ and $\kappa=0.228$ in the Market frame. The absolute difference is thus larger for the aheadness aversion estimates than for the Kantian morality estimates, and the difference is statistically different only for the aheadness aversion estimates ($\beta$) as shown in the third column. 

The estimates based on the full sample and the Core 2 samples, reported in Tables \ref{tab:rep_agent_full}-\ref{tab:rep_agent_c2} in Appendix \ref{appendixA}, are qualitatively similar, except that for the Core 2 sample, for which the estimated aheadness aversion parameter is not significantly different under the Market than under the Neutral frame.\footnote{This result is partly driven by the clustering of the standard errors at the individual level, as they generate wider confidence intervals compared to non-clustered standard errors.}
\clearpage

\subsubsection{Finite-mixture estimates}

Table \ref{tab:finmix_c1} presents the finite mixture estimates using a two-type model ($K=2$ in \eqref{eq:loglik_mixture}), as well as the shares of subjects who are classified either as Type 1 or as Type 2, depending on the type that is more aligned with their choices.\footnote{The estimates based on the full sample and the Core 2 samples are reported in Tables \ref{tab:finmix_f}-\ref{tab:finmix_c2} in Appendix \ref{appendixA}.} It is worth underscoring that both types in either frame show some degree of Kantian morality, which is in line with the significant VOI treatment effects presented in Section \ref{sec:hyptest}.

\begin{table}[ht!]
\centering
\begin{threeparttable}
\caption{\label{tab:finmix_c1}Two-type Finite Mixture Model estimates, Core 1 sample}
\begin{tabularx}{\textwidth}{>{\hsize=1.15\hsize}X *{4}{>{\hsize=0.6\hsize}X}}
\toprule
& \multicolumn{2}{c}{Neutral} & \multicolumn{2}{c}{Market} \\
\cmidrule(lr){2-3} \cmidrule(lr){4-5}
& Type 1 & Type 2 & Type 1 & Type 2 \\
\midrule
$\beta$: Aheadness aversion   & 0.327$^{***}$ & -0.065 & 0.203$^{***}$ & -0.143 \\
 & \textit{(0.082)} & \textit{(0.258)}  & \textit{(0.015)} & \textit{(0.117)} \\ 
 & [0.023]  & [0.049]  & [0.007] & [0.046] \\ 
$\kappa$: Degree of morality & 0.116$^{**}$ & 0.342$^{*}$ & 0.153$^{***}$ & 0.325$^{***}$ \\
& \textit{(0.049)} & \textit{(0.199)} & \textit{(0.023)} & \textit{(0.072)} \\ 
 & [0.029]  & [0.070]  & [0.017] & [0.056] \\ 
$\sigma$: Choice sensitivity  & 0.046$^{***}$ & 0.025$^{**}$ & 0.068$^{***}$ & 0.030$^{***}$ \\
& \textit{(0.009) }& \textit{(0.012)} & \textit{(0.013)} & \textit{(0.010)}\\
 & [0.005]  & [0.004]  & [0.005] & [0.004] \\ 
Share & 0.554$^{***}$ & 0.446$^{***}$ & 0.560$^{***}$ & 0.440$^{***}$ \\
& \textit{(0.173)} & \textit{(0.173) }& \textit{(0.056)}& \textit{(0.056)} \\ 
 & [0.058]  & [0.058]  & [0.052] & [0.052] \\ 
\midrule
Observations & \multicolumn{2}{c}{3,920} & \multicolumn{2}{c}{4,000} \\ 
Number of subjects & \multicolumn{2}{c}{98} & \multicolumn{2}{c}{100} \\ 
Log likelihood & \multicolumn{2}{c}{-2,155.558} & \multicolumn{2}{c}{-2,193.214} \\ 
\bottomrule
\end{tabularx}
\begin{tablenotes}
\item Notes: Standard errors clustered at the individual level in parenthesis, non-clustered standard errors in brackets. $^{***}$ Significant at 1\%, $^{**}$ Significant at 5\%, $^{*}$ Significant at 10\% using clustered standard errors.
\end{tablenotes}
\end{threeparttable}
\end{table}

Starting with the Neutral frame, Type 1, which accounts for 55.4\% of the sample,  displays aheadness aversion that is almost twice that of the representative agent (0.327 compared to 0.194), but a Kantian morality that is only about half as large as the representative agent's (0.116 compared to 0.238). By contrast, Type 2 (with a share of 44.6\%) seems to be indifferent to advantageous inequality (notice that the -0.065 estimate is not significant at any of the usual confidence levels) and exhibits a much larger Kantian morality estimate than the representative agent (0.342 versus 0.258). 

Turning now to the Market frame, a similar pattern emerges. We again see a type (Type 1) who is about twice as aheadness averse as the representative agent --0.203 compared with 0.099-- but less morally concerned, with a Kantian morality estimate of 0.153 (versus 0.228 for the representative agent). In addition, we estimates that 56\% of subjects in the Market frame belong to Type 1. Type 2 appears to not display aheadness liking nor aversion (the negative $\beta$ estimate of -0.143 is once more not statistically significant at the usual levels). Meanwhile, the Kantian morality estimate of this second type is larger than that of the representative agent (0.325 compared to 0.228). Here, 44\% of the subjects are classified as this latter type. 

The estimated parameter values described above allow us to endogenously classify each subject into the preference type that best fits her observed behaviour. More precisely, using Bayes' rule, a subject $i$'s posterior probability of belonging to Type $k = 1,2$ is given by:
\begin{equation} \label{eq:tau}
\tau_{ik} = \frac{\hat{\pi}_k f(\hat{\beta}_k,\hat{\kappa}_k,\hat{\sigma}_k;\bold{x}_i)}{\sum_{m=1}^2 \hat{\pi}_m f(\hat{\beta}_m,\hat{\kappa}_m,\hat{\sigma}_m;\bold{x}_i)}.
\end{equation}

\begin{figure}[ht!]
    \centering
    \begin{subfigure}[b]{0.45\textwidth}
        \centering
        \includegraphics[width=\textwidth]{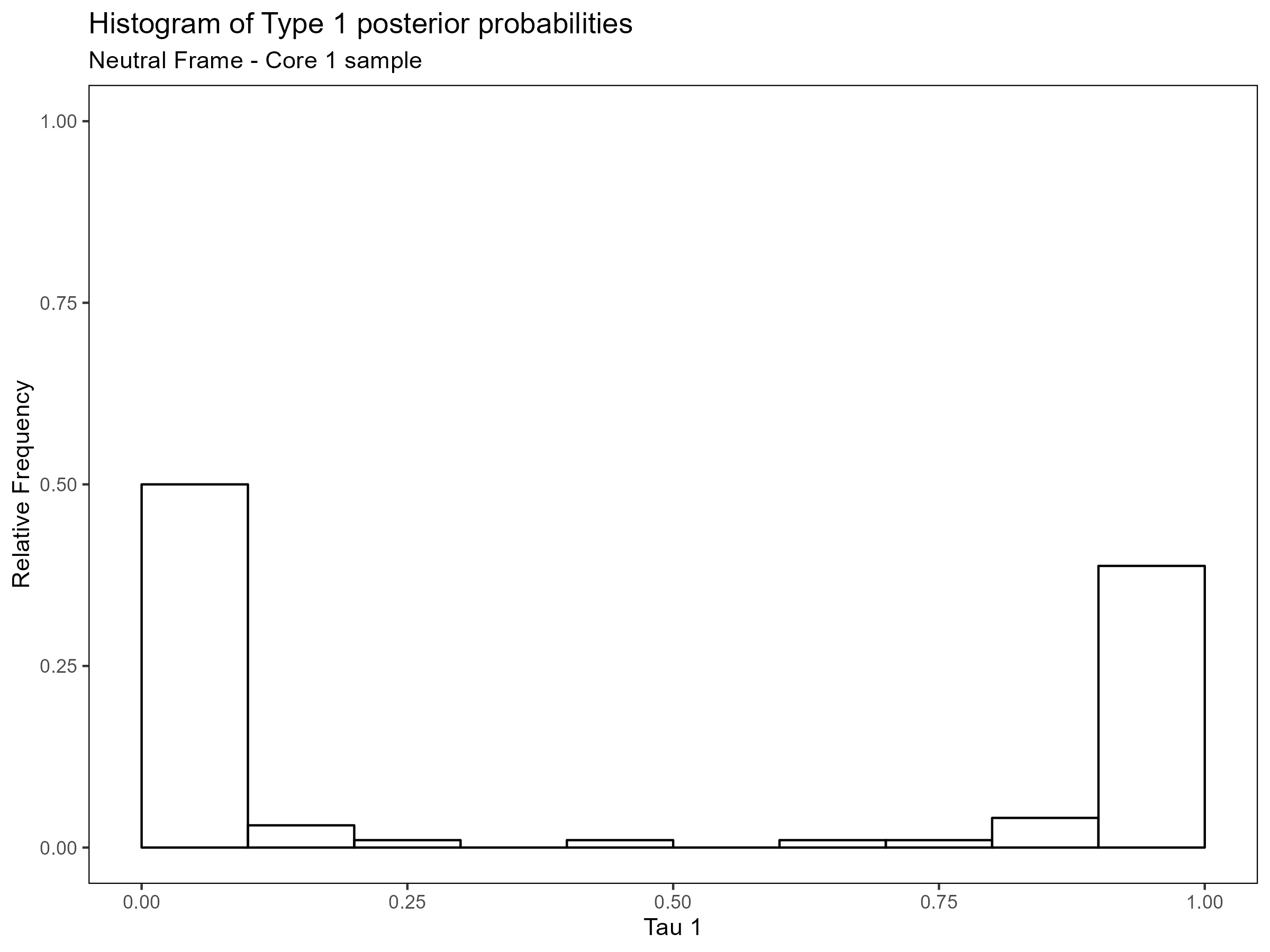}
        
        \label{fig:sub1}
    \end{subfigure}
    \hfill
    \begin{subfigure}[b]{0.45\textwidth}
        \centering
        \includegraphics[width=\textwidth]{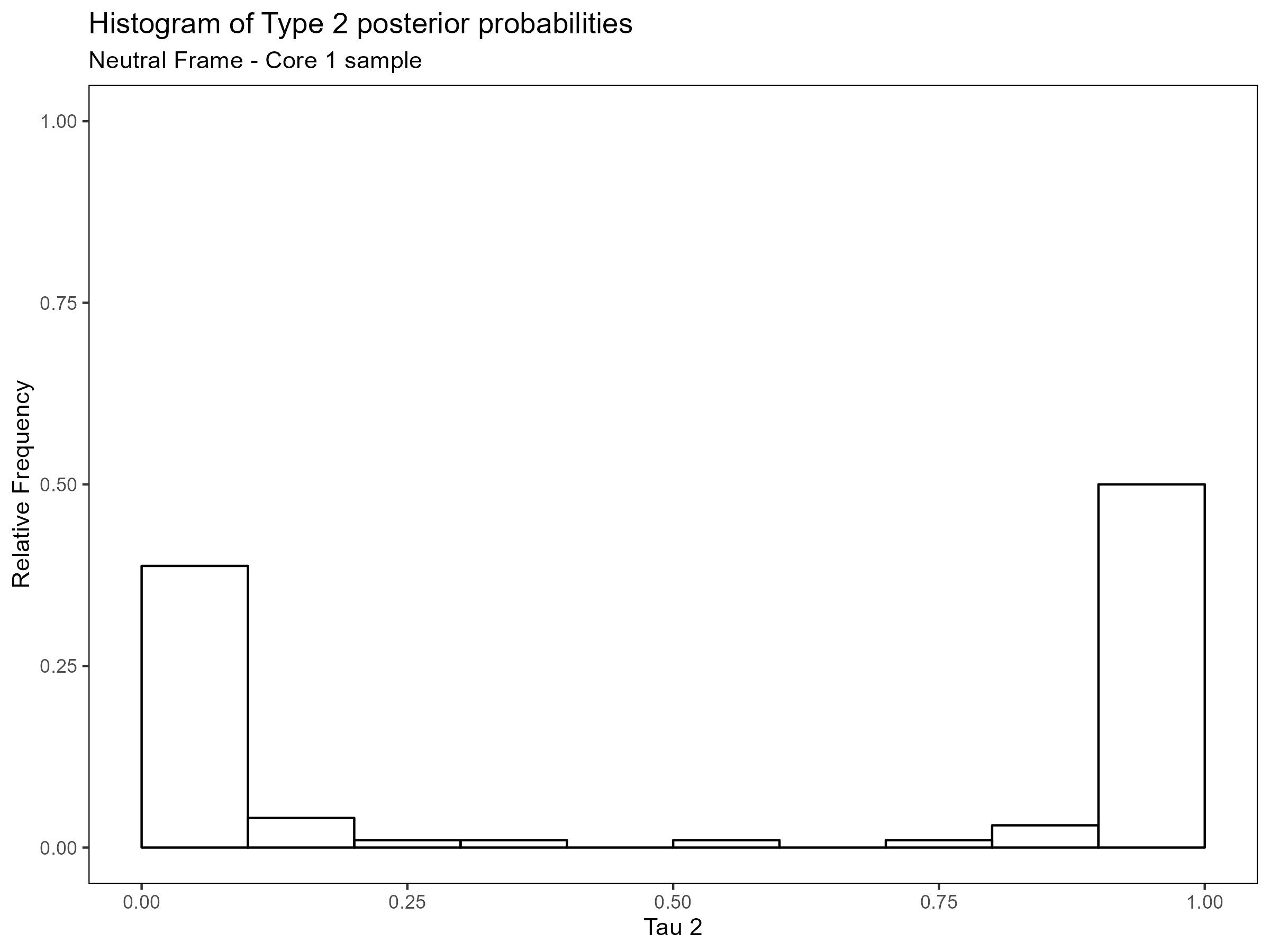}
        
        \label{fig:sub2}
    \end{subfigure}
    \\
    \begin{subfigure}[b]{0.45\textwidth}
        \centering
        \includegraphics[width=\textwidth]{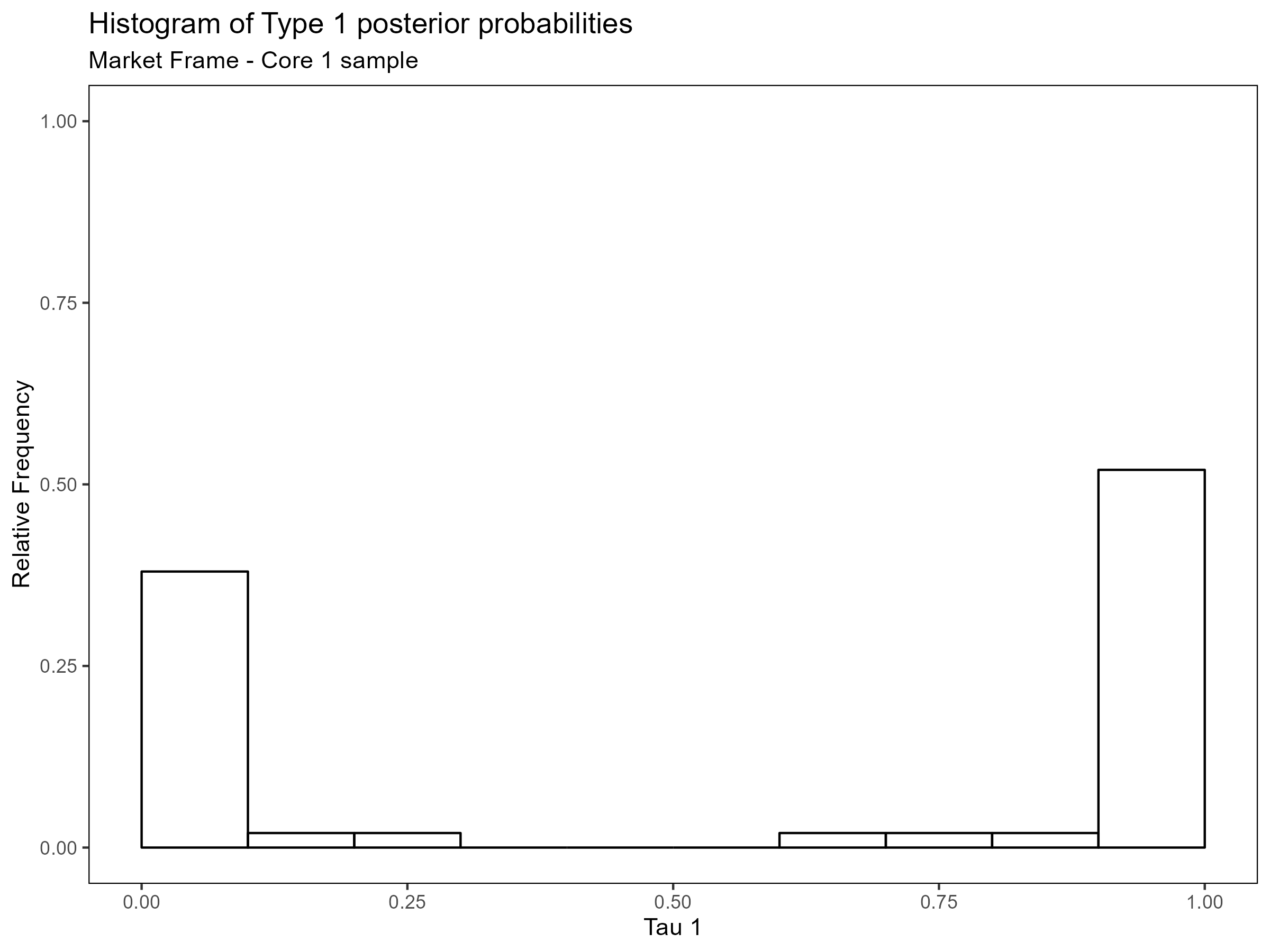}
        
        \label{fig:sub3}
    \end{subfigure}
    \hfill
    \begin{subfigure}[b]{0.45\textwidth}
        \centering
        \includegraphics[width=\textwidth]{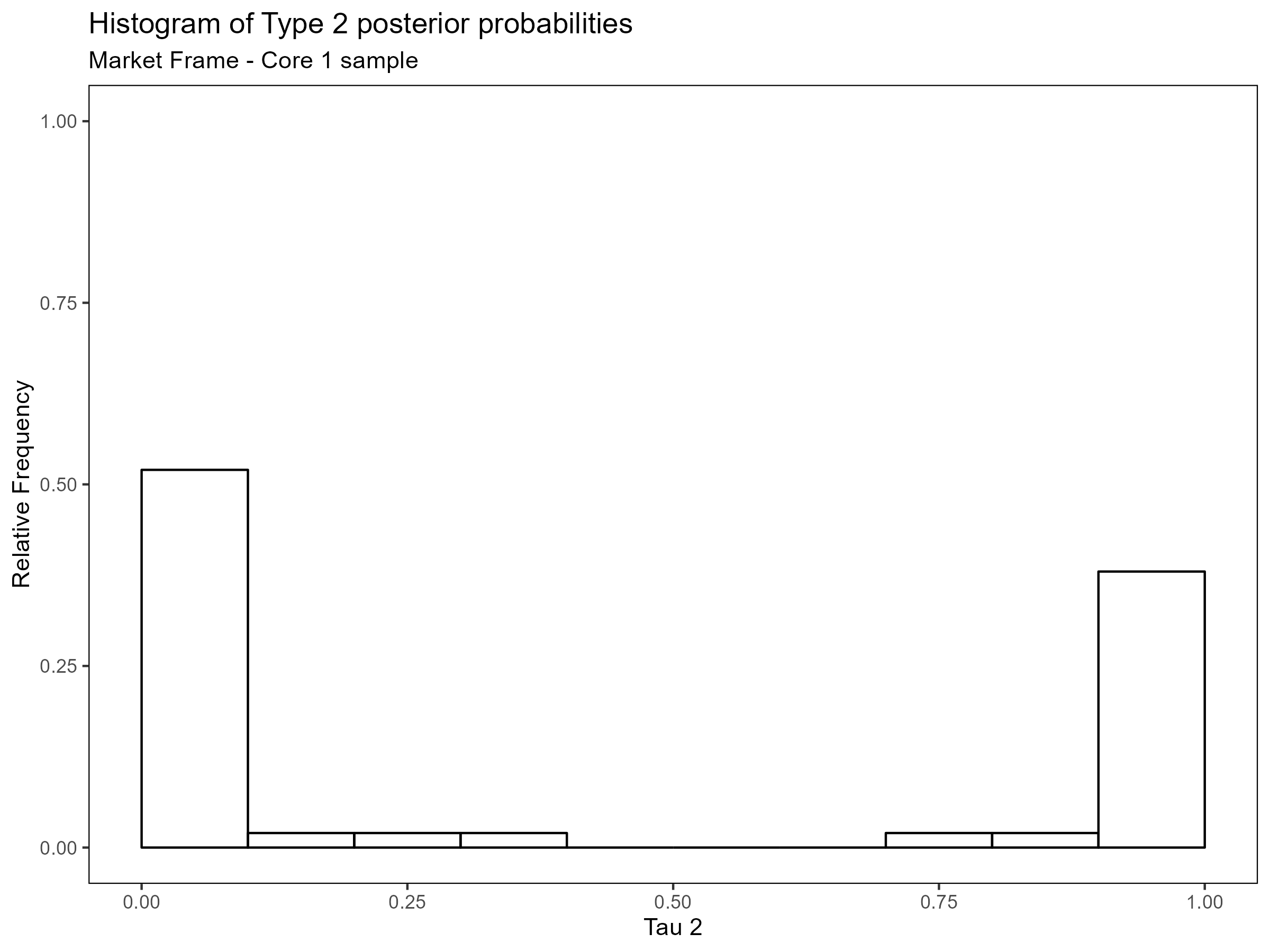}
        
        \label{fig:sub4}
    \end{subfigure}
    \caption{Distribution of posterior probabilities of individual type-membership in Neutral (first row) and Market frames (second row). Core 1 sample estimates.}
    \label{fig:taus_c1}
\end{figure}

These probabilities indicate the type with which the participant's behavior is the most compatible. A model that manages to capture the preferences and share of the underlying types should give way to most posterior probabilities being very close to 0 or to 1. Moreover, the share of subjects classified as belonging to a given type according to these probabilities should be similar to that same type's share estimated by the model. Figure \ref{fig:taus_c1} shows that this is indeed the case here. 

\begin{figure}[ht!]
    \centering
    \begin{subfigure}{\textwidth}
        \centering
        \includegraphics[width=0.8\linewidth]{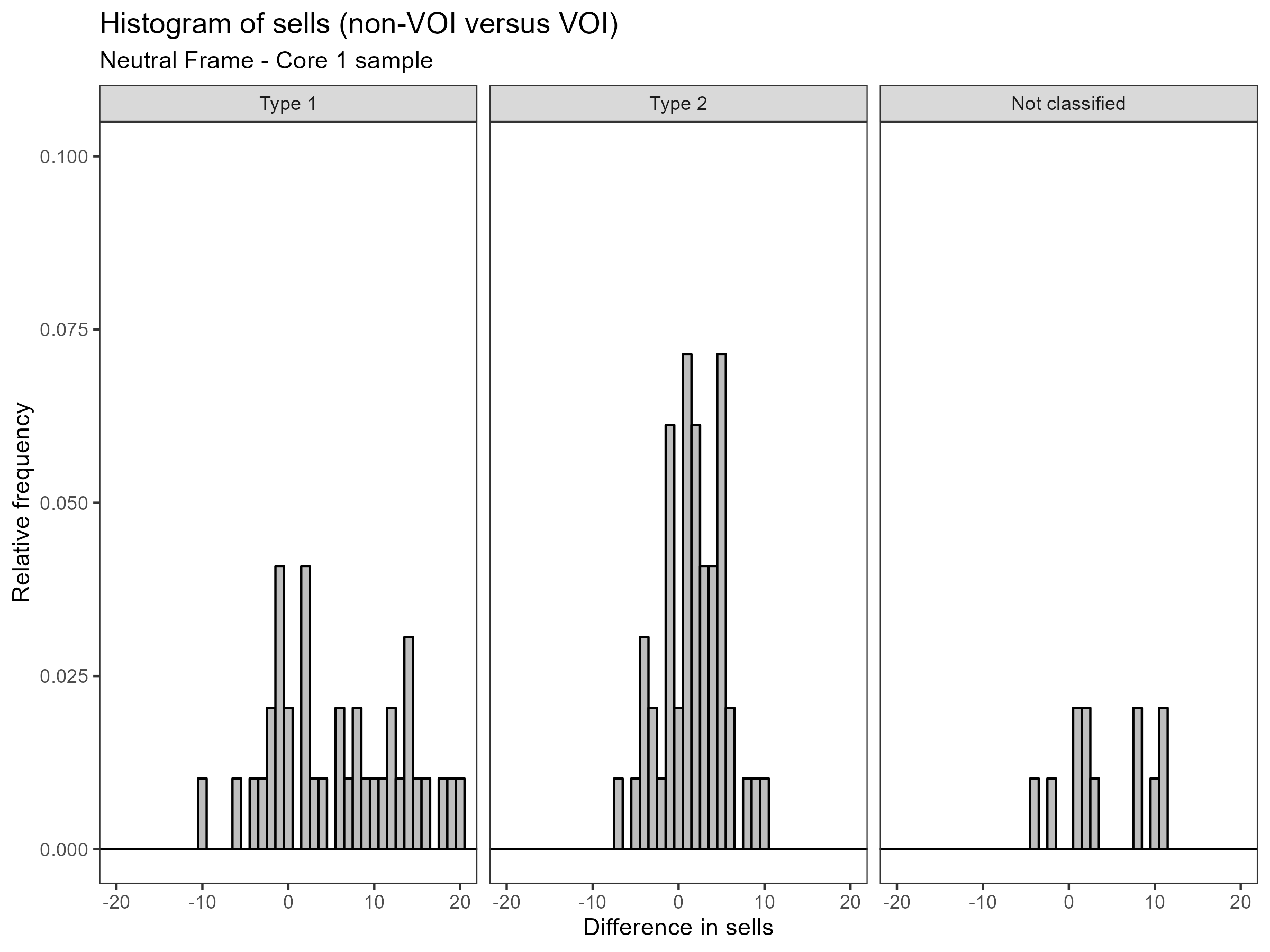}
        \caption{Neutral frame}
        \label{fig:subfiga}
    \end{subfigure}
    
    \begin{subfigure}{\textwidth}
        \centering
        \includegraphics[width=0.8\linewidth]{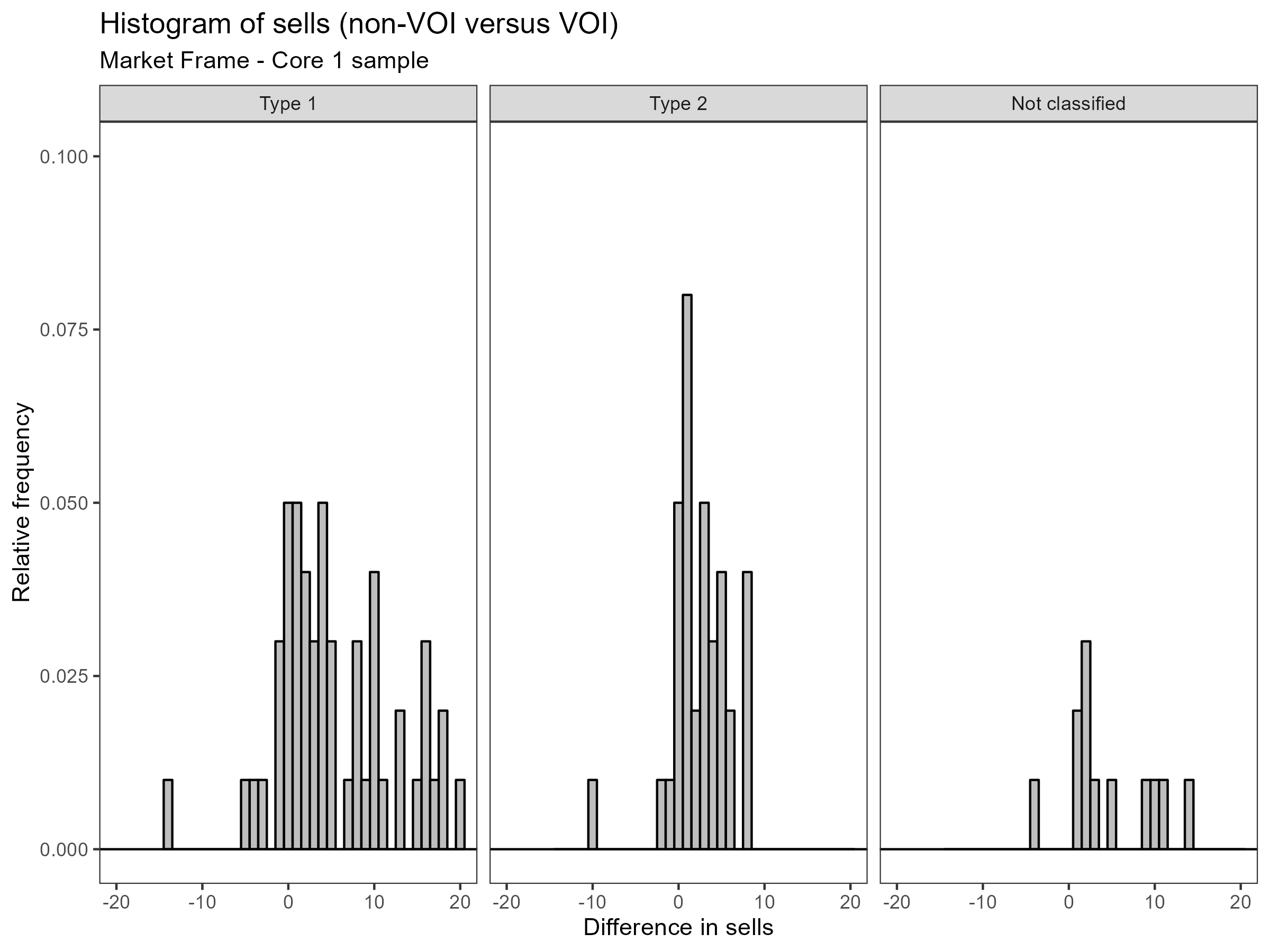} 
        \caption{Market frame}
        \label{fig:subfigb}
    \end{subfigure}
    \caption{Difference in selfish actions non-VOI versus VOI. Core 1 sample. Subjects classified as Type 1 if $\tau_1 > 0.95$ or Type 2 if $\tau_1 < 0.05$}
    \label{fig:diffsell_c1}
\end{figure}

According to our theory, subjects who are classified as Type 1 should display behavior that is more consistent across the non-VOI and VOI-treatments, as indicated by the relatively low value of $\kappa$. By contrast, those who are classified as Type 2 ought to exhibit less consistent behavior across the non-VOI and VOI-treatments, as suggested by the high value of $\kappa$. This is clearly seen in Figure \ref{fig:diffsell_c1}, where we show that the difference in the number of selfish actions in the non-VOI and VOI treatments is more concentrated around zero for participants classified as Type 1.\footnote{We refer to Figures \ref{fig:taus_f}-\ref{fig:diffsell_c2} in Appendix \ref{appendixA} for the corresponding analysis for the full and Core 2 samples.
}

We do not report estimates based on more than two types, since our experimental design is better suited for estimation at the aggregate level (i.e., treating the observations as if they emanated from one or a small number of preference types) than for estimation of each individual's preference type $(\beta_i,\kappa_i)$ (we refer to Appendix \ref{app:individual} for a detailed explanation).

\clearpage

\section{Concluding remarks}

In order to issue effective and desirable policy recommendations, economists need to achieve a realistic understanding of the motivations of human behavior. Indeed, insights about the functioning of markets and institutions derived in theoretical models where humans are driven solely by material self-interest may be misleading. We designed an experiment aimed at evaluating the willingness of subjects to sell a lemon -- i.e., a good that makes the buyer worse off -- and to disentangle the role of other-regarding concerns and Kantian moral concerns in this decision. Our design further allows us to structurally estimate preference parameters, and to test whether the obtained estimates differ between decisions taken in a frame that describes the situation as the sale of a lemon -- the market frame -- and those taken in a standard neutral frame. 

The subjects in our experiment were on average much more prone to ``sell lemons'' in the market frame than they were to select the payoff-equivalent decisions in the neutral frame. The propensity to sell lemons was, however, strongly mitigated when the instructions made explicit reference to the arbitrariness in the role distribution. This result is in line with our theoretical model, according to which such instructions are expected to awaken or strengthen subjects' Kantian moral concerns. As per our structural estimations of this Kantian moral concern and the aversion towards being ahead materially compared to the other subject in a match, the difference in behavior between the market and the neutral frame was driven mainly by a weaker aheadness aversion in the market frame. 

Our results are in line with those of other studies that have also shown that preferences that combine a Kantian moral concern with other-regard significantly enhances the explanatory power compared to preferences without the Kantian moral concern \citep*{weibull2019horserace,vanleeuwen23}. Like \citet{vanleeuwen23}, who also estimate aheadness aversion and Kantian moral preference parameters, we find estimates which indicate that some individuals appear to be driven by a combination of altruism (a positive $\beta$) and a Kantian moral concern, and all the estimates of the Kantian moral concern parameter are positive and significantly so. Interestingly, under the Market frame some individuals appear to be driven by a combination of spite (a negative  $\beta$) and a Kantian moral concern. This suggests that in our experiment the Market frame may have triggered more competitive preferences.

Our study is the first to explore -- and find support for -- the hypothesis that the Kantian moral concern is (at least somewhat) muted in situations where role uncertainty is not made explicit. If externally valid, this result would mean that the mere mention of role uncertainty in situations outside of the experimental laboratory could be used to awaken Kantian moral concerns, and thus trigger behavioral changes. In this line, \cite{bowles2023} suggests that indeed, in order to improve the functioning of markets, policies ought to combine both incentives and \textit{moral messages}, leveraging synergies between the two \citep{kranton2019}.
Field experiments might be appropriate to examine this question in the future. 

Our results further strongly suggest that both frames in our experiment matter. On average decisions are more selfish under the Market than under the Neutral frame, and the difference is large. However, the VOI wording also has a stronger dampening effect on selfish actions under the Market frame. Accordingly the estimates of the preferences based on subjects' decisions in market interactions differ from those based on decisions in interactions with similar payoff consequences but presented with a neutral frame.  

Such disparities in preferences across different social situations would be in line with the theory of preference evolution, which predicts that as long as humans in our evolutionary past could discriminate between classes of interactions (say, market vs non-market), preferences would be expected to depend on the specifics of each class of interaction (for surveys, see \citet{alger2019annual} and \citet{Alger2023}).  
As such, they strengthen the more general case made elsewhere for further research on how the language used to describe a task matters for subjects' decisions \citep*{ALEKSEEV2017,capraro2024}.

As a final note, it is worth mentioning that recent theoretical studies highlight the role that Kantian moral concerns can play in mitigating inefficiencies in equilibrium outcomes in bilateral trade situations plagued by information asymmetries  \citep*{wildemauwe2023,wildemauwe2023a}. While our study focuses solely on the behavior of potential ``lemon'' sellers, by shutting down the buyers' decisions, these references find that morality decreases or even eliminates the exchange of ``lemons'' \textit{in equilibrium}. Moreover, they document that altruism has effects that go in the same direction, although they are generally weaker. It would thus be worth extending our present work to capture interactions between both sides of the market. With both sellers and buyers being active, beliefs should also play a role; the evidence that preferences may be affected by framing can help explain why beliefs would also be affected by framing
\citep*{DUFWENBERG2011}. 
 %A potentially promising line of research would consist in comparing behavior in payoff-equivalent decisions under several different frames, in order to establish whether systematic patterns emerge.

\newpage 

\bibliography{biblio_lemons}

\begin{thebibliography}{}

\bibitem [\protect \citeauthoryear {%
Abeler%
, Nosenzo%
\BCBL {}\ \BBA {} Raymond%
}{%
Abeler%
\ \protect \BOthers {.}}{%
{\protect \APACyear {2019}}%
}]{%
Abeler2019}
\APACinsertmetastar {%
Abeler2019}%
\begin{APACrefauthors}%
Abeler, J.%
, Nosenzo, D.%
\BCBL {}\ \BBA {} Raymond, C.%
\end{APACrefauthors}%
\unskip\
\newblock
\APACrefYearMonthDay{2019}{}{}.
\newblock
{\BBOQ}\APACrefatitle {Preferences for Truth-Telling} {Preferences for truth-telling}.{\BBCQ}
\newblock
\APACjournalVolNumPages{Econometrica}{87}{4}{1115--1153}.
\PrintBackRefs{\CurrentBib}

\bibitem [\protect \citeauthoryear {%
Agneman%
\ \BBA {} Chevrot-Bianco%
}{%
Agneman%
\ \BBA {} Chevrot-Bianco%
}{%
{\protect \APACyear {2022}}%
}]{%
agneman2022}
\APACinsertmetastar {%
agneman2022}%
\begin{APACrefauthors}%
Agneman, G.%
\BCBT {}\ \BBA {} Chevrot-Bianco, E.%
\end{APACrefauthors}%
\unskip\
\newblock
\APACrefYearMonthDay{2022}{}{}.
\newblock
{\BBOQ}\APACrefatitle {Market Participation and Moral Decision-Making: Experimental Evidence from {G}reenland} {Market participation and moral decision-making: Experimental evidence from {G}reenland}.{\BBCQ}
\newblock
\APACjournalVolNumPages{Economic Journal}{133}{650}{537-581}.
\PrintBackRefs{\CurrentBib}

\bibitem [\protect \citeauthoryear {%
Akerlof%
}{%
Akerlof%
}{%
{\protect \APACyear {1970}}%
}]{%
Akerlof1970}
\APACinsertmetastar {%
Akerlof1970}%
\begin{APACrefauthors}%
Akerlof, G\BPBI A.%
\end{APACrefauthors}%
\unskip\
\newblock
\APACrefYearMonthDay{1970}{}{}.
\newblock
{\BBOQ}\APACrefatitle {The Market for ``Lemons''': Quality Uncertainty and the Market Mechanism} {The market for ``lemons''': Quality uncertainty and the market mechanism}.{\BBCQ}
\newblock
\APACjournalVolNumPages{Quarterly Journal of Economics}{84}{3}{488--500}.
\PrintBackRefs{\CurrentBib}

\bibitem [\protect \citeauthoryear {%
Alekseev%
, Charness%
\BCBL {}\ \BBA {} Gneezy%
}{%
Alekseev%
\ \protect \BOthers {.}}{%
{\protect \APACyear {2017}}%
}]{%
ALEKSEEV2017}
\APACinsertmetastar {%
ALEKSEEV2017}%
\begin{APACrefauthors}%
Alekseev, A.%
, Charness, G.%
\BCBL {}\ \BBA {} Gneezy, U.%
\end{APACrefauthors}%
\unskip\
\newblock
\APACrefYearMonthDay{2017}{}{}.
\newblock
{\BBOQ}\APACrefatitle {Experimental methods: When and why contextual instructions are important} {Experimental methods: When and why contextual instructions are important}.{\BBCQ}
\newblock
\APACjournalVolNumPages{Journal of Economic Behavior and Organization}{134}{}{48-59}.
\PrintBackRefs{\CurrentBib}

\bibitem [\protect \citeauthoryear {%
Alger%
}{%
Alger%
}{%
{\protect \APACyear {2023}}%
}]{%
Alger2023}
\APACinsertmetastar {%
Alger2023}%
\begin{APACrefauthors}%
Alger, I.%
\end{APACrefauthors}%
\unskip\
\newblock
\APACrefYearMonthDay{2023}{}{}.
\newblock
{\BBOQ}\APACrefatitle {Evolutionarily stable preferences} {Evolutionarily stable preferences}.{\BBCQ}
\newblock
\APACjournalVolNumPages{Philosophical Transactions B}{378}{}{20210505}.
\PrintBackRefs{\CurrentBib}

\bibitem [\protect \citeauthoryear {%
Alger%
\ \BBA {} Weibull%
}{%
Alger%
\ \BBA {} Weibull%
}{%
{\protect \APACyear {2013}}%
}]{%
alger2013homo}
\APACinsertmetastar {%
alger2013homo}%
\begin{APACrefauthors}%
Alger, I.%
\BCBT {}\ \BBA {} Weibull, J\BPBI W.%
\end{APACrefauthors}%
\unskip\
\newblock
\APACrefYearMonthDay{2013}{}{}.
\newblock
{\BBOQ}\APACrefatitle {Homo moralis—preference evolution under incomplete information and assortative matching} {Homo moralis—preference evolution under incomplete information and assortative matching}.{\BBCQ}
\newblock
\APACjournalVolNumPages{Econometrica}{81}{6}{2269--2302}.
\PrintBackRefs{\CurrentBib}

\bibitem [\protect \citeauthoryear {%
Alger%
\ \BBA {} Weibull%
}{%
Alger%
\ \BBA {} Weibull%
}{%
{\protect \APACyear {2019}}%
}]{%
alger2019annual}
\APACinsertmetastar {%
alger2019annual}%
\begin{APACrefauthors}%
Alger, I.%
\BCBT {}\ \BBA {} Weibull, J\BPBI W.%
\end{APACrefauthors}%
\unskip\
\newblock
\APACrefYearMonthDay{2019}{}{}.
\newblock
{\BBOQ}\APACrefatitle {Evolutionary Models of Preference Formation} {Evolutionary models of preference formation}.{\BBCQ}
\newblock
\APACjournalVolNumPages{Annual Review of Economics}{11}{}{329--354}.
\PrintBackRefs{\CurrentBib}

\bibitem [\protect \citeauthoryear {%
Alger%
, Weibull%
\BCBL {}\ \BBA {} Lehmann%
}{%
Alger%
\ \protect \BOthers {.}}{%
{\protect \APACyear {2020}}%
}]{%
alger2019evolution}
\APACinsertmetastar {%
alger2019evolution}%
\begin{APACrefauthors}%
Alger, I.%
, Weibull, J\BPBI W.%
\BCBL {}\ \BBA {} Lehmann, L.%
\end{APACrefauthors}%
\unskip\
\newblock
\APACrefYearMonthDay{2020}{}{}.
\newblock
{\BBOQ}\APACrefatitle {Evolution of preferences in structured populations: {G}enes, guns, and culture} {Evolution of preferences in structured populations: {G}enes, guns, and culture}.{\BBCQ}
\newblock
\APACjournalVolNumPages{Journal of Economic Theory}{185}{}{104951}.
\PrintBackRefs{\CurrentBib}

\bibitem [\protect \citeauthoryear {%
Andreoni%
\ \BBA {} Miller%
}{%
Andreoni%
\ \BBA {} Miller%
}{%
{\protect \APACyear {1993}}%
}]{%
Andreoni1993}
\APACinsertmetastar {%
Andreoni1993}%
\begin{APACrefauthors}%
Andreoni, J.%
\BCBT {}\ \BBA {} Miller, J\BPBI H.%
\end{APACrefauthors}%
\unskip\
\newblock
\APACrefYearMonthDay{1993}{}{}.
\newblock
{\BBOQ}\APACrefatitle {Rational Cooperation in the Finitely Repeated Prisoner's Dilemma: Experimental Evidence} {Rational cooperation in the finitely repeated prisoner's dilemma: Experimental evidence}.{\BBCQ}
\newblock
\APACjournalVolNumPages{Economic Journal}{103}{418}{570--585}.
\PrintBackRefs{\CurrentBib}

\bibitem [\protect \citeauthoryear {%
Bartling%
\ \BBA {} {\"{O}}zdemir%
}{%
Bartling%
\ \BBA {} {\"{O}}zdemir%
}{%
{\protect \APACyear {2023}}%
}]{%
Bartling2023replacement}
\APACinsertmetastar {%
Bartling2023replacement}%
\begin{APACrefauthors}%
Bartling, B.%
\BCBT {}\ \BBA {} {\"{O}}zdemir, Y.%
\end{APACrefauthors}%
\unskip\
\newblock
\APACrefYearMonthDay{2023}{}{}.
\newblock
{\BBOQ}\APACrefatitle {{The limits to moral erosion in markets: Social norms and the replacement excuse}} {{The limits to moral erosion in markets: Social norms and the replacement excuse}}.{\BBCQ}
\newblock
\APACjournalVolNumPages{Games and Economic Behavior}{138}{}{143--160}.
\PrintBackRefs{\CurrentBib}

\bibitem [\protect \citeauthoryear {%
Becker%
}{%
Becker%
}{%
{\protect \APACyear {1974}}%
}]{%
Becker1974}
\APACinsertmetastar {%
Becker1974}%
\begin{APACrefauthors}%
Becker, G\BPBI S.%
\end{APACrefauthors}%
\unskip\
\newblock
\APACrefYearMonthDay{1974}{}{}.
\newblock
{\BBOQ}\APACrefatitle {A Theory of Social Interactions} {A theory of social interactions}.{\BBCQ}
\newblock
\APACjournalVolNumPages{Journal of Political Economy}{82}{6}{1063--1093}.
\PrintBackRefs{\CurrentBib}

\bibitem [\protect \citeauthoryear {%
B{\'e}nabou%
, Falk%
\BCBL {}\ \BBA {} Henkel%
}{%
B{\'e}nabou%
\ \protect \BOthers {.}}{%
{\protect \APACyear {2024}}%
}]{%
benabou22b}
\APACinsertmetastar {%
benabou22b}%
\begin{APACrefauthors}%
B{\'e}nabou, R.%
, Falk, A.%
\BCBL {}\ \BBA {} Henkel, L.%
\end{APACrefauthors}%
\unskip\
\newblock
\APACrefYearMonthDay{2024}{}{}.
\newblock
\APACrefbtitle {Ends versus Means: {K}antians, Utilitarians, and Moral Decisions} {Ends versus means: {K}antians, utilitarians, and moral decisions}\ \APACbVolEdTR {}{{NBER Working Paper 32073}}.
\PrintBackRefs{\CurrentBib}

\bibitem [\protect \citeauthoryear {%
B{\'e}nabou%
, Falk%
, Henkel%
\BCBL {}\ \BBA {} Tirole%
}{%
B{\'e}nabou%
\ \protect \BOthers {.}}{%
{\protect \APACyear {2023}}%
}]{%
benabou22a}
\APACinsertmetastar {%
benabou22a}%
\begin{APACrefauthors}%
B{\'e}nabou, R.%
, Falk, A.%
, Henkel, L.%
\BCBL {}\ \BBA {} Tirole, J.%
\end{APACrefauthors}%
\unskip\
\newblock
\APACrefYearMonthDay{2023}{}{}.
\newblock
\APACrefbtitle {Eliciting Moral Preferences: Theory and Experiment} {Eliciting moral preferences: Theory and experiment}\ \APACbVolEdTR {}{{CRC TR 224 Discussion Paper Series}}.
\PrintBackRefs{\CurrentBib}

\bibitem [\protect \citeauthoryear {%
B{\'e}nabou%
\ \BBA {} Tirole%
}{%
B{\'e}nabou%
\ \BBA {} Tirole%
}{%
{\protect \APACyear {2006}}%
}]{%
benabou2006image}
\APACinsertmetastar {%
benabou2006image}%
\begin{APACrefauthors}%
B{\'e}nabou, R.%
\BCBT {}\ \BBA {} Tirole, J.%
\end{APACrefauthors}%
\unskip\
\newblock
\APACrefYearMonthDay{2006}{}{}.
\newblock
{\BBOQ}\APACrefatitle {Incentives and Prosocial Behavior} {Incentives and prosocial behavior}.{\BBCQ}
\newblock
\APACjournalVolNumPages{American Economic Review}{96}{5}{1652-1678}.
\PrintBackRefs{\CurrentBib}

\bibitem [\protect \citeauthoryear {%
Bowles%
}{%
Bowles%
}{%
{\protect \APACyear {1998}}%
}]{%
Bowles1998b}
\APACinsertmetastar {%
Bowles1998b}%
\begin{APACrefauthors}%
Bowles, S.%
\end{APACrefauthors}%
\unskip\
\newblock
\APACrefYearMonthDay{1998}{}{}.
\newblock
{\BBOQ}\APACrefatitle {{Endogenous preferences: The cultural consequences of markets and other economic institutions}} {{Endogenous preferences: The cultural consequences of markets and other economic institutions}}.{\BBCQ}
\newblock
\APACjournalVolNumPages{Journal of Economic Literature}{36}{1}{75--111}.
\PrintBackRefs{\CurrentBib}

\bibitem [\protect \citeauthoryear {%
Bowles%
}{%
Bowles%
}{%
{\protect \APACyear {2023}}%
}]{%
bowles2023}
\APACinsertmetastar {%
bowles2023}%
\begin{APACrefauthors}%
Bowles, S.%
\end{APACrefauthors}%
\unskip\
\newblock
\APACrefYearMonthDay{2023}{}{}.
\newblock
{\BBOQ}\APACrefatitle {Moral economics} {Moral economics}.{\BBCQ}
\newblock
\APACjournalVolNumPages{Fiscal Studies}{44}{2}{151-160}.
\PrintBackRefs{\CurrentBib}

\bibitem [\protect \citeauthoryear {%
Brandts%
\ \BBA {} Charness%
}{%
Brandts%
\ \BBA {} Charness%
}{%
{\protect \APACyear {2011}}%
}]{%
Brandts2011}
\APACinsertmetastar {%
Brandts2011}%
\begin{APACrefauthors}%
Brandts, J.%
\BCBT {}\ \BBA {} Charness, G.%
\end{APACrefauthors}%
\unskip\
\newblock
\APACrefYearMonthDay{2011}{}{}.
\newblock
{\BBOQ}\APACrefatitle {{The strategy versus the direct-response method: {A} first survey of experimental comparisons}} {{The strategy versus the direct-response method: {A} first survey of experimental comparisons}}.{\BBCQ}
\newblock
\APACjournalVolNumPages{Experimental Economics}{14}{3}{375--398}.
\PrintBackRefs{\CurrentBib}

\bibitem [\protect \citeauthoryear {%
Bruhin%
, Fehr%
\BCBL {}\ \BBA {} Schunk%
}{%
Bruhin%
\ \protect \BOthers {.}}{%
{\protect \APACyear {2018}}%
}]{%
bruhin2018many}
\APACinsertmetastar {%
bruhin2018many}%
\begin{APACrefauthors}%
Bruhin, A.%
, Fehr, E.%
\BCBL {}\ \BBA {} Schunk, D.%
\end{APACrefauthors}%
\unskip\
\newblock
\APACrefYearMonthDay{2018}{}{}.
\newblock
{\BBOQ}\APACrefatitle {The many faces of human sociality: Uncovering the distribution and stability of social preferences} {The many faces of human sociality: Uncovering the distribution and stability of social preferences}.{\BBCQ}
\newblock
\APACjournalVolNumPages{Journal of the European Economic Association}{17}{4}{1025--1069}.
\PrintBackRefs{\CurrentBib}

\bibitem [\protect \citeauthoryear {%
Bursztyn%
, Fiorin%
, Gottlieb%
\BCBL {}\ \BBA {} Kanz%
}{%
Bursztyn%
\ \protect \BOthers {.}}{%
{\protect \APACyear {2019}}%
}]{%
Bursztyn2019}
\APACinsertmetastar {%
Bursztyn2019}%
\begin{APACrefauthors}%
Bursztyn, L.%
, Fiorin, S.%
, Gottlieb, D.%
\BCBL {}\ \BBA {} Kanz, M.%
\end{APACrefauthors}%
\unskip\
\newblock
\APACrefYearMonthDay{2019}{}{}.
\newblock
{\BBOQ}\APACrefatitle {{Moral incentives in credit card debt repayment: Evidence from a field experiment}} {{Moral incentives in credit card debt repayment: Evidence from a field experiment}}.{\BBCQ}
\newblock
\APACjournalVolNumPages{Journal of Political Economy}{127}{4}{1641--1683}.
\PrintBackRefs{\CurrentBib}

\bibitem [\protect \citeauthoryear {%
Cabrales%
, Miniaci%
, Piovesan%
\BCBL {}\ \BBA {} Ponti%
}{%
Cabrales%
\ \protect \BOthers {.}}{%
{\protect \APACyear {2010}}%
}]{%
Cabrales2010}
\APACinsertmetastar {%
Cabrales2010}%
\begin{APACrefauthors}%
Cabrales, A.%
, Miniaci, R.%
, Piovesan, M.%
\BCBL {}\ \BBA {} Ponti, G.%
\end{APACrefauthors}%
\unskip\
\newblock
\APACrefYearMonthDay{2010}{}{}.
\newblock
{\BBOQ}\APACrefatitle {{Social preferences and strategic uncertainty: An experiment on markets and contracts}} {{Social preferences and strategic uncertainty: An experiment on markets and contracts}}.{\BBCQ}
\newblock
\APACjournalVolNumPages{American Economic Review}{100}{5}{2261--2278}.
\PrintBackRefs{\CurrentBib}

\bibitem [\protect \citeauthoryear {%
Capraro%
, Halpern%
\BCBL {}\ \BBA {} Perc%
}{%
Capraro%
\ \protect \BOthers {.}}{%
{\protect \APACyear {2024}}%
}]{%
capraro2024}
\APACinsertmetastar {%
capraro2024}%
\begin{APACrefauthors}%
Capraro, V.%
, Halpern, J\BPBI Y.%
\BCBL {}\ \BBA {} Perc, M.%
\end{APACrefauthors}%
\unskip\
\newblock
\APACrefYearMonthDay{2024}{March}{}.
\newblock
{\BBOQ}\APACrefatitle {From Outcome-Based to Language-Based Preferences} {From outcome-based to language-based preferences}.{\BBCQ}
\newblock
\APACjournalVolNumPages{Journal of Economic Literature}{62}{1}{115-54}.
\PrintBackRefs{\CurrentBib}

\bibitem [\protect \citeauthoryear {%
Capraro%
\ \BBA {} Rand%
}{%
Capraro%
\ \BBA {} Rand%
}{%
{\protect \APACyear {2018}}%
}]{%
capraro2018right}
\APACinsertmetastar {%
capraro2018right}%
\begin{APACrefauthors}%
Capraro, V.%
\BCBT {}\ \BBA {} Rand, D\BPBI G.%
\end{APACrefauthors}%
\unskip\
\newblock
\APACrefYearMonthDay{2018}{}{}.
\newblock
{\BBOQ}\APACrefatitle {Do the Right Thing: Experimental evidence that preferences for moral behavior, rather than equity or efficiency per se, drive human prosociality} {Do the right thing: Experimental evidence that preferences for moral behavior, rather than equity or efficiency per se, drive human prosociality}.{\BBCQ}
\newblock
\APACjournalVolNumPages{Judgment and Decision Making}{}{}{}.
\PrintBackRefs{\CurrentBib}

\bibitem [\protect \citeauthoryear {%
Charness%
\ \BBA {} Rabin%
}{%
Charness%
\ \BBA {} Rabin%
}{%
{\protect \APACyear {2002}}%
}]{%
charness2002understanding}
\APACinsertmetastar {%
charness2002understanding}%
\begin{APACrefauthors}%
Charness, G.%
\BCBT {}\ \BBA {} Rabin, M.%
\end{APACrefauthors}%
\unskip\
\newblock
\APACrefYearMonthDay{2002}{}{}.
\newblock
{\BBOQ}\APACrefatitle {Understanding social preferences with simple tests} {Understanding social preferences with simple tests}.{\BBCQ}
\newblock
\APACjournalVolNumPages{Quarterly Journal of Economics}{117}{3}{817--869}.
\PrintBackRefs{\CurrentBib}

\bibitem [\protect \citeauthoryear {%
Chen%
\ \BBA {} Schonger%
}{%
Chen%
\ \BBA {} Schonger%
}{%
{\protect \APACyear {2022}}%
}]{%
Chen2022}
\APACinsertmetastar {%
Chen2022}%
\begin{APACrefauthors}%
Chen, D\BPBI L.%
\BCBT {}\ \BBA {} Schonger, M.%
\end{APACrefauthors}%
\unskip\
\newblock
\APACrefYearMonthDay{2022}{}{}.
\newblock
{\BBOQ}\APACrefatitle {{Social preferences or sacred values? Theory and evidence of deontological motivations}} {{Social preferences or sacred values? Theory and evidence of deontological motivations}}.{\BBCQ}
\newblock
\APACjournalVolNumPages{Science Advances}{8}{19}{}.
\PrintBackRefs{\CurrentBib}

\bibitem [\protect \citeauthoryear {%
Daley%
\ \BBA {} Sadowski%
}{%
Daley%
\ \BBA {} Sadowski%
}{%
{\protect \APACyear {2017}}%
}]{%
Daley2017}
\APACinsertmetastar {%
Daley2017}%
\begin{APACrefauthors}%
Daley, B.%
\BCBT {}\ \BBA {} Sadowski, P.%
\end{APACrefauthors}%
\unskip\
\newblock
\APACrefYearMonthDay{2017}{}{}.
\newblock
{\BBOQ}\APACrefatitle {{Magical thinking: A representation result}} {{Magical thinking: A representation result}}.{\BBCQ}
\newblock
\APACjournalVolNumPages{Theoretical Economics}{12}{2}{909--956}.
\PrintBackRefs{\CurrentBib}

\bibitem [\protect \citeauthoryear {%
Dreber%
, Ellingsen%
, Johannesson%
\BCBL {}\ \BBA {} Rand%
}{%
Dreber%
\ \protect \BOthers {.}}{%
{\protect \APACyear {2013}}%
}]{%
Dreber2013}
\APACinsertmetastar {%
Dreber2013}%
\begin{APACrefauthors}%
Dreber, A.%
, Ellingsen, T.%
, Johannesson, M.%
\BCBL {}\ \BBA {} Rand, D\BPBI G.%
\end{APACrefauthors}%
\unskip\
\newblock
\APACrefYearMonthDay{2013}{}{}.
\newblock
{\BBOQ}\APACrefatitle {{Do people care about social context? Framing effects in dictator games}} {{Do people care about social context? Framing effects in dictator games}}.{\BBCQ}
\newblock
\APACjournalVolNumPages{Experimental Economics}{16}{3}{349--371}.
\PrintBackRefs{\CurrentBib}

\bibitem [\protect \citeauthoryear {%
Dufwenberg%
, G\"{a}chter%
\BCBL {}\ \BBA {} Hennig-Schmidt%
}{%
Dufwenberg%
\ \protect \BOthers {.}}{%
{\protect \APACyear {2011}}%
}]{%
DUFWENBERG2011}
\APACinsertmetastar {%
DUFWENBERG2011}%
\begin{APACrefauthors}%
Dufwenberg, M.%
, G\"{a}chter, S.%
\BCBL {}\ \BBA {} Hennig-Schmidt, H.%
\end{APACrefauthors}%
\unskip\
\newblock
\APACrefYearMonthDay{2011}{}{}.
\newblock
{\BBOQ}\APACrefatitle {The framing of games and the psychology of play} {The framing of games and the psychology of play}.{\BBCQ}
\newblock
\APACjournalVolNumPages{Games and Economic Behavior}{73}{2}{459-478}.
\PrintBackRefs{\CurrentBib}

\bibitem [\protect \citeauthoryear {%
Dufwenberg%
, Johansson-Stenman%
, Kirchler%
, Lindner%
\BCBL {}\ \BBA {} Schwaiger%
}{%
Dufwenberg%
\ \protect \BOthers {.}}{%
{\protect \APACyear {2022}}%
}]{%
Dufwenberg2022}
\APACinsertmetastar {%
Dufwenberg2022}%
\begin{APACrefauthors}%
Dufwenberg, M.%
, Johansson-Stenman, O.%
, Kirchler, M.%
, Lindner, F.%
\BCBL {}\ \BBA {} Schwaiger, R.%
\end{APACrefauthors}%
\unskip\
\newblock
\APACrefYearMonthDay{2022}{}{}.
\newblock
{\BBOQ}\APACrefatitle {{Mean markets or kind commerce?}} {{Mean markets or kind commerce?}}{\BBCQ}
\newblock
\APACjournalVolNumPages{Journal of Public Economics}{209}{}{104648}.
\PrintBackRefs{\CurrentBib}

\bibitem [\protect \citeauthoryear {%
C.~Engel%
}{%
C.~Engel%
}{%
{\protect \APACyear {2011}}%
}]{%
Engel2011}
\APACinsertmetastar {%
Engel2011}%
\begin{APACrefauthors}%
Engel, C.%
\end{APACrefauthors}%
\unskip\
\newblock
\APACrefYearMonthDay{2011}{}{}.
\newblock
{\BBOQ}\APACrefatitle {{Dictator games: A meta study}} {{Dictator games: A meta study}}.{\BBCQ}
\newblock
\APACjournalVolNumPages{Experimental Economics}{14}{4}{583--610}.
\PrintBackRefs{\CurrentBib}

\bibitem [\protect \citeauthoryear {%
J.~Engel%
\ \BBA {} Szech%
}{%
J.~Engel%
\ \BBA {} Szech%
}{%
{\protect \APACyear {2020}}%
}]{%
Engel2020}
\APACinsertmetastar {%
Engel2020}%
\begin{APACrefauthors}%
Engel, J.%
\BCBT {}\ \BBA {} Szech, N.%
\end{APACrefauthors}%
\unskip\
\newblock
\APACrefYearMonthDay{2020}{}{}.
\newblock
{\BBOQ}\APACrefatitle {{A little good is good enough: Ethical consumption, cheap excuses, and moral self-licensing}} {{A little good is good enough: Ethical consumption, cheap excuses, and moral self-licensing}}.{\BBCQ}
\newblock
\APACjournalVolNumPages{PLoS ONE}{15}{1}{1--19}.
\PrintBackRefs{\CurrentBib}

\bibitem [\protect \citeauthoryear {%
Feess%
, Kerzenmacher%
\BCBL {}\ \BBA {} Timofeyev%
}{%
Feess%
\ \protect \BOthers {.}}{%
{\protect \APACyear {2022}}%
}]{%
Feess2022}
\APACinsertmetastar {%
Feess2022}%
\begin{APACrefauthors}%
Feess, E.%
, Kerzenmacher, F.%
\BCBL {}\ \BBA {} Timofeyev, Y.%
\end{APACrefauthors}%
\unskip\
\newblock
\APACrefYearMonthDay{2022}{}{}.
\newblock
{\BBOQ}\APACrefatitle {{Utilitarian or deontological models of moral behavior—What predicts morally questionable decisions?}} {{Utilitarian or deontological models of moral behavior—What predicts morally questionable decisions?}}{\BBCQ}
\newblock
\APACjournalVolNumPages{European Economic Review}{149}{September 2021}{104264}.
\PrintBackRefs{\CurrentBib}

\bibitem [\protect \citeauthoryear {%
Fehr%
\ \BBA {} Schmidt%
}{%
Fehr%
\ \BBA {} Schmidt%
}{%
{\protect \APACyear {1999}}%
}]{%
fehr1999theory}
\APACinsertmetastar {%
fehr1999theory}%
\begin{APACrefauthors}%
Fehr, E.%
\BCBT {}\ \BBA {} Schmidt, K\BPBI M.%
\end{APACrefauthors}%
\unskip\
\newblock
\APACrefYearMonthDay{1999}{}{}.
\newblock
{\BBOQ}\APACrefatitle {A theory of fairness, competition, and cooperation} {A theory of fairness, competition, and cooperation}.{\BBCQ}
\newblock
\APACjournalVolNumPages{Quarterly Journal of Economics}{114}{3}{817--868}.
\PrintBackRefs{\CurrentBib}

\bibitem [\protect \citeauthoryear {%
Fisman%
, Kariv%
\BCBL {}\ \BBA {} Markovits%
}{%
Fisman%
\ \protect \BOthers {.}}{%
{\protect \APACyear {2007}}%
}]{%
fisman2007individual}
\APACinsertmetastar {%
fisman2007individual}%
\begin{APACrefauthors}%
Fisman, R.%
, Kariv, S.%
\BCBL {}\ \BBA {} Markovits, D.%
\end{APACrefauthors}%
\unskip\
\newblock
\APACrefYearMonthDay{2007}{}{}.
\newblock
{\BBOQ}\APACrefatitle {Individual preferences for giving} {Individual preferences for giving}.{\BBCQ}
\newblock
\APACjournalVolNumPages{American Economic Review}{97}{5}{1858--1876}.
\PrintBackRefs{\CurrentBib}

\bibitem [\protect \citeauthoryear {%
Garc{\'{i}}a-Pola%
, Iriberri%
\BCBL {}\ \BBA {} Kov{\'{a}}\v{r}{\'{i}}k%
}{%
Garc{\'{i}}a-Pola%
\ \protect \BOthers {.}}{%
{\protect \APACyear {2020}}%
}]{%
Garcia-Pola2020}
\APACinsertmetastar {%
Garcia-Pola2020}%
\begin{APACrefauthors}%
Garc{\'{i}}a-Pola, B.%
, Iriberri, N.%
\BCBL {}\ \BBA {} Kov{\'{a}}\v{r}{\'{i}}k, J.%
\end{APACrefauthors}%
\unskip\
\newblock
\APACrefYearMonthDay{2020}{}{}.
\newblock
{\BBOQ}\APACrefatitle {{Hot versus cold behavior in centipede games}} {{Hot versus cold behavior in centipede games}}.{\BBCQ}
\newblock
\APACjournalVolNumPages{Journal of the Economic Science Association}{6}{2}{226--238}.
\PrintBackRefs{\CurrentBib}

\bibitem [\protect \citeauthoryear {%
Gneezy%
, Rockenbach%
\BCBL {}\ \BBA {} Serra-Garcia%
}{%
Gneezy%
\ \protect \BOthers {.}}{%
{\protect \APACyear {2013}}%
}]{%
Gneezy2013}
\APACinsertmetastar {%
Gneezy2013}%
\begin{APACrefauthors}%
Gneezy, U.%
, Rockenbach, B.%
\BCBL {}\ \BBA {} Serra-Garcia, M.%
\end{APACrefauthors}%
\unskip\
\newblock
\APACrefYearMonthDay{2013}{}{}.
\newblock
{\BBOQ}\APACrefatitle {{Measuring lying aversion}} {{Measuring lying aversion}}.{\BBCQ}
\newblock
\APACjournalVolNumPages{Journal of Economic Behavior and Organization}{93}{}{293--300}.
\PrintBackRefs{\CurrentBib}

\bibitem [\protect \citeauthoryear {%
Grech%
, Nax%
\BCBL {}\ \BBA {} Soos%
}{%
Grech%
\ \protect \BOthers {.}}{%
{\protect \APACyear {2022}}%
}]{%
Grech2022}
\APACinsertmetastar {%
Grech2022}%
\begin{APACrefauthors}%
Grech, P\BPBI D.%
, Nax, H\BPBI H.%
\BCBL {}\ \BBA {} Soos, A.%
\end{APACrefauthors}%
\unskip\
\newblock
\APACrefYearMonthDay{2022}{}{}.
\newblock
{\BBOQ}\APACrefatitle {{Incentivization matters: A meta-perspective on dictator games}} {{Incentivization matters: A meta-perspective on dictator games}}.{\BBCQ}
\newblock
\APACjournalVolNumPages{Journal of the Economic Science Association}{8}{1}{34--44}.
\PrintBackRefs{\CurrentBib}

\bibitem [\protect \citeauthoryear {%
Henrich%
\ \protect \BOthers {.}}{%
Henrich%
\ \protect \BOthers {.}}{%
{\protect \APACyear {2005}}%
}]{%
Henrich2005}
\APACinsertmetastar {%
Henrich2005}%
\begin{APACrefauthors}%
Henrich, J.%
, Boyd, R.%
, Bowles, S.%
, Camerer, C.%
, Fehr, E.%
, Gintis, H.%
\BDBL {}Tracer, D.%
\end{APACrefauthors}%
\unskip\
\newblock
\APACrefYearMonthDay{2005}{}{}.
\newblock
{\BBOQ}\APACrefatitle {{“Economic man” in cross-cultural perspective: Behavioral experiments in 15 small-scale societies}} {{“Economic man” in cross-cultural perspective: Behavioral experiments in 15 small-scale societies}}.{\BBCQ}
\newblock
\APACjournalVolNumPages{Behavioral and Brain Sciences}{28}{06}{795--855}.
\PrintBackRefs{\CurrentBib}

\bibitem [\protect \citeauthoryear {%
Henrich%
\ \protect \BOthers {.}}{%
Henrich%
\ \protect \BOthers {.}}{%
{\protect \APACyear {2016}}%
}]{%
Henrich2016}
\APACinsertmetastar {%
Henrich2016}%
\begin{APACrefauthors}%
Henrich, J.%
, Ensminger, J.%
, McElreath, R.%
, Barr, A.%
, Bolyanatz, A.%
, Cardenas, J\BPBI C.%
\BDBL {}Ziker, J.%
\end{APACrefauthors}%
\unskip\
\newblock
\APACrefYearMonthDay{2016}{}{}.
\newblock
{\BBOQ}\APACrefatitle {Markets , Religion , Community Size , and the Evolution of Fairness and Punishment} {Markets , religion , community size , and the evolution of fairness and punishment}.{\BBCQ}
\newblock
\APACjournalVolNumPages{Science}{327}{5972}{1480--1484}.
\PrintBackRefs{\CurrentBib}

\bibitem [\protect \citeauthoryear {%
Herne%
, Hietanen%
, Lappalainen%
\BCBL {}\ \BBA {} Palosaari%
}{%
Herne%
\ \protect \BOthers {.}}{%
{\protect \APACyear {2022}}%
}]{%
Herne2022}
\APACinsertmetastar {%
Herne2022}%
\begin{APACrefauthors}%
Herne, K.%
, Hietanen, J\BPBI K.%
, Lappalainen, O.%
\BCBL {}\ \BBA {} Palosaari, E.%
\end{APACrefauthors}%
\unskip\
\newblock
\APACrefYearMonthDay{2022}{}{}.
\newblock
{\BBOQ}\APACrefatitle {{The influence of role awareness, empathy induction and trait empathy on dictator game giving}} {{The influence of role awareness, empathy induction and trait empathy on dictator game giving}}.{\BBCQ}
\newblock
\APACjournalVolNumPages{PLoS ONE}{17}{3}{1--19}.
\PrintBackRefs{\CurrentBib}

\bibitem [\protect \citeauthoryear {%
Huang%
, Greene%
\BCBL {}\ \BBA {} Bazerman%
}{%
Huang%
\ \protect \BOthers {.}}{%
{\protect \APACyear {2019}}%
}]{%
Huang2019}
\APACinsertmetastar {%
Huang2019}%
\begin{APACrefauthors}%
Huang, K.%
, Greene, J\BPBI D.%
\BCBL {}\ \BBA {} Bazerman, M.%
\end{APACrefauthors}%
\unskip\
\newblock
\APACrefYearMonthDay{2019}{}{}.
\newblock
{\BBOQ}\APACrefatitle {{Veil-of-ignorance reasoning favors the greater good}} {{Veil-of-ignorance reasoning favors the greater good}}.{\BBCQ}
\newblock
\APACjournalVolNumPages{Proceedings of the National Academy of Sciences of the United States of America}{116}{48}{23989--23995}.
\PrintBackRefs{\CurrentBib}

\bibitem [\protect \citeauthoryear {%
Iriberri%
\ \BBA {} Rey-Biel%
}{%
Iriberri%
\ \BBA {} Rey-Biel%
}{%
{\protect \APACyear {2011}}%
}]{%
Iriberri2011}
\APACinsertmetastar {%
Iriberri2011}%
\begin{APACrefauthors}%
Iriberri, N.%
\BCBT {}\ \BBA {} Rey-Biel, P.%
\end{APACrefauthors}%
\unskip\
\newblock
\APACrefYearMonthDay{2011}{}{}.
\newblock
{\BBOQ}\APACrefatitle {{The role of role uncertainty in modified dictator games}} {{The role of role uncertainty in modified dictator games}}.{\BBCQ}
\newblock
\APACjournalVolNumPages{Experimental Economics}{14}{2}{160--180}.
\PrintBackRefs{\CurrentBib}

\bibitem [\protect \citeauthoryear {%
Kay%
\ \BBA {} Ross%
}{%
Kay%
\ \BBA {} Ross%
}{%
{\protect \APACyear {2003}}%
}]{%
KAY2003}
\APACinsertmetastar {%
KAY2003}%
\begin{APACrefauthors}%
Kay, A\BPBI C.%
\BCBT {}\ \BBA {} Ross, L.%
\end{APACrefauthors}%
\unskip\
\newblock
\APACrefYearMonthDay{2003}{}{}.
\newblock
{\BBOQ}\APACrefatitle {The perceptual push: The interplay of implicit cues and explicit situational construals on behavioral intentions in the Prisoner’s Dilemma} {The perceptual push: The interplay of implicit cues and explicit situational construals on behavioral intentions in the prisoner’s dilemma}.{\BBCQ}
\newblock
\APACjournalVolNumPages{Journal of Experimental Social Psychology}{39}{6}{634-643}.
\PrintBackRefs{\CurrentBib}

\bibitem [\protect \citeauthoryear {%
Kirman%
\ \BBA {} Teschl%
}{%
Kirman%
\ \BBA {} Teschl%
}{%
{\protect \APACyear {2010}}%
}]{%
Kirman2010}
\APACinsertmetastar {%
Kirman2010}%
\begin{APACrefauthors}%
Kirman, A.%
\BCBT {}\ \BBA {} Teschl, M.%
\end{APACrefauthors}%
\unskip\
\newblock
\APACrefYearMonthDay{2010}{}{}.
\newblock
{\BBOQ}\APACrefatitle {{Do markets foster selfishness?}} {{Do markets foster selfishness?}}{\BBCQ}
\newblock
\APACjournalVolNumPages{Revue de philosophie {\'{e}}conomique}{Vol. 11}{1}{113--140}.
\PrintBackRefs{\CurrentBib}

\bibitem [\protect \citeauthoryear {%
Kranton%
}{%
Kranton%
}{%
{\protect \APACyear {2019}}%
}]{%
kranton2019}
\APACinsertmetastar {%
kranton2019}%
\begin{APACrefauthors}%
Kranton, R.%
\end{APACrefauthors}%
\unskip\
\newblock
\APACrefYearMonthDay{2019}{}{}.
\newblock
{\BBOQ}\APACrefatitle {The Devil Is in the Details: Implications of {S}amuel {B}owles’s \emph{{T}he {M}oral {E}conomy} for Economics and Policy Research} {The devil is in the details: Implications of {S}amuel {B}owles’s \emph{{T}he {M}oral {E}conomy} for economics and policy research}.{\BBCQ}
\newblock
\APACjournalVolNumPages{Journal of Economic Literature}{57}{1}{147--160}.
\PrintBackRefs{\CurrentBib}

\bibitem [\protect \citeauthoryear {%
D\BPBI K.~Levine%
}{%
D\BPBI K.~Levine%
}{%
{\protect \APACyear {1998}}%
}]{%
Levine1998}
\APACinsertmetastar {%
Levine1998}%
\begin{APACrefauthors}%
Levine, D\BPBI K.%
\end{APACrefauthors}%
\unskip\
\newblock
\APACrefYearMonthDay{1998}{}{}.
\newblock
{\BBOQ}\APACrefatitle {Modeling Altruism and Spitefulness in Experiments} {Modeling altruism and spitefulness in experiments}.{\BBCQ}
\newblock
\APACjournalVolNumPages{Review of Economic Dynamics}{1}{3}{593--622}.
\PrintBackRefs{\CurrentBib}

\bibitem [\protect \citeauthoryear {%
S.~Levine%
, Kleiman-Weiner%
, Schulz%
, Tenenbaum%
\BCBL {}\ \BBA {} Cushman%
}{%
S.~Levine%
\ \protect \BOthers {.}}{%
{\protect \APACyear {2020}}%
}]{%
Levine2020}
\APACinsertmetastar {%
Levine2020}%
\begin{APACrefauthors}%
Levine, S.%
, Kleiman-Weiner, M.%
, Schulz, L.%
, Tenenbaum, J.%
\BCBL {}\ \BBA {} Cushman, F.%
\end{APACrefauthors}%
\unskip\
\newblock
\APACrefYearMonthDay{2020}{}{}.
\newblock
{\BBOQ}\APACrefatitle {{The logic of universalization guides moral judgment}} {{The logic of universalization guides moral judgment}}.{\BBCQ}
\newblock
\APACjournalVolNumPages{Proceedings of the National Academy of Sciences of the United States of America}{117}{42}{26158--26169}.
\PrintBackRefs{\CurrentBib}

\bibitem [\protect \citeauthoryear {%
Liberman%
, Samuels%
\BCBL {}\ \BBA {} Ross%
}{%
Liberman%
\ \protect \BOthers {.}}{%
{\protect \APACyear {2004}}%
}]{%
Liberman2004}
\APACinsertmetastar {%
Liberman2004}%
\begin{APACrefauthors}%
Liberman, V.%
, Samuels, S\BPBI M.%
\BCBL {}\ \BBA {} Ross, L.%
\end{APACrefauthors}%
\unskip\
\newblock
\APACrefYearMonthDay{2004}{}{}.
\newblock
{\BBOQ}\APACrefatitle {{The name of the game: Predictive power of reputations versus situational labels in determining Prisoner's Dilemma game moves}} {{The name of the game: Predictive power of reputations versus situational labels in determining Prisoner's Dilemma game moves}}.{\BBCQ}
\newblock
\APACjournalVolNumPages{Personality and Social Psychology Bulletin}{30}{9}{1175--1185}.
\PrintBackRefs{\CurrentBib}

\bibitem [\protect \citeauthoryear {%
Miettinen%
, Kosfeld%
, Fehr%
\BCBL {}\ \BBA {} Weibull%
}{%
Miettinen%
\ \protect \BOthers {.}}{%
{\protect \APACyear {2020}}%
}]{%
weibull2019horserace}
\APACinsertmetastar {%
weibull2019horserace}%
\begin{APACrefauthors}%
Miettinen, T.%
, Kosfeld, M.%
, Fehr, E.%
\BCBL {}\ \BBA {} Weibull, J\BPBI W.%
\end{APACrefauthors}%
\unskip\
\newblock
\APACrefYearMonthDay{2020}{}{}.
\newblock
{\BBOQ}\APACrefatitle {Revealed preferences in a sequential prisoners' dilemma: {A} horse-race between six utility functions} {Revealed preferences in a sequential prisoners' dilemma: {A} horse-race between six utility functions}.{\BBCQ}
\newblock
\APACjournalVolNumPages{Journal of Economic Behavior and Organization}{173}{}{1-25}.
\PrintBackRefs{\CurrentBib}

\bibitem [\protect \citeauthoryear {%
Ortiz-Riomalo%
, Koessler%
\BCBL {}\ \BBA {} Engel%
}{%
Ortiz-Riomalo%
\ \protect \BOthers {.}}{%
{\protect \APACyear {2021}}%
}]{%
Ortiz-Riomalo2021}
\APACinsertmetastar {%
Ortiz-Riomalo2021}%
\begin{APACrefauthors}%
Ortiz-Riomalo, J\BPBI F.%
, Koessler, A\BPBI K.%
\BCBL {}\ \BBA {} Engel, S.%
\end{APACrefauthors}%
\unskip\
\newblock
\APACrefYearMonthDay{2021}{}{}.
\newblock
{\BBOQ}\APACrefatitle {{Inducing perspective-taking for prosocial behaviour in natural resource management}} {{Inducing perspective-taking for prosocial behaviour in natural resource management}}.{\BBCQ}
\newblock
\APACjournalVolNumPages{Journal of Environmental Economics and Management}{110}{August}{102513}.
\PrintBackRefs{\CurrentBib}

\bibitem [\protect \citeauthoryear {%
Rivero-Wildemauwe%
}{%
Rivero-Wildemauwe%
}{%
{\protect \APACyear {2023}}%
{\protect \APACexlab {{\protect \BCnt {1}}}}}]{%
wildemauwe2023}
\APACinsertmetastar {%
wildemauwe2023}%
\begin{APACrefauthors}%
Rivero-Wildemauwe, J\BPBI I.%
\end{APACrefauthors}%
\unskip\
\newblock
\APACrefYearMonthDay{2023{\protect \BCnt {1}}}{}{}.
\newblock
\APACrefbtitle {Moral motivations in sequential buyer-seller interactions with adverse selection} {Moral motivations in sequential buyer-seller interactions with adverse selection}\ \APACbVolEdTR{}{\BTR{}}.
\newblock
\APACaddressInstitution{}{THEMA (TH{\'e}orie Economique, Mod{\'e}lisation et Applications), CY Cergy Paris Université}.
\PrintBackRefs{\CurrentBib}

\bibitem [\protect \citeauthoryear {%
Rivero-Wildemauwe%
}{%
Rivero-Wildemauwe%
}{%
{\protect \APACyear {2023}}%
{\protect \APACexlab {{\protect \BCnt {2}}}}}]{%
wildemauwe2023a}
\APACinsertmetastar {%
wildemauwe2023a}%
\begin{APACrefauthors}%
Rivero-Wildemauwe, J\BPBI I.%
\end{APACrefauthors}%
\unskip\
\newblock
\APACrefYearMonthDay{2023{\protect \BCnt {2}}}{}{}.
\newblock
\APACrefbtitle {Trade among moral agents with information asymmetries} {Trade among moral agents with information asymmetries}\ \APACbVolEdTR{}{\BTR{}}.
\newblock
\APACaddressInstitution{}{THEMA (TH{\'e}orie Economique, Mod{\'e}lisation et Applications), CY Cergy Paris Université}.
\PrintBackRefs{\CurrentBib}

\bibitem [\protect \citeauthoryear {%
Roemer%
}{%
Roemer%
}{%
{\protect \APACyear {2010}}%
}]{%
Roemer2010}
\APACinsertmetastar {%
Roemer2010}%
\begin{APACrefauthors}%
Roemer, J\BPBI E.%
\end{APACrefauthors}%
\unskip\
\newblock
\APACrefYearMonthDay{2010}{}{}.
\newblock
{\BBOQ}\APACrefatitle {Kantian Equilibrium} {Kantian equilibrium}.{\BBCQ}
\newblock
\APACjournalVolNumPages{Scandinavian Journal of Economics}{112}{1}{1--24}.
\PrintBackRefs{\CurrentBib}

\bibitem [\protect \citeauthoryear {%
Roth%
\ \BBA {} Schoumaker%
}{%
Roth%
\ \BBA {} Schoumaker%
}{%
{\protect \APACyear {1983}}%
}]{%
Roth1983}
\APACinsertmetastar {%
Roth1983}%
\begin{APACrefauthors}%
Roth, B\BPBI A\BPBI E.%
\BCBT {}\ \BBA {} Schoumaker, F.%
\end{APACrefauthors}%
\unskip\
\newblock
\APACrefYearMonthDay{1983}{}{}.
\newblock
{\BBOQ}\APACrefatitle {Expectations and Reputations in Bargaining: {A}n Experimental Study} {Expectations and reputations in bargaining: {A}n experimental study}.{\BBCQ}
\newblock
\APACjournalVolNumPages{American Economic Review}{73}{3}{362--372}.
\PrintBackRefs{\CurrentBib}

\bibitem [\protect \citeauthoryear {%
Shafir%
\ \BBA {} Tversky%
}{%
Shafir%
\ \BBA {} Tversky%
}{%
{\protect \APACyear {1992}}%
}]{%
Shafir1992}
\APACinsertmetastar {%
Shafir1992}%
\begin{APACrefauthors}%
Shafir, E.%
\BCBT {}\ \BBA {} Tversky, A.%
\end{APACrefauthors}%
\unskip\
\newblock
\APACrefYearMonthDay{1992}{}{}.
\newblock
{\BBOQ}\APACrefatitle {{Thinking through uncertainty: Nonconsequential reasoning and choice}} {{Thinking through uncertainty: Nonconsequential reasoning and choice}}.{\BBCQ}
\newblock
\APACjournalVolNumPages{Cognitive Psychology}{24}{4}{449--474}.
\PrintBackRefs{\CurrentBib}

\bibitem [\protect \citeauthoryear {%
Sutter%
, Huber%
, Kirchler%
, Stefan%
\BCBL {}\ \BBA {} Walzl%
}{%
Sutter%
\ \protect \BOthers {.}}{%
{\protect \APACyear {2020}}%
}]{%
Sutter2020}
\APACinsertmetastar {%
Sutter2020}%
\begin{APACrefauthors}%
Sutter, M.%
, Huber, J.%
, Kirchler, M.%
, Stefan, M.%
\BCBL {}\ \BBA {} Walzl, M.%
\end{APACrefauthors}%
\unskip\
\newblock
\APACrefYearMonthDay{2020}{}{}.
\newblock
{\BBOQ}\APACrefatitle {Where to look for the morals in markets?} {Where to look for the morals in markets?}{\BBCQ}
\newblock
\APACjournalVolNumPages{Experimental Economics}{23}{1}{30--52}.
\PrintBackRefs{\CurrentBib}

\bibitem [\protect \citeauthoryear {%
Sutter%
\ \BBA {} Weck-Hannemann%
}{%
Sutter%
\ \BBA {} Weck-Hannemann%
}{%
{\protect \APACyear {2003}}%
}]{%
Sutter2003}
\APACinsertmetastar {%
Sutter2003}%
\begin{APACrefauthors}%
Sutter, M.%
\BCBT {}\ \BBA {} Weck-Hannemann, H.%
\end{APACrefauthors}%
\unskip\
\newblock
\APACrefYearMonthDay{2003}{}{}.
\newblock
{\BBOQ}\APACrefatitle {{Taxation and the veil of ignorance - A real effort experiment on the Laffer curve}} {{Taxation and the veil of ignorance - A real effort experiment on the Laffer curve}}.{\BBCQ}
\newblock
\APACjournalVolNumPages{Public Choice}{115}{1-2}{217--240}.
\PrintBackRefs{\CurrentBib}

\bibitem [\protect \citeauthoryear {%
Th\"{o}ni%
\ \BBA {} G\"{a}chter%
}{%
Th\"{o}ni%
\ \BBA {} G\"{a}chter%
}{%
{\protect \APACyear {2015}}%
}]{%
THONI2015}
\APACinsertmetastar {%
THONI2015}%
\begin{APACrefauthors}%
Th\"{o}ni, C.%
\BCBT {}\ \BBA {} G\"{a}chter, S.%
\end{APACrefauthors}%
\unskip\
\newblock
\APACrefYearMonthDay{2015}{}{}.
\newblock
{\BBOQ}\APACrefatitle {Peer effects and social preferences in voluntary cooperation: A theoretical and experimental analysis} {Peer effects and social preferences in voluntary cooperation: A theoretical and experimental analysis}.{\BBCQ}
\newblock
\APACjournalVolNumPages{Journal of Economic Psychology}{48}{}{72-88}.
\PrintBackRefs{\CurrentBib}

\bibitem [\protect \citeauthoryear {%
Van~Leeuwen%
\ \BBA {} Alger%
}{%
Van~Leeuwen%
\ \BBA {} Alger%
}{%
{\protect \APACyear {{\protect \BIP {}}}}%
}]{%
vanleeuwen23}
\APACinsertmetastar {%
vanleeuwen23}%
\begin{APACrefauthors}%
Van~Leeuwen, B.%
\BCBT {}\ \BBA {} Alger, I.%
\end{APACrefauthors}%
\unskip\
\newblock
\APACrefYearMonthDay{{\protect \BIP {}}}{}{}.
\newblock
{\BBOQ}\APACrefatitle {Estimating Social Preferences and {K}antian Morality in Strategic Interactions} {Estimating social preferences and {K}antian morality in strategic interactions}.{\BBCQ}
\newblock
\APACjournalVolNumPages{Journal of Political Economy Microeconomics}{}{}{}.
\PrintBackRefs{\CurrentBib}

\bibitem [\protect \citeauthoryear {%
Zelmer%
}{%
Zelmer%
}{%
{\protect \APACyear {2003}}%
}]{%
Zelmer2003}
\APACinsertmetastar {%
Zelmer2003}%
\begin{APACrefauthors}%
Zelmer, J.%
\end{APACrefauthors}%
\unskip\
\newblock
\APACrefYearMonthDay{2003}{}{}.
\newblock
{\BBOQ}\APACrefatitle {{Linear public goods experiments: A meta-analysis}} {{Linear public goods experiments: A meta-analysis}}.{\BBCQ}
\newblock
\APACjournalVolNumPages{Experimental Economics}{6}{3}{299--310}.
\PrintBackRefs{\CurrentBib}

\end{thebibliography}

\begin{appendices}

\renewcommand{\thesection}{A\arabic{section}}
\setcounter{table}{0} \renewcommand{\thetable}{A.\arabic{table}} 
\setcounter{figure}{0} \renewcommand{\thefigure}{A.\arabic{figure}} 

\newpage

\section{Additional tables and figures}\label{appendixA}
%\section*{Appendix A: Descriptive statistics (tables and figures)}

\begin{table}[ht!]
	\centering
	\renewcommand*{\arraystretch}{1.1}
	\caption{Participants' socio-demographic characteristics}
	\label{tab:subj_desc_table_age}
	\begin{tabular}{p{0.3\textwidth}r r p{0.25\textwidth}r r}
		\hline
		\hline
		Variable 	&	 N 	&	 Mean 	&	Variable 	&	 N 	&	 Mean 	\\
		\hline											
		\textbf{Age} 	&	453	&	21	&	\textbf{Education} 	&	413	&		\\
		\textbf{Gender}	&	453	&		&	... Highschool 	&	37	&	 8\% 	\\
		... Female 	&	193	&	 43\% 	&	... Undergrad. 	&	272	&	 61\% 	\\
		... Male 	&	260	&	 57\%	&	... Engineer 	&	5	&	 1\% 	 \\
		\textbf{Background} 	&	418	&		&	... Master 1 	&	68	&	 15\% 	  \\
		... Economics 	&	339	&	 81\% 	&	... Master 2 	&	44	&	 10\% 	\\
		... Sciences, Eng. 	&	10	&	 2\% 	&	... PhD 	&	3	&	 1\% 	\\
		... Political Science 	&	47	&	 11\% 	&	... Other 	&	19	&	 4\%	\\
		... Languages 	&	8	&	 2\% 	&	\textbf{Experience} 	&	453	&	  	\\
		... Literature, Philo. 	&	11	&	 3\% 	&	... Yes 	&	140	&	 31\% 	\\
		... History, Geog. 	&	3	&	 1\%	&	... No 	&	313	&	 69\%	 \\
		\textbf{Nationality} 	&	453	&		&	 \textbf{Couple} 	&	453	&		 \\
		... French 	&	362	&	 80\% 	&	 ... Yes 	&	66	&	 15\% 	\\
		... Other 	&	91	&	 20\%   	&	... No 	&	387	&	 85\%	 \\
		
		\hline
		\hline
	\end{tabular}
	\begin{tablenotes}
		\small
		\item For Age,  Std. Dev. =  3.9,  Min = 18,  Pctl. 25 =  19, Pctl. 75 = 22, Max = 54. Five participants did not report their education level, while 35 did not report their background.
	\end{tablenotes}
\end{table}
\begin{table}[ht!] 
\centering \renewcommand*{\arraystretch}{1.1}\caption{Participants' socio-demographic characteristics}\label{tab:subj_desc_table_treatment_age}\resizebox{\textwidth}{!}{
		\begin{tabular}{p{0.3\textwidth}wc{0.1\textwidth}wc{0.1\textwidth}wc{0.1\textwidth}wc{0.1\textwidth}wc{0.1\textwidth}wc{0.1\textwidth}wc{0.1\textwidth}wc{0.1\textwidth}wc{0.1\textwidth}}
			\hline\hline
			Treatment              &              \multicolumn{2}{c}{N}               &              \multicolumn{2}{c}{M}               &              \multicolumn{2}{c}{A}               &              \multicolumn{2}{c}{B}               &                            \\
			Variable               & \multicolumn{1}{c}{N} & \multicolumn{1}{c}{Mean} & \multicolumn{1}{c}{N} & \multicolumn{1}{c}{Mean} & \multicolumn{1}{c}{N} & \multicolumn{1}{c}{Mean} & \multicolumn{1}{c}{N} & \multicolumn{1}{c}{Mean} & \multicolumn{1}{c}{Test}   \\ \hline
			\textbf{Age}                    & 108                   &           21.2           & 112                   &            22            & 121                   &           20.5           & 112                   &           20.3           & F$:0.003^{***}$            \\
			\textbf{Gender}                 & 108                   &                          & 112                   &                          & 121                   &                          & 112                   &                          & $\chi^2:0.29^{}$           \\
			... Female             & 53                    &           49\%           & 44                    &           39\%           & 54                    &           45\%           & 42                    &           38\%           &                            \\
			... Male               & 55                    &           51\%           & 68                    &           61\%           & 67                    &           55\%           & 70                    &           62\%           &                            \\
			\textbf{Education}              & 108                   &                          & 112                   &                          & 118                   &                          & 110                   &                          & $\chi^2:0.12^{}$           \\
			... Highschool         & 10                    &           9\%            & 7                     &           6\%            & 11                    &           9\%            & 9                     &           8\%            &                            \\
			... Undergrad.         & 60                    &           56\%           & 67                    &           60\%           & 73                    &           62\%           & 72                    &           65\%           &                            \\
			... Engineer           & 1                     &           1\%            & 2                     &           2\%            & 1                     &           1\%            & 1                     &           1\%            &                            \\
			... Master 1           & 25                    &           23\%           & 18                    &           16\%           & 17                    &           14\%           & 8                     &           7\%            &                            \\
			... Master 2           & 10                    &           9\%            & 13                    &           12\%           & 11                    &           9\%            & 10                    &           9\%            &                            \\
			... PhD                & 1                     &           1\%            & 0                     &           0\%            & 2                     &           2\%            & 0                     &           0\%            &                            \\
			... Other              & 1                     &           1\%            & 5                     &           4\%            & 3                     &           3\%            & 10                    &           9\%            &                            \\
			\textbf{Background}             & 103                   &                          & 99                    &                          & 111                   &                          & 105                   &                          & $\chi^2:0.034^{**}$        \\
			... Economics          & 84                    &           82\%           & 78                    &           79\%           & 98                    &           88\%           & 79                    &           75\%           &                            \\
			... Sciences, Eng.     & 3                     &           3\%            & 2                     &           2\%            & 4                     &           4\%            & 1                     &           1\%            &                            \\
			... Political Science  & 9                     &           9\%            & 16                    &           16\%           & 7                     &           6\%            & 15                    &           14\%           &                            \\
			... Languages          & 1                     &           1\%            & 0                     &           0\%            & 1                     &           1\%            & 6                     &           6\%            &                            \\
			... Literature, Philo. & 5                     &           5\%            & 2                     &           2\%            & 0                     &           0\%            & 4                     &           4\%            &                            \\
			... History, Geog.     & 1                     &           1\%            & 1                     &           1\%            & 1                     &           1\%            & 0                     &           0\%            &                            \\
			\textbf{Experience}             & 108                   &                          & 112                   &                          & 121                   &                          & 112                   &                          & $\chi^2:0.023^{**}$        \\
			... Yes                & 28                    &           26\%           & 44                    &           39\%           & 28                    &           23\%           & 40                    &           36\%           &                            \\
			... No                 & 80                    &           74\%           & 68                    &           61\%           & 93                    &           77\%           & 72                    &           64\%           &                            \\
			\textbf{Couple}                 & 108                   &                          & 112                   &                          & 121                   &                          & 112                   &                          & $\chi^2:0.551^{}$          \\
			... Yes                & 15                    &           14\%           & 13                    &           12\%           & 22                    &           18\%           & 16                    &           14\%           &                            \\
			... No                 & 93                    &           86\%           & 99                    &           88\%           & 99                    &           82\%           & 96                    &           86\%           &                            \\
			\textbf{Nationality}            & 108                   &                          & 112                   &                          & 121                   &                          & 112                   &                          & $\chi^2:0.818^{}$          \\
			... French             & 83                    &           77\%           & 90                    &           80\%           & 99                    &           82\%           & 90                    &           80\%           &                            \\
			... Other              & 25                    &           23\%           & 22                    &           20\%           & 22                    &           18\%           & 22                    &           20\%           &                            \\ \hline\hline
		\end{tabular}
	}
	\begin{tablenotes}
		\small
		\item Standard deviations for Age are respectively: 3.6, 5.6, 2.9 and 2.4. The last column reports the p-value of an independence test between each variable and the treatment arm. An F test is used for the ``Age'' variable and ``$\chi^2$'' for the remaining ones. Statistical significance markers: * $p<0.1$; ** $p<0.05$; *** $p<0.01$.  Five participants did not report their education level, while 35 did not report their background.
	\end{tablenotes}
\end{table}

\begin{figure}[ht!]
\begin{center}
\includegraphics[scale=0.2]{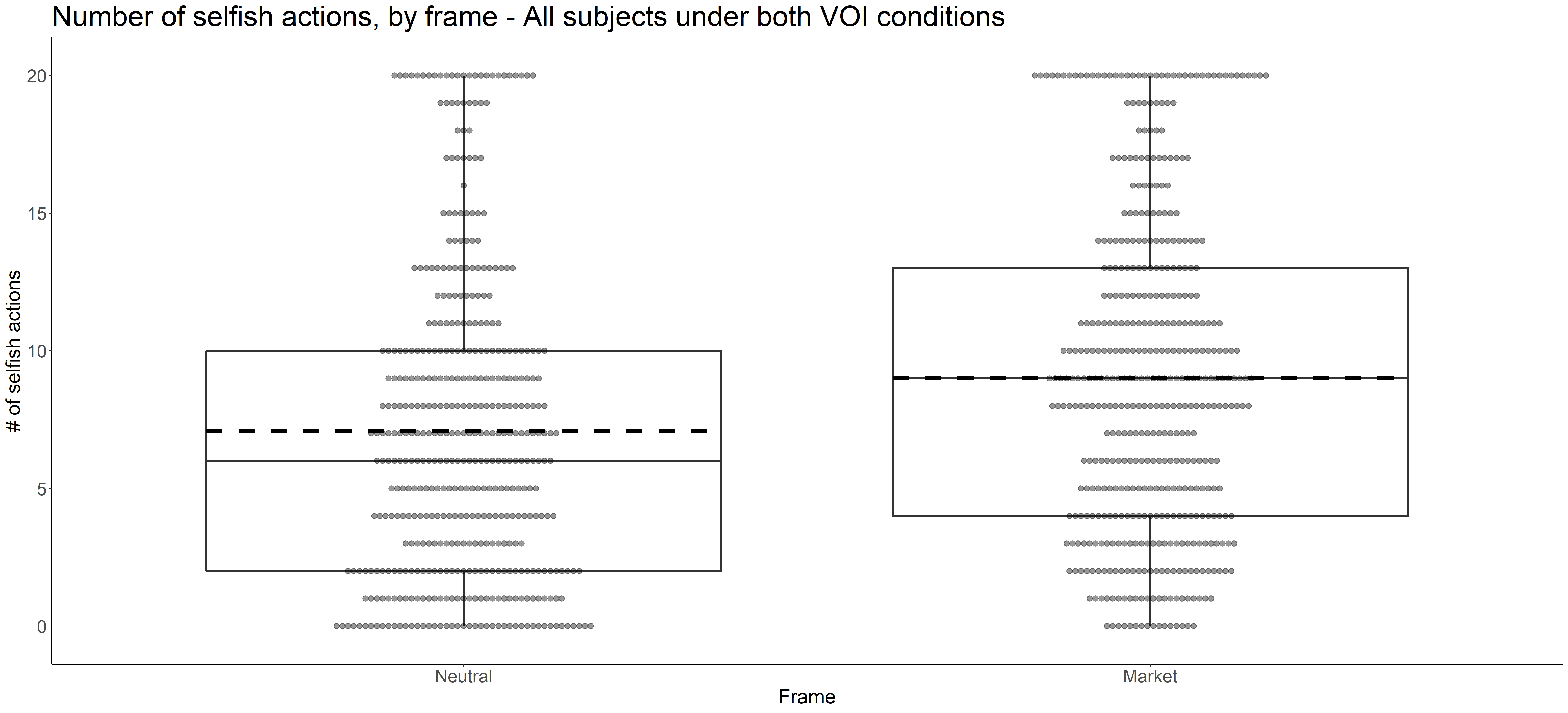}
\caption{Each dot represents one observation, defined as the total number of times that one subject selected the selfish option in one sequence of 20 decisions. Each subject was presented with two such sequences, hence each subject is represented by two observations. The left part shows the $n=453$ observations collected under the Neutral frame in sequences $N1$, $N2$, $A1$, and $B2$, while the right part shows the $n=453$ observations collected under the Market frame in sequences $M1$, $M2$, $A2$, and $B1$. The two-sided t-test for the difference in means results in a $p$-value of $0.0000004$. The Wilcoxon rank-sum test for the difference in medians results in a $p$-value of $0.0000004$.}
\label{Fig:frame}
\end{center}
\end{figure}

\begin{figure}[ht!]
\begin{center}
\includegraphics[scale=0.2]{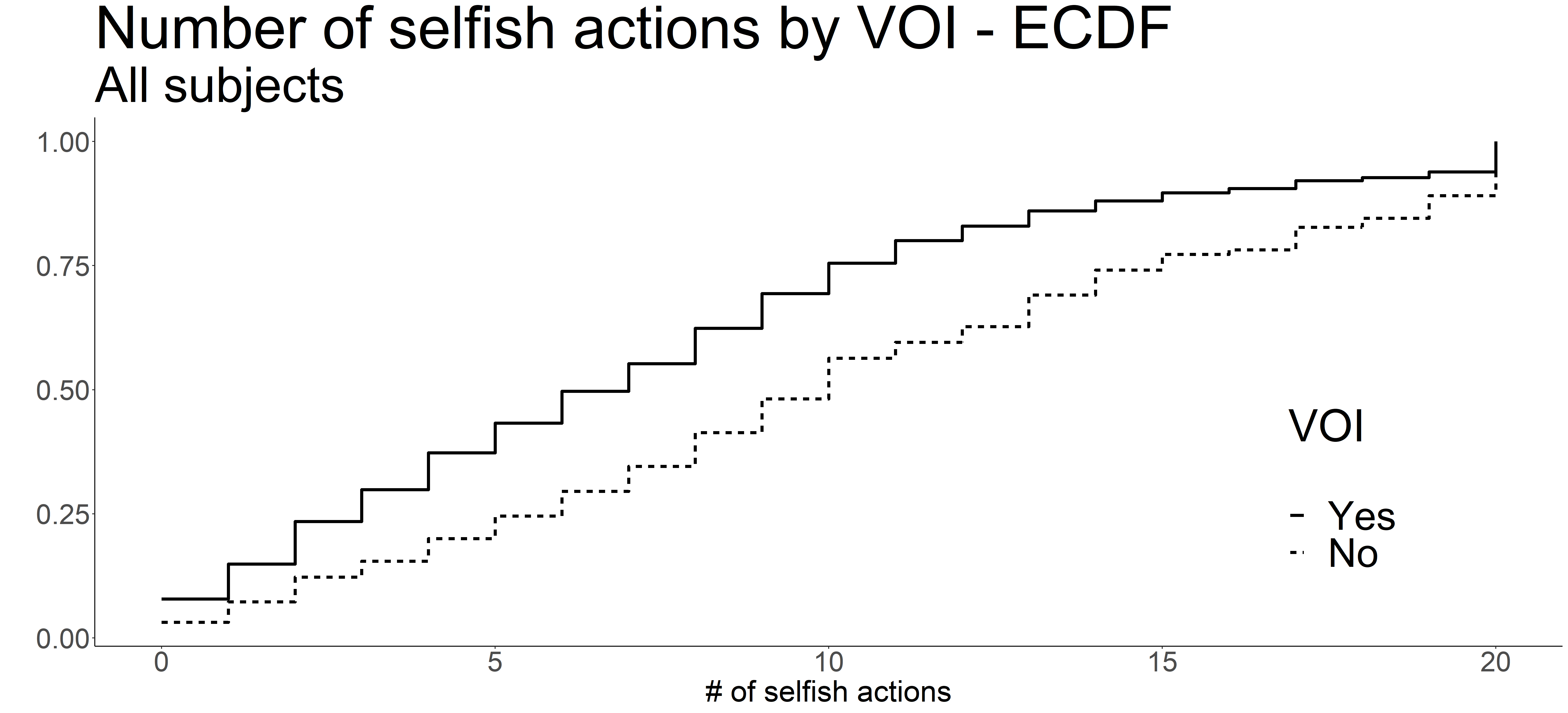}
\caption{Cumulative distributions of observations---defined as the total number of times that one subject selected the selfish option in one sequence of 20 decisions---for decisions collected in the VOI sequences (solid line, for the $n=686$ observations in  sequences $N2$, $M2$, $A1$, $A2$, $B1$, and $B2$) and those collected in the non-VOI sequences (dashed line, for the $n=220$ observations collected in sequences $N1$ and $M1$). The Kolmogorov–Smirnov test's $D$ statistic was 0.21206, while the \textit{p-value} was .0000006231. \label{Fig:voi}}
\label{Fig:voiandnonvoi}
\end{center}
\end{figure}
\begin{figure}[ht!]
\begin{center}
\includegraphics[scale=0.2]{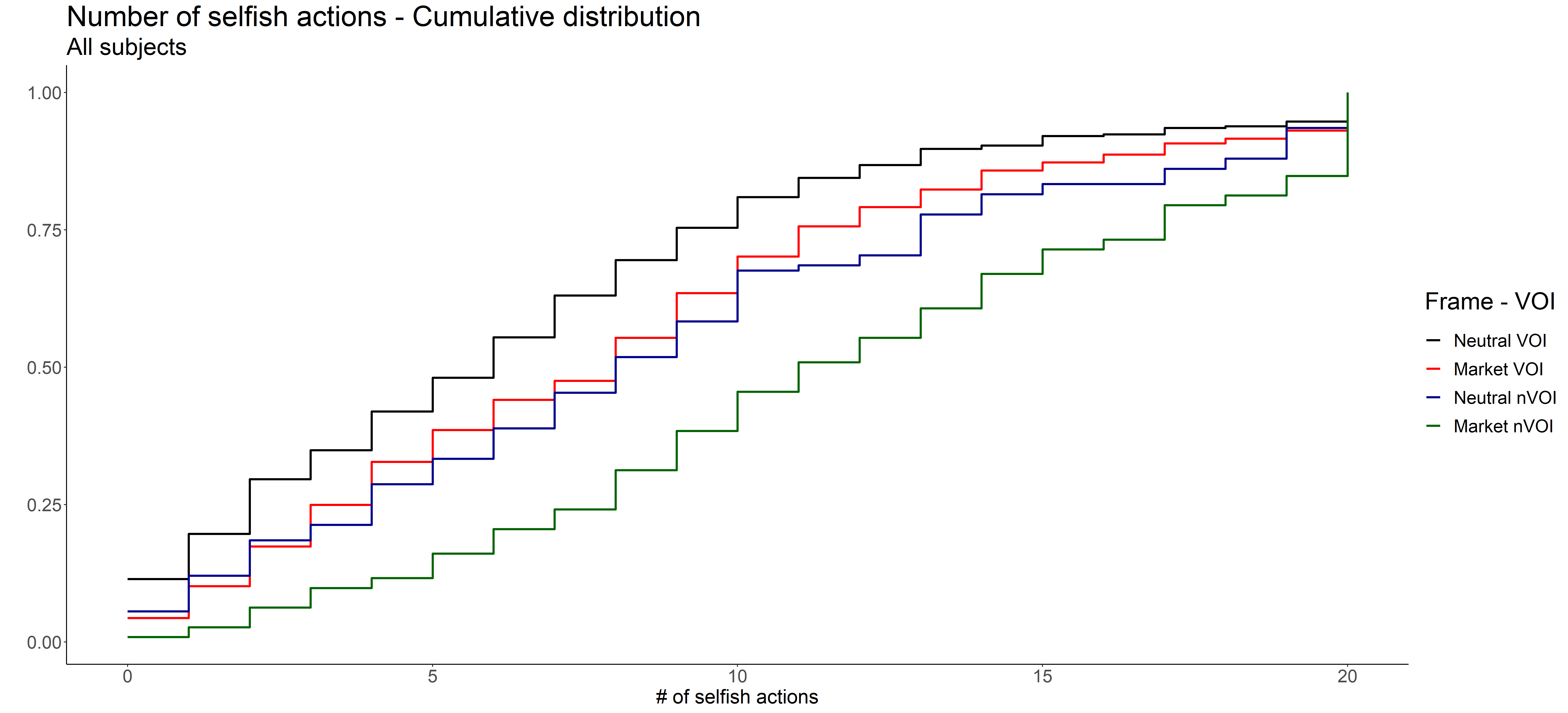}
\caption{Cumulative distributions of observations---defined as the total number of times that one subject selected the selfish option in one sequence of 20 decisions---for decisions collected in the Neutral VOI sequences $N2$, $A1$, and $B2$ (black line, $n=341$), the Market VOI sequences $M2$, $A2$, and $B1$ (red line, $n=345$), the Neutral non-VOI sequence $N1$ (blue line, $n=108$), and the Market non-VOI sequences $M1$ (green line, $n=112$). All pairwise Kolmogorov-Smirnoff tests result in $p$-values at or below 0.01, with the exception of Market VOI versus Neutral non-VOI, which gives a $p$-value of 0.6.}
\label{Fig:voiandmarket}
\end{center}
\end{figure}

\begin{table}[ht!]
		\captionsetup{justification=centering}
	\caption{Share of subjects choosing selfish action - By Frame}
	\label{tab:sellshare_frame_all}
 \centering
	\begin{tabular}{|c|cc|cc|cc|}
		\hline
		             & \multicolumn{2}{c|}{n-VOI - between (N1-M1)} & \multicolumn{2}{c|}{VOI - between (A1-B1)} & \multicolumn{2}{c|}{VOI - within (A1-A2-B1-B2)} \\ \hline
		Payoff & Neutral &                Market                & Neutral &              Market              & Neutral &               Market                \\ \hline
		     1       &  0.69   &                 0.84                 &  0.59   &               0.80               &  0.65   &                0.77                 \\
		     2       &  0.69   &                 0.82                 &  0.55   &               0.58               &  0.60   &                0.65                 \\
		     3       &  0.64   &                 0.74                 &  0.54   &               0.56               &  0.62   &                0.66                 \\
		     4       &  0.49   &                 0.72                 &  0.48   &               0.49               &  0.50   &                0.50                 \\
		     5       &  0.53   &                 0.72                 &  0.45   &               0.61               &  0.45   &                0.57                 \\
		     6       &  0.53   &                 0.67                 &  0.42   &               0.44               &  0.40   &                0.48                 \\
		     7       &  0.45   &                 0.66                 &  0.37   &               0.47               &  0.34   &                0.48                 \\
		     8       &  0.51   &                 0.60                 &  0.36   &               0.34               &  0.31   &                0.39                 \\
		     9       &  0.52   &                 0.73                 &  0.38   &               0.53               &  0.35   &                0.52                 \\
		     10      &  0.39   &                 0.60                 &  0.31   &               0.25               &  0.23   &                0.32                 \\
		     11      &  0.38   &                 0.58                 &  0.29   &               0.30               &  0.27   &                0.33                 \\
		     12      &  0.36   &                 0.50                 &  0.32   &               0.32               &  0.26   &                0.34                 \\
		     13      &  0.42   &                 0.46                 &  0.32   &               0.42               &  0.36   &                0.40                 \\
		     14      &  0.34   &                 0.39                 &  0.24   &               0.29               &  0.25   &                0.31                 \\
		     15      &  0.31   &                 0.54                 &  0.36   &               0.30               &  0.28   &                0.30                 \\
		     16      &  0.36   &                 0.46                 &  0.30   &               0.21               &  0.23   &                0.24                 \\
		     17      &  0.24   &                 0.39                 &  0.24   &               0.34               &  0.19   &                0.33                 \\
		     18      &  0.44   &                 0.60                 &  0.25   &               0.24               &  0.19   &                0.26                 \\
		     19      &  0.29   &                 0.35                 &  0.35   &               0.27               &  0.26   &                0.29                 \\
		     20      &  0.29   &                 0.30                 &  0.26   &               0.24               &  0.21   &                0.25                 \\ \hline
		    Mean     &  0.44   &                 0.58                 &  0.37   &               0.40               &  0.35   &                0.42                 \\ \hline
	\end{tabular}
\end{table}
% \input{Figures/Tables/reg_frame_bet_nVOI.tex}

% %\input{Figures/Tables/tab:sellshare_frame_bet_nvoi}

% \input{Figures/Tables/reg_frame_bet_VOI_A1B1}

% %\input{Figures/Tables/tab:sellshare_frame_bet_voi}

% \input{Figures/Tables/reg_frame_with_VOI.tex}

% %\input{Figures/Tables/tab:sellshare_frame_with_voi}

% %Effect of frame under VOI conditions, between subjects: $t\in\{A1,B1\}$.
\newgeometry{left=0.5cm}

\begin{table}[ht!]

	\centering
	\captionsetup{justification=centering}
	\caption{Share of subjects choosing selfish action - By VOI}
	\label{tab:sellshare_voi_all}
 %\resizebox{1.1\textwidth}{!}{
	\begin{tabular}{|c|cc|cc|cc|cc|}
		\hline
		       & \multicolumn{2}{c|}{Neutral, bet. (N1-A1)} & \multicolumn{2}{c|}{Neutral, within (N1-N2)} & \multicolumn{2}{c|}{Market, within (M1-M2)} & \multicolumn{2}{c|}{Market, bet. (M1-B1)} \\ \hline
		Payoff & non-VOI &               VOI                & non-VOI &                VOI                 & non-VOI &                VOI                & non-VOI &               VOI               \\ \hline
		  1    &  0.69   &               0.59               &  0.69   &                0.60                &  0.84   &               0.83                &  0.84   &              0.80               \\
		  2    &  0.69   &               0.55               &  0.69   &                0.42                &  0.82   &               0.74                &  0.82   &              0.58               \\
		  3    &  0.64   &               0.54               &  0.64   &                0.60                &  0.74   &               0.83                &  0.74   &              0.56               \\
		  4    &  0.49   &               0.48               &  0.49   &                0.40                &  0.72   &               0.61                &  0.72   &              0.49               \\
		  5    &  0.53   &               0.45               &  0.53   &                0.40                &  0.72   &               0.53                &  0.72   &              0.61               \\
		  6    &  0.53   &               0.42               &  0.53   &                0.31                &  0.67   &               0.46                &  0.67   &              0.44               \\
		  7    &  0.45   &               0.37               &  0.45   &                0.30                &  0.66   &               0.41                &  0.66   &              0.47               \\
		  8    &  0.51   &               0.36               &  0.51   &                0.28                &  0.60   &               0.34                &  0.60   &              0.34               \\
		  9    &  0.52   &               0.38               &  0.52   &                0.30                &  0.73   &               0.39                &  0.73   &              0.53               \\
		  10   &  0.39   &               0.31               &  0.39   &                0.23                &  0.60   &               0.29                &  0.60   &              0.25               \\
		  11   &  0.38   &               0.29               &  0.38   &                0.26                &  0.58   &               0.31                &  0.58   &              0.30               \\
		  12   &  0.36   &               0.32               &  0.36   &                0.19                &  0.50   &               0.27                &  0.50   &              0.32               \\
		  13   &  0.42   &               0.32               &  0.42   &                0.29                &  0.46   &               0.32                &  0.46   &              0.42               \\
		  14   &  0.34   &               0.24               &  0.34   &                0.23                &  0.39   &               0.26                &  0.39   &              0.29               \\
		  15   &  0.31   &               0.36               &  0.31   &                0.23                &  0.54   &               0.26                &  0.54   &              0.30               \\
		  16   &  0.36   &               0.30               &  0.36   &                0.19                &  0.46   &               0.21                &  0.46   &              0.21               \\
		  17   &  0.24   &               0.24               &  0.24   &                0.17                &  0.39   &               0.27                &  0.39   &              0.34               \\
		  18   &  0.44   &               0.25               &  0.44   &                0.15                &  0.60   &               0.25                &  0.60   &              0.24               \\
		  19   &  0.29   &               0.35               &  0.29   &                0.15                &  0.35   &               0.18                &  0.35   &              0.27               \\
		  20   &  0.29   &               0.26               &  0.29   &                0.12                &  0.30   &               0.20                &  0.30   &              0.24               \\ \hline
		 Mean  &  0.44   &               0.37               &  0.44   &                0.29                &  0.58   &               0.40                &  0.58   &              0.40               \\ \hline
	\end{tabular}
 %}

\end{table}
\restoregeometry

% \input{Figures/Tables/reg_VOI_with_neutral.tex}

% %\input{Figures/Tables/tab:sellshare_voi_with_neut}

% \input{Figures/Tables/reg_VOI_bet_neutral.tex}

% %\input{Figures/Tables/tab:sellshare_voi_bet_neut}

% \input{Figures/Tables/reg_VOI_with_mkt.tex}

% %\input{Figures/Tables/tab:sellshare_voi_with_mkt}

% \input{Figures/Tables/reg_VOI_bet_mkt.tex}

% %\input{Figures/Tables/tab:sellshare_voi_bet_mkt}

\begin{table}[htb!]
 \hspace*{-1.8cm}  \centering
   \begin{threeparttable}[b]
      \caption{\label{reg_VOI_with_frame_bet} Effects of VOI (within) and Market frame (between) on Selling (N1-N2-M1-M2)}
      \begin{tabular}{lccccc}
         \tabularnewline \midrule \midrule
         Dependent Variable: & \multicolumn{5}{c}{Sell (Yes or No)}\\
         Model:              & (1)             & (2)             & (3)             & (4)             & (5)\\  
         \midrule
         \emph{Variables}\\
         VOI                 & -0.1528$^{***}$ & -0.1528$^{***}$ & -0.1528$^{***}$ & -0.1528$^{***}$ & -0.1528$^{***}$\\   
                             & (0.0147)        & (0.0141)        & (0.0276)        & (0.0277)        & (0.0294)\\   
         Market               & 0.1413$^{***}$  & 0.1413$^{***}$  & 0.1413$^{***}$  & 0.1413$^{***}$  & 0.1411$^{***}$\\   
                             & (0.0146)        & (0.0140)        & (0.0396)        & (0.0396)        & (0.0403)\\   
         VOI x Market         & -0.0338         & -0.0338$^{*}$   & -0.0338         & -0.0338         & -0.0338\\   
                             & (0.0206)        & (0.0197)        & (0.0382)        & (0.0383)        & (0.0412)\\   
         z                   &                 & 0.9079$^{***}$  & 0.9079$^{***}$  &                 &   \\   
                             &                 & (0.0315)        & (0.0623)        &                 &   \\   
         Socio-dem. controls & No              & No              & No              & No              & Yes\\  
         Clustering of SE    & No              & No              & Subject         & Subject         & Subject and Payoff\\  
         \midrule
         \emph{Fixed-effects}\\
         Payoff              &                 &                 &                 & Yes             & Yes\\  
         \midrule
         \emph{Fit statistics}\\
         Observations        & 8,800           & 8,800           & 8,800           & 8,800           & 8,800\\  
         R$^2$               & 0.04555         & 0.12801         & 0.12801         & 0.13370         & 0.13965\\  
         Within R$^2$        &                 &                 &                 & 0.04995         & 0.05648\\  
         \midrule \midrule
         \multicolumn{6}{l}{\emph{Signif. Codes: ***: 0.01, **: 0.05, *: 0.1}}\\
      \end{tabular}
      
      \begin{tablenotes}\item Note: Column (3) includes clustering of SE's at the Subject level, column (4) includes Payoff fixed effects with clustering of SE's at the Subject level,
                                        column (5) includes Payoff fixed effects with clustering of SE's at the Subject and Payoff level. Socio-demographic controls are: nationality, gender, couple status, 
                                        previous experience in experiments and attendance to a private school.
      \end{tablenotes}
   \end{threeparttable}
\end{table}

\begin{table}[ht!]
    \begin{threeparttable}
    \centering
    \caption{\label{tab:rep_agent_full}Estimated preferences for the rep. agent in Neutral and Market frames - Full sample}
    \begin{tabular}{lccc}
    \toprule
    \toprule
    & \multicolumn{1}{p{3cm}}{\centering \vspace{0.1cm} Neutral} & \multicolumn{1}{p{3cm}}{\centering  \vspace{0.1cm} Market} & \multicolumn{1}{p{3cm}}{\centering $H_0$: Neutral = Market\tnote{+}}\\
    \midrule
    \multirow{2}{*}{$\beta$: Aheadness aversion} & 0.230$^{***}$ & 0.106$^{***}$ & 0.003 \\ 
    & (0.034) - \textit{[0.013]} & (0.023) - \textit{[0.008]} &  \\ 
    \multirow{2}{*}{$\kappa$: Degree of morality} & 0.223$^{***}$ & 0.196$^{***}$ & 0.738 \\ 
    & (0.071) - \textit{[0.030]}  & (0.035) - \textit{[0.019]}&  \\ 
    \multirow{2}{*}{$\sigma$: Choice sensitivity} & 0.024$^{***}$ & 0.034$^{***}$ & 0.331 \\ 
    &  (0.007) - \textit{[0.002]}  & (0.007) - \textit{[0.002]} &  \\ 
    \midrule
    Number of Observations & 4320 & 4480 &  \\ 
    Number of subjects & 108 & 112 &  \\ 
    Log likelihood & -2665.795 & -2845.855 &  \\ 
    \hline
    \end{tabular}
    \begin{tablenotes}
    \item Notes: Standard errors clustered at the individual level in parentheses. Non-clustered standard errors in brackets. $^{***}$ Significant at 1\% using clustered standard errors.
    \item[+] $p$-value of $z$-test, using clustered standard errors.
    \end{tablenotes}
    \end{threeparttable}
\end{table}
	
\begin{table}[ht!]
    \begin{threeparttable}
    \centering
    \caption{\label{tab:rep_agent_c2} Estimated preferences for the representative agent in Neutral and Market frames - Core 2 sample}
    \begin{tabular}{lccc}
    \toprule
    \toprule
    & \multicolumn{1}{p{3cm}}{\centering \vspace{0.1cm} Neutral} & \multicolumn{1}{p{3cm}}{\centering  \vspace{0.1cm} Market} & \multicolumn{1}{p{3cm}}{\centering $H_0$: Neutral = Market\tnote{+}}\\
    \midrule
    \multirow{2}{*}{$\beta$: Aheadness aversion} & 0.147$^{***}$ & 0.089$^{***}$ & 0.159 \\ 
    & (0.035) - \textit{[0.013]} & (0.022) - \textit{[0.008]} &  \\ 
    \multirow{2}{*}{$\kappa$: Degree of morality} & 0.441$^{***}$ & 0.255$^{***}$ & 0.120 \\ 
    & (0.112) - \textit{[0.058]}  & (0.041) - \textit{[0.019]}&  \\ 
    \multirow{2}{*}{$\sigma$: Choice sensitivity} & 0.025$^{***}$ & 0.045$^{***}$ & 0.254 \\ 
    &  (0.011) - \textit{[0.003]}  & (0.013) - \textit{[0.003]} &  \\ 
    \midrule
    Number of Observations & 2,720 & 3,160 &  \\ 
    Number of subjects & 68 & 79 &  \\ 
    Log likelihood & -1.576.849 & -1,846.436 &  \\ 
    \hline
    \end{tabular}
    \begin{tablenotes}
    \item Notes: Standard errors clustered at the individual level in parentheses. Non-clustered standard errors in brackets. $^{***}$ Significant at 1\% using clustered standard errors.
    \item[+] $p$-value of $z$-test, using clustered standard errors.
    \end{tablenotes}
    \end{threeparttable}
\end{table}
\newpage

\begin{table}[ht!]
\centering
\begin{threeparttable}
\caption{\label{tab:finmix_f}Two-type Finite Mixture Model estimates, Full sample}
\begin{tabularx}{\textwidth}{>{\hsize=1\hsize}X *{4}{>{\hsize=0.625\hsize}X}}
\toprule
& \multicolumn{2}{c}{Neutral} & \multicolumn{2}{c}{Market} \\
\cmidrule(lr){2-3} \cmidrule(lr){4-5}
& Type 1 & Type 2 & Type 1 & Type 2 \\
\midrule
$\beta$: Aheadness aversion & 0.373$^{***}$ & -0.057$^{}$  &0.188$^{***}$ &-0.104$^{}$ \\
& \textit{(0.086)} &\textit{ (0.159)} &\textit{(0.012) }&\textit{(0.072)} \\ 
 & [0.026]  & [0.036]  & [0.007] & [0.034] \\ 
$\kappa$: Degree of morality & 0.125$^{***}$ & 0.309$^{***}$ &0.229$^{***}$ & 0.136$^{***}$ \\
& \textit{(0.064)} &\textit{ (0.054)} &\textit{(0.020) }&\textit{(0.040)} \\ 
& [0.034] & [0.053] & [0.020] & [0.033] \\ 
$\sigma$: Choice sensitivity  & 0.043$^{***}$ & 0.026$^{***}$ & 0.055$^{***}$& 0.037$^{***}$ \\
& \textit{(0.009)} &\textit{ (0.003)} &\textit{(0.011) }&\textit{(0.011)} \\ 
& [0.005] &[0.004] &[0.004] & [0.005]\\ 
Share & 0.553$^{***}$ &0.447$^{***}$ & 0.643$^{***}$ & 0.357 $^{***}$ \\
& \textit{(0.134)} &\textit{ (0.134)} &\textit{(0.026) }&\textit{(0.026)} \\ 
& [0.053] & [0.053] & [0.046]& [0.046] \\ 
\midrule
Number of Observations & \multicolumn{2}{c}{4,320} & \multicolumn{2}{c}{4,480} \\ 
Number of subjects & \multicolumn{2}{c}{108} & \multicolumn{2}{c}{112} \\ 
Log likelihood & \multicolumn{2}{c}{-2297.761} & \multicolumn{2}{c}{-2403.156} \\ 
\bottomrule
\end{tabularx}
\begin{tablenotes}
\item Notes: Standard errors clustered at the individual level in parenthesis, non-clustered standard errors in brackets. $^{***}$ Significant at 1\%, $^{**}$ Significant at 5\%, $^{*}$ Significant at 10\% using clustered standard errors.
\end{tablenotes}
\end{threeparttable}
\end{table}
\newpage

\begin{figure}[ht!]
    \centering
    \begin{subfigure}[b]{0.45\textwidth}
        \centering
        \includegraphics[width=\textwidth]{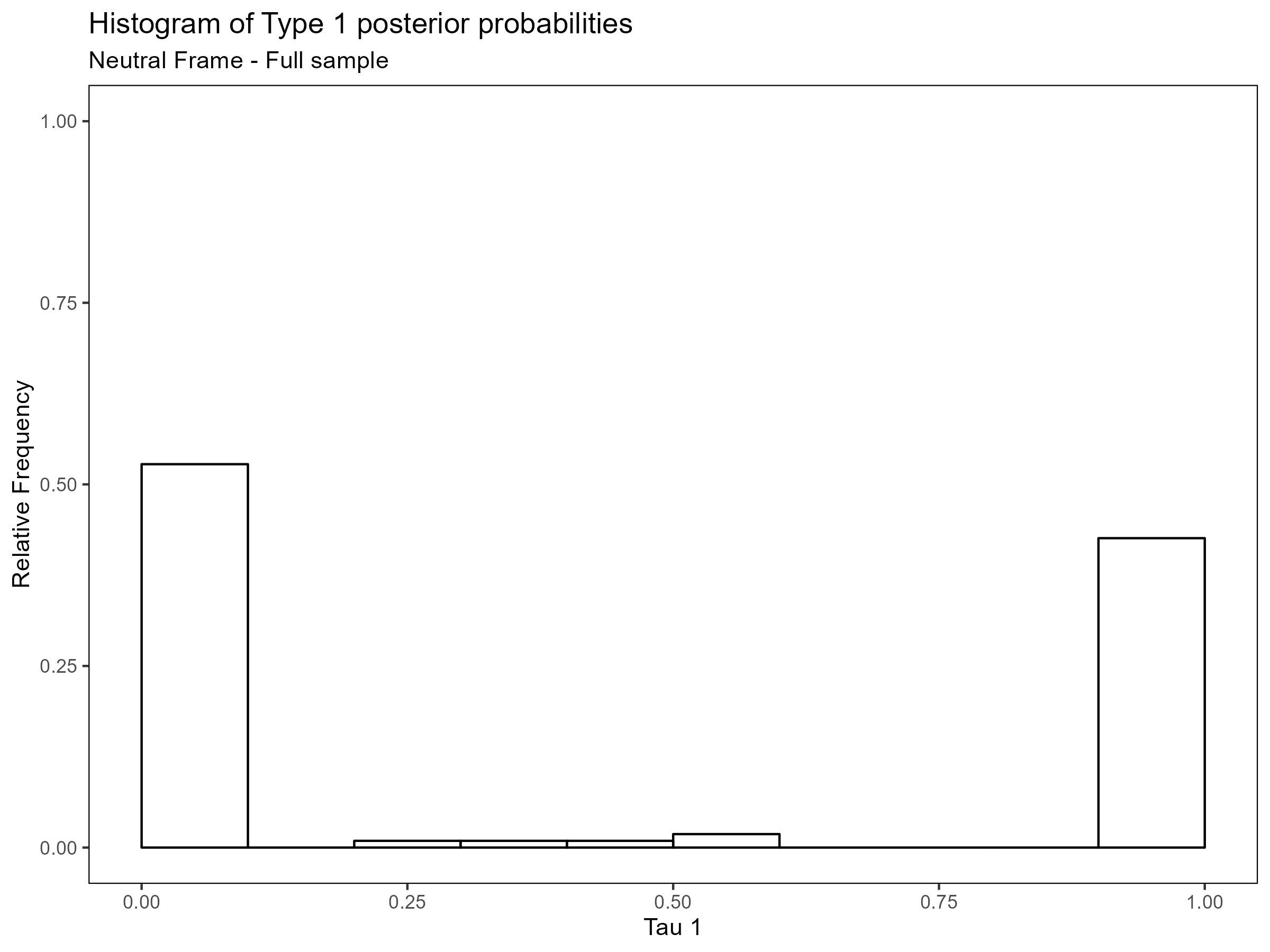}
        
        \label{fig:sub1}
    \end{subfigure}
    \hfill
    \begin{subfigure}[b]{0.45\textwidth}
        \centering
        \includegraphics[width=\textwidth]{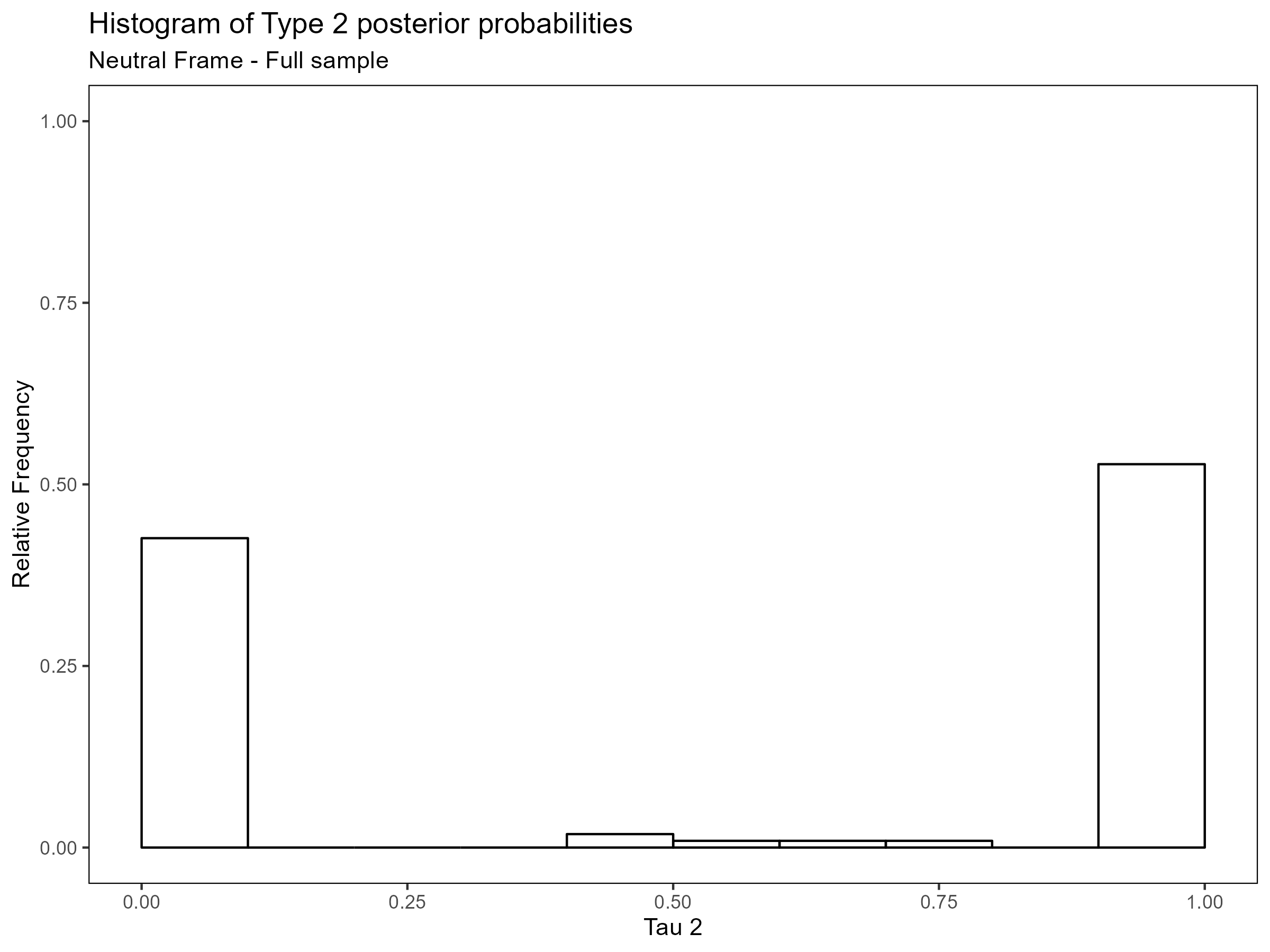}
        
        \label{fig:sub2}
    \end{subfigure}
    \\
    \begin{subfigure}[b]{0.45\textwidth}
        \centering
        \includegraphics[width=\textwidth]{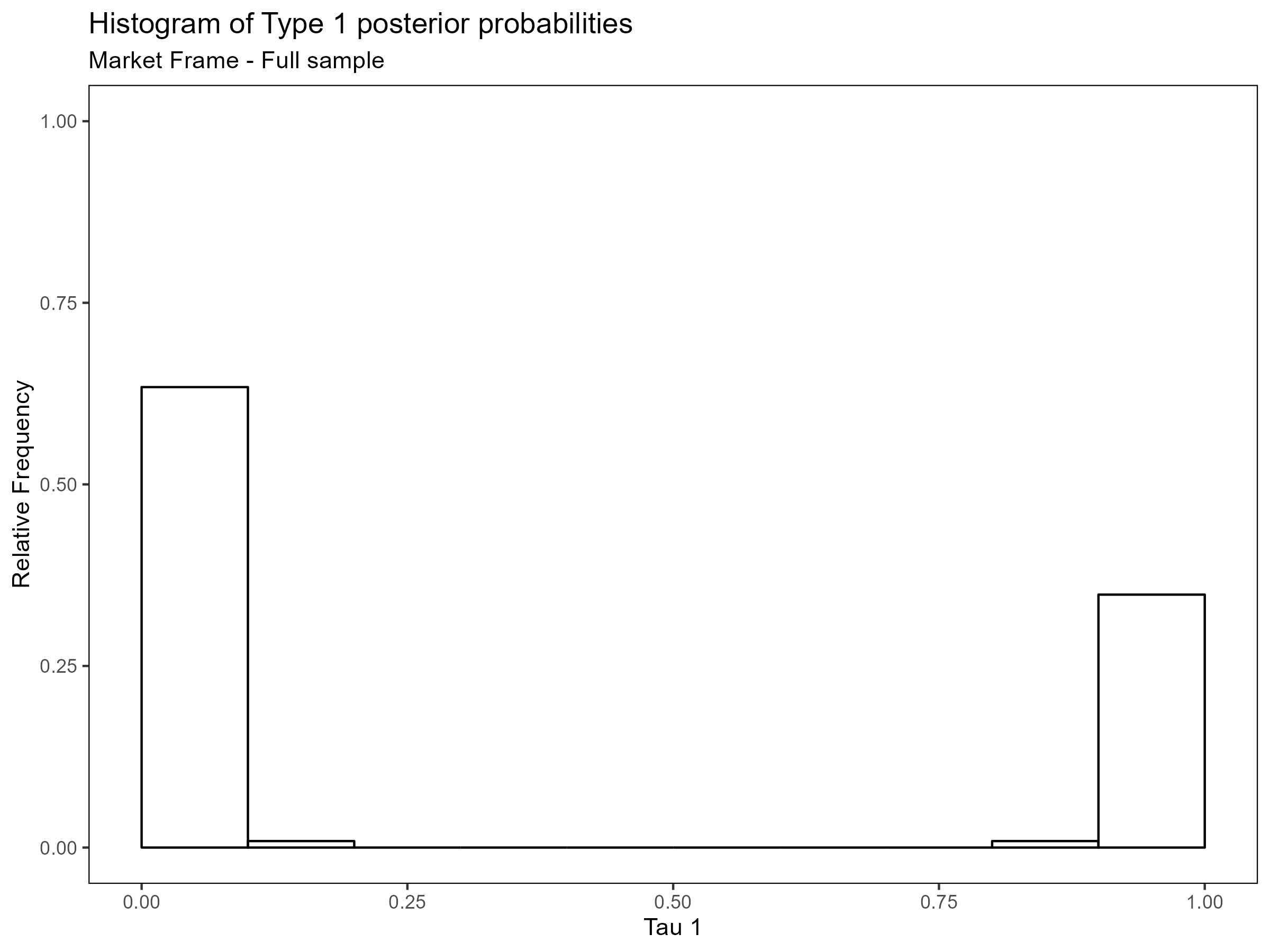}
        
        \label{fig:sub3}
    \end{subfigure}
    \hfill
    \begin{subfigure}[b]{0.45\textwidth}
        \centering
        \includegraphics[width=\textwidth]{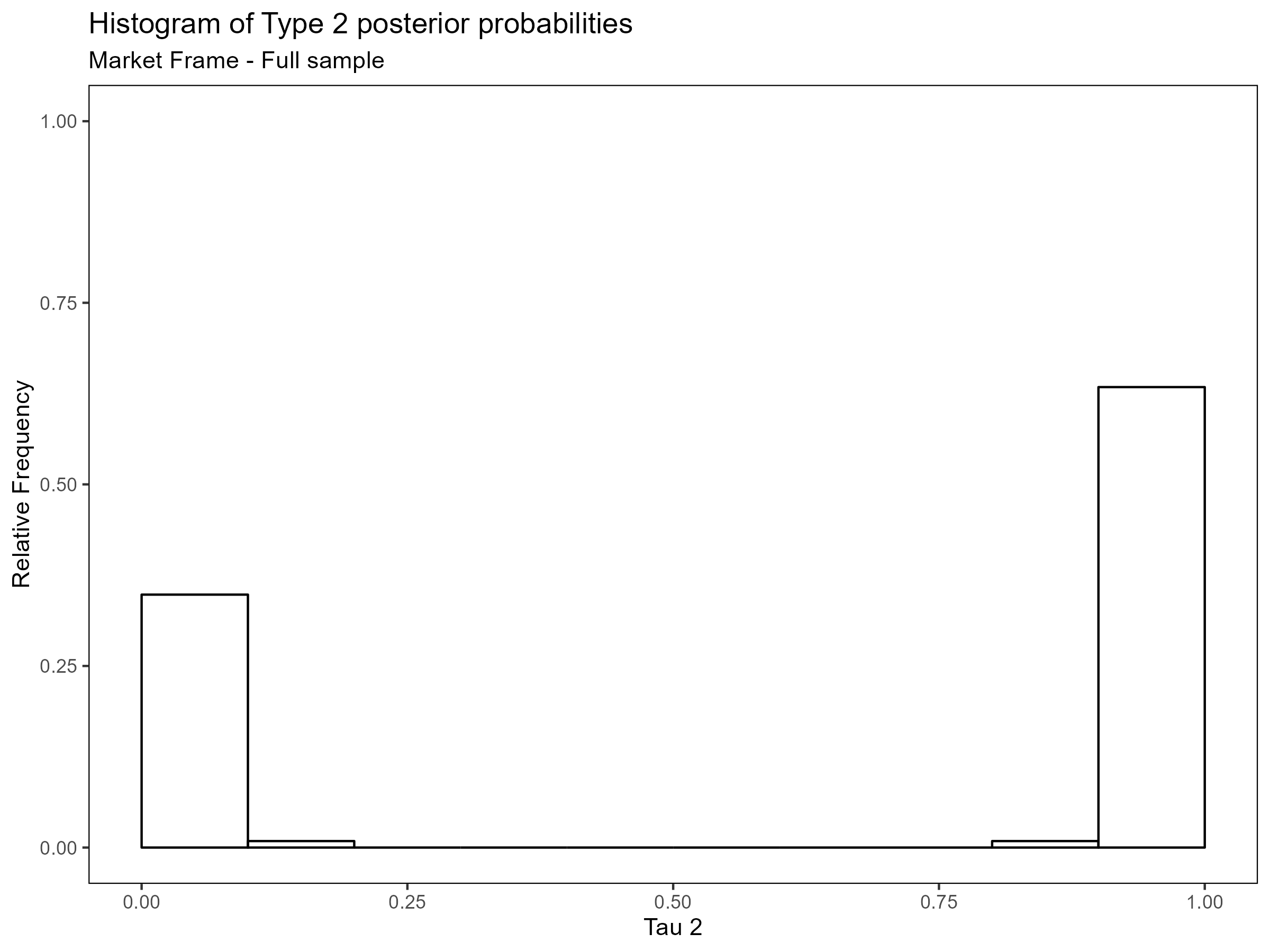}
        
        \label{fig:sub4}
    \end{subfigure}
    \caption{Distribution of posterior probabilities of individual type-membership in Neutral (first row) and Market frames (second row). Full sample estimates.}
    \label{fig:taus_f}
\end{figure}
\newpage

\begin{figure}[ht!]
    \centering
    \begin{subfigure}{\textwidth}
        \centering
        \includegraphics[width=0.8\linewidth]{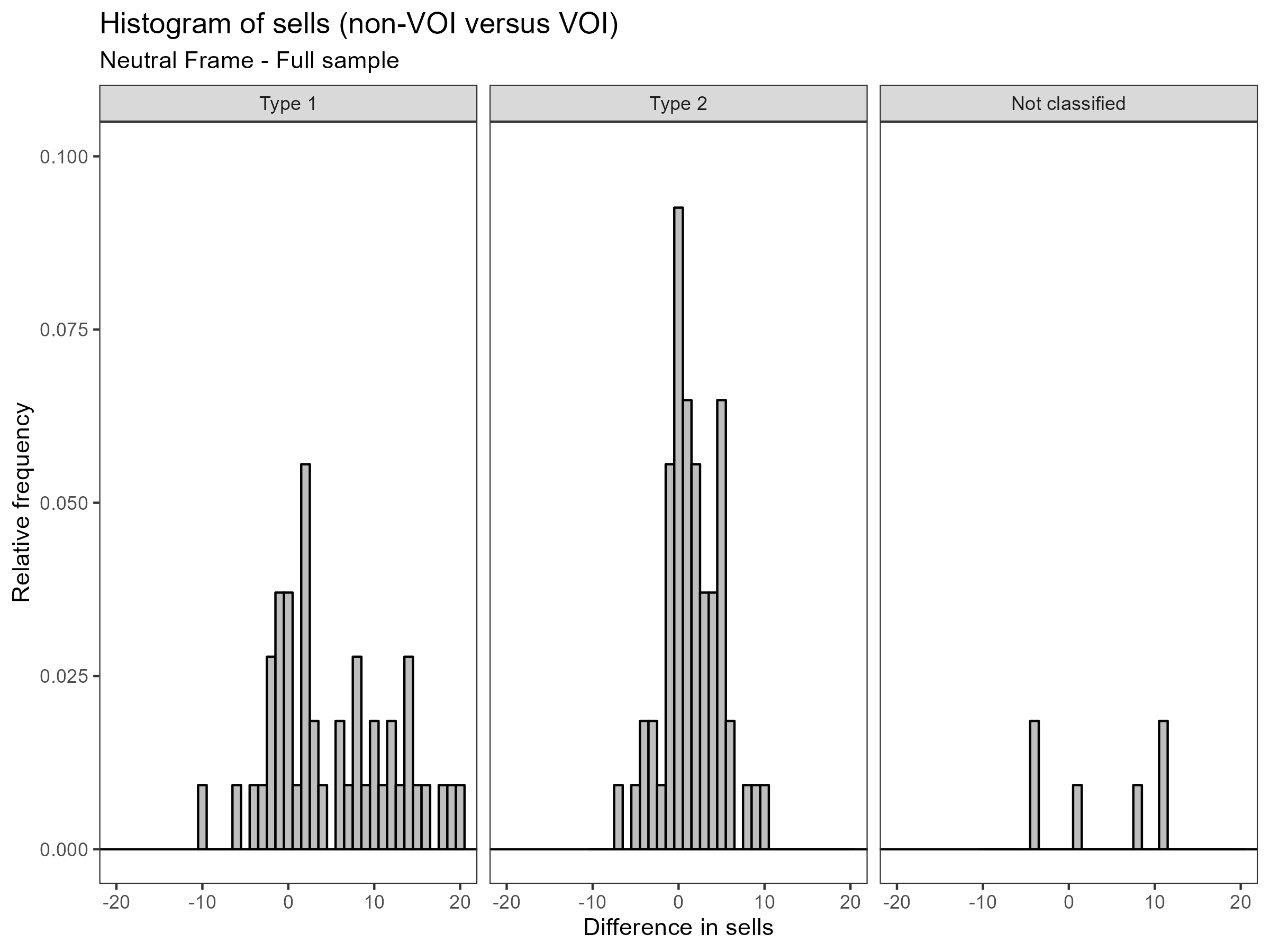}
        \caption{Neutral frame}
        \label{fig:subfiga}
    \end{subfigure}
    
    \begin{subfigure}{\textwidth}
        \centering
        \includegraphics[width=0.8\linewidth]{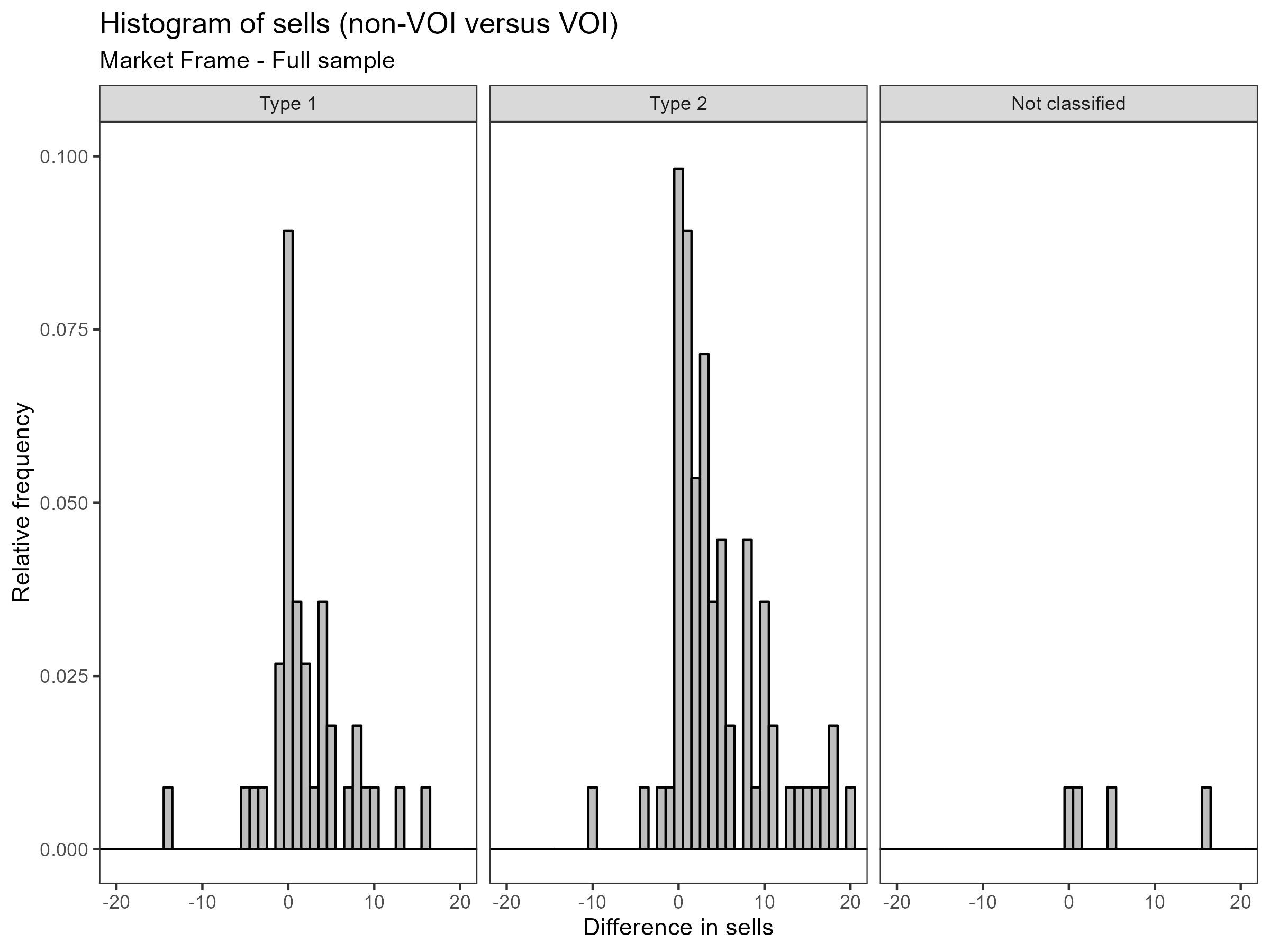} 
        \caption{Market frame}
        \label{fig:subfigb}
    \end{subfigure}
    \caption{Difference in selfish actions non-VOI versus VOI. Full sample. Subjects classified as Type 1 if $\tau_1 > 0.95$ or Type 2 if $\tau_1 < 0.05$}
    \label{fig:diffsell_f}
\end{figure}

\newpage

\begin{table}[ht!]
\centering
\begin{threeparttable}
\caption{\label{tab:finmix_c2}Two-type Finite Mixture Model estimates, Core 2 sample}
\begin{tabularx}{\textwidth}{>{\hsize=1.5\hsize}X *{4}{>{\hsize=0.625\hsize}X}}
\toprule
& \multicolumn{2}{c}{Neutral} & \multicolumn{2}{c}{Market} \\
\cmidrule(lr){2-3} \cmidrule(lr){4-5}
& Type 1 & Type 2 & Type 1 & Type 2 \\
\midrule
$\beta$: Aheadness aversion & 0.287$^{***}$ & -0.145$^{}$ & 0.18$^{***}$ & -0.305$^{}$ \\
& \textit{(0.024)} & \textit{(0.194)} & \textit{(0.012)} & \textit{(0.283)} \\ 
& [0.020] & [0.066] & [0.007] & [0.109] \\ 
$\kappa$: Degree of morality & 0.344$^{***}$ & 0.595$^{***}$ & 0.176$^{***}$ & 0.597$^{**}$ \\
& \textit{(0.085)} & \textit{(0.146)} & \textit{(0.020)} & \textit{(0.297)} \\ 
& [0.061] & [0.124] & [0.016] & [0.143] \\ 
$\sigma$: Choice sensitivity  & 0.047$^{***}$ & 0.023$^{***}$ & 0.073$^{***}$ & 0.029$^{*}$ \\
& \textit{(0.018)} & \textit{(0.006)} & \textit{(0.012)} & \textit{(0.015)} \\ 
& [0.007] & [0.005] & [0.005] & [0.006] \\ 
Share & 0.522$^{***}$ & 0.478$^{***}$ & 0.623$^{***}$ & 0.377$^{***}$ \\
& \textit{(0.084)} & \textit{(0.084)} & \textit{(0.028)} & \textit{(0.028)} \\ 
& [0.065] & [0.065] & [0.056] & [0.056] \\ 
\midrule
Number of Observations & \multicolumn{2}{c}{2,720} & \multicolumn{2}{c}{3,160} \\ 
Number of subjects & \multicolumn{2}{c}{68} & \multicolumn{2}{c}{79} \\ 
Log likelihood & \multicolumn{2}{c}{-1384.981} & \multicolumn{2}{c}{-1617.195} \\ 
\bottomrule
\end{tabularx}
\begin{tablenotes}
\item Notes: Standard errors clustered at the individual level in parenthesis, non-clustered standard errors in brackets. $^{***}$ Significant at 1\%, $^{**}$ Significant at 5\%, $^{*}$ Significant at 10\% using clustered standard errors.
\end{tablenotes}
\end{threeparttable}
\end{table}

\newpage

\begin{figure}[ht!]
    \centering
    \begin{subfigure}[b]{0.45\textwidth}
        \centering
        \includegraphics[width=\textwidth]{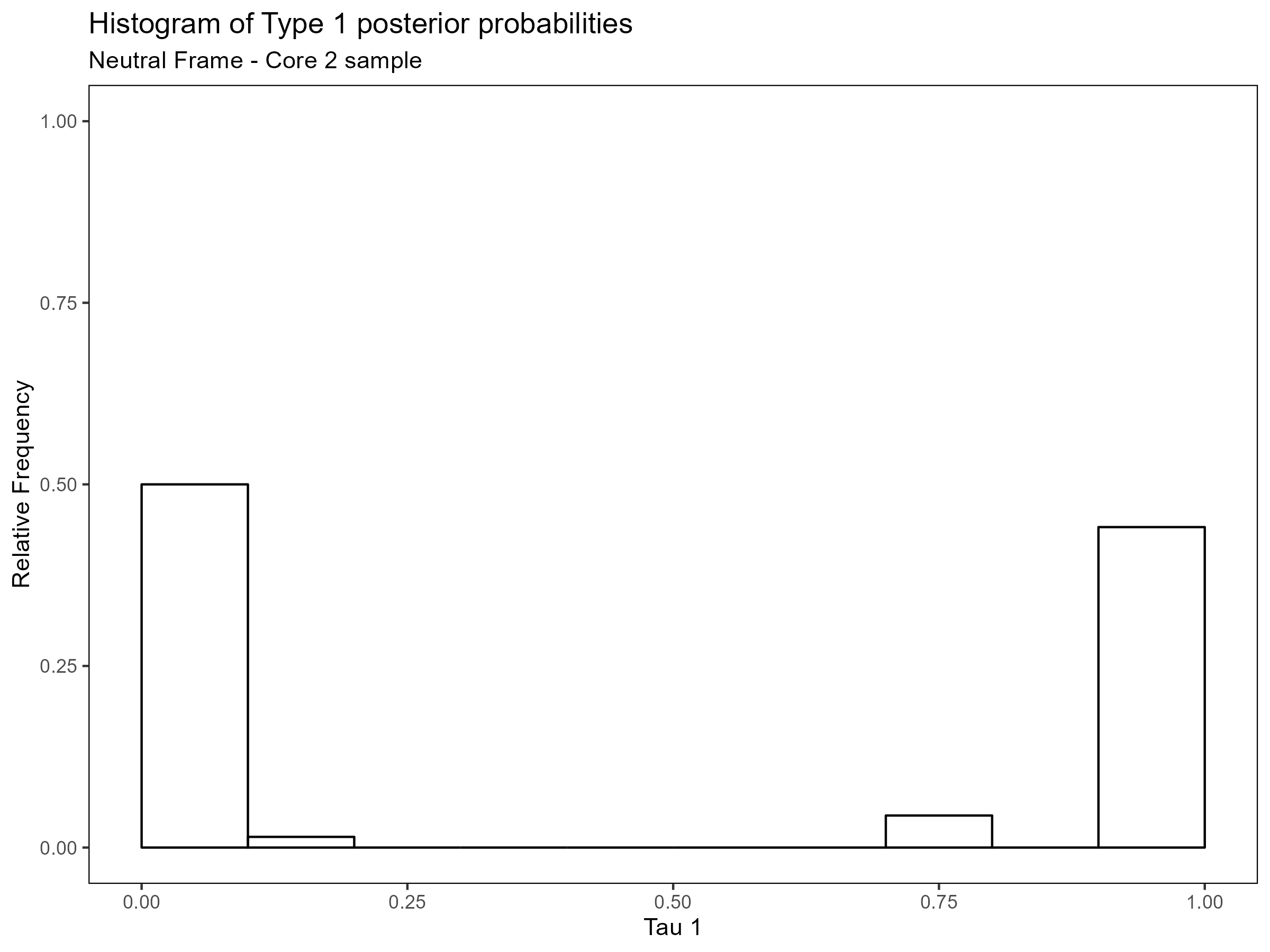}
        
        \label{fig:sub1}
    \end{subfigure}
    \hfill
    \begin{subfigure}[b]{0.45\textwidth}
        \centering
        \includegraphics[width=\textwidth]{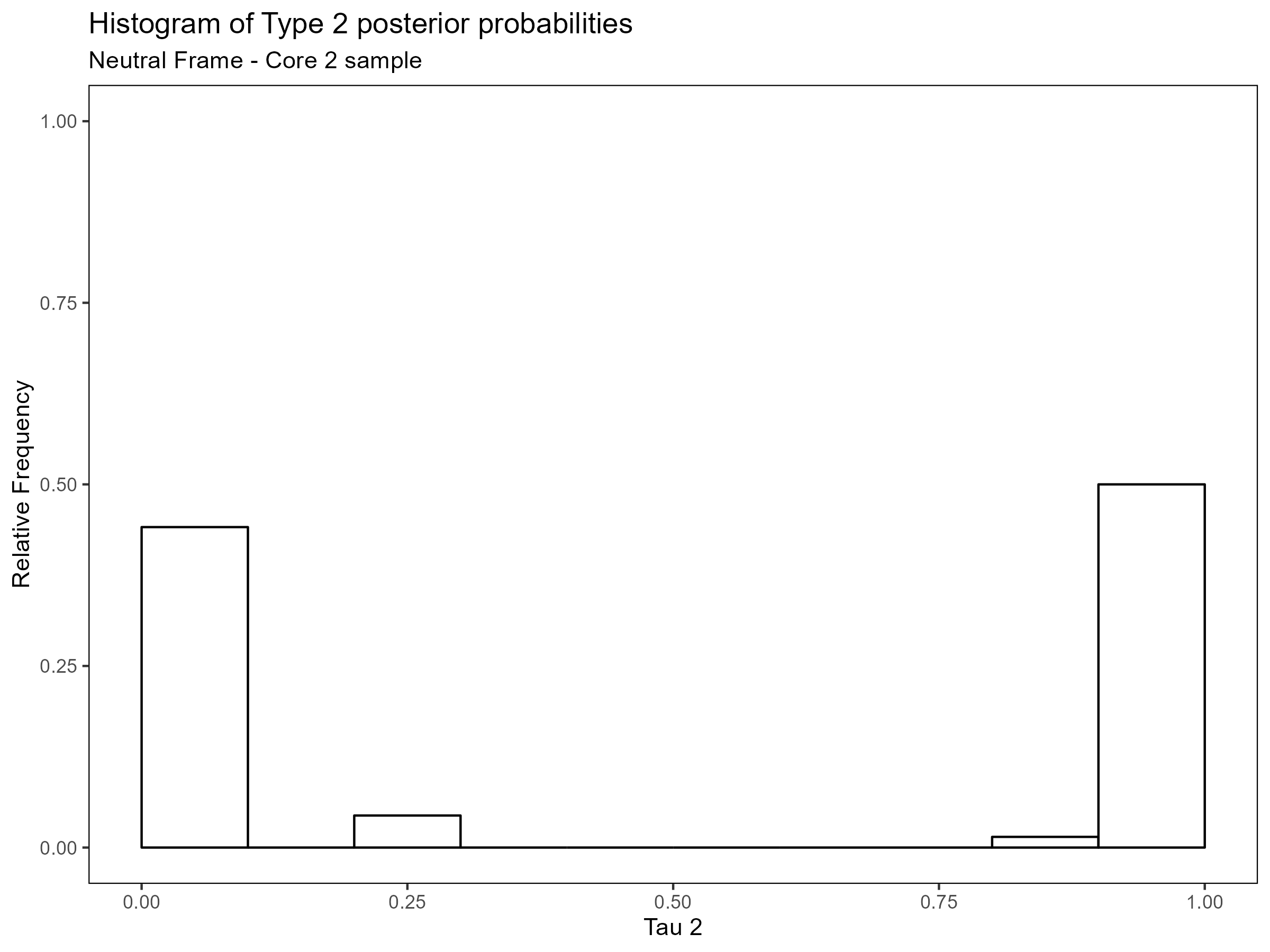}
        
        \label{fig:sub2}
    \end{subfigure}
    \\
    \begin{subfigure}[b]{0.45\textwidth}
        \centering
        \includegraphics[width=\textwidth]{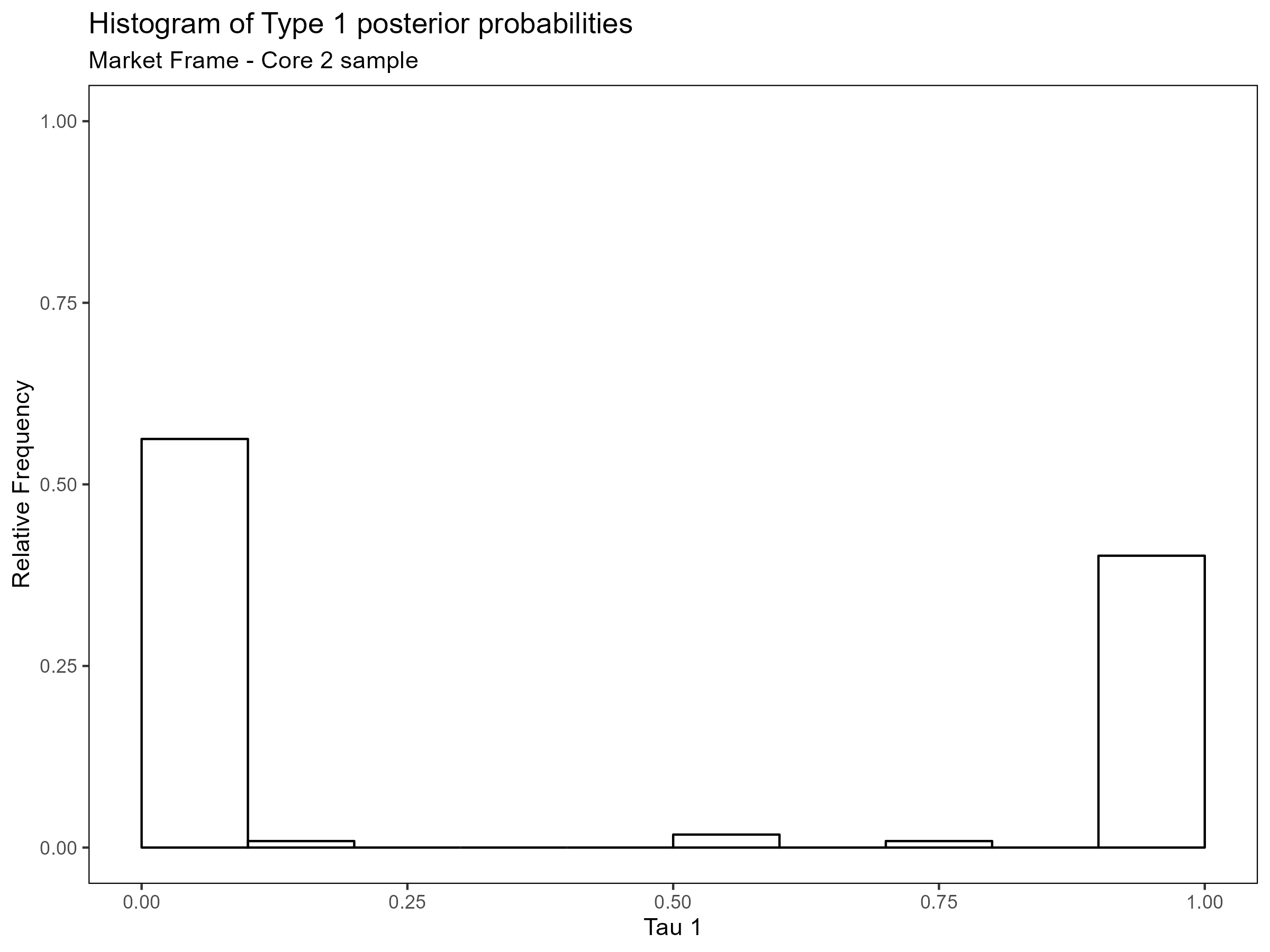}
        
        \label{fig:sub3}
    \end{subfigure}
    \hfill
    \begin{subfigure}[b]{0.45\textwidth}
        \centering
        \includegraphics[width=\textwidth]{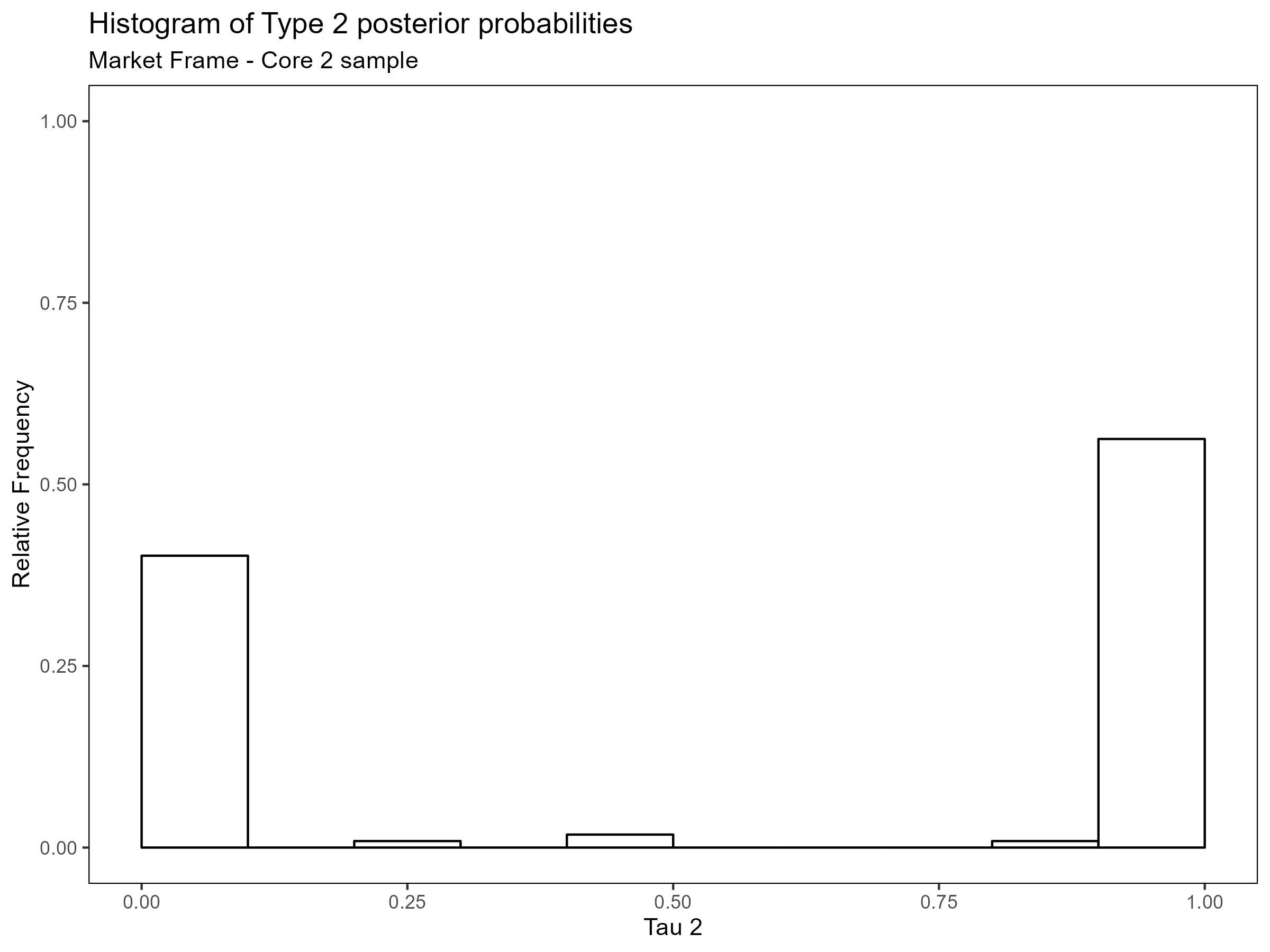}
        
        \label{fig:sub4}
    \end{subfigure}
    \caption{Distribution of posterior probabilities of individual type-membership in Neutral (first row) and Market frames (second row). Core 2 sample estimates.}
    \label{fig:taus_c2}
\end{figure}
\newpage

\begin{figure}[ht!]
    \centering
    \begin{subfigure}{\textwidth}
        \centering
        \includegraphics[width=0.8\linewidth]{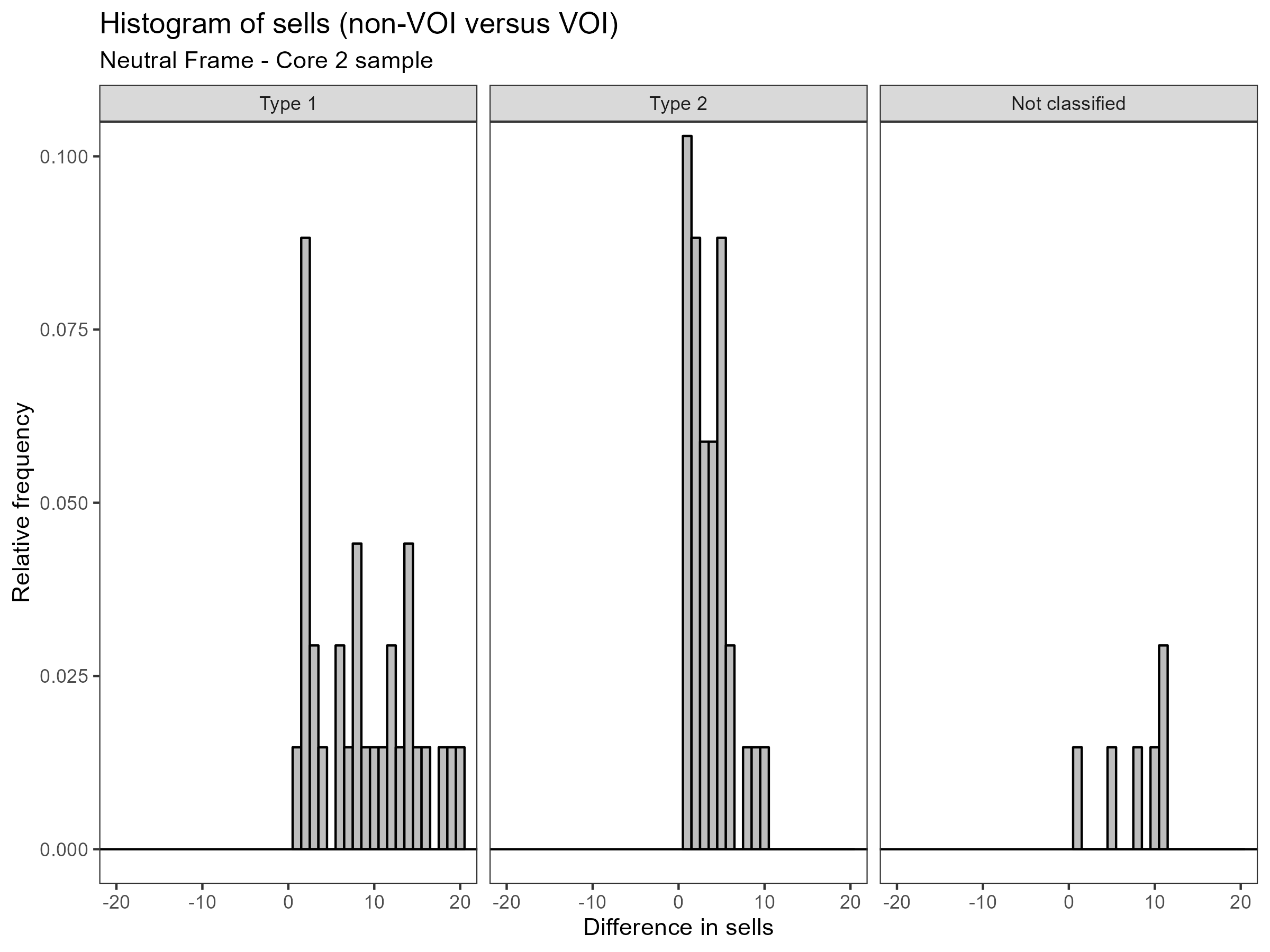}
        \caption{Neutral frame}
        \label{fig:subfiga}
    \end{subfigure}
    
    \begin{subfigure}{\textwidth}
        \centering
        \includegraphics[width=0.8\linewidth]{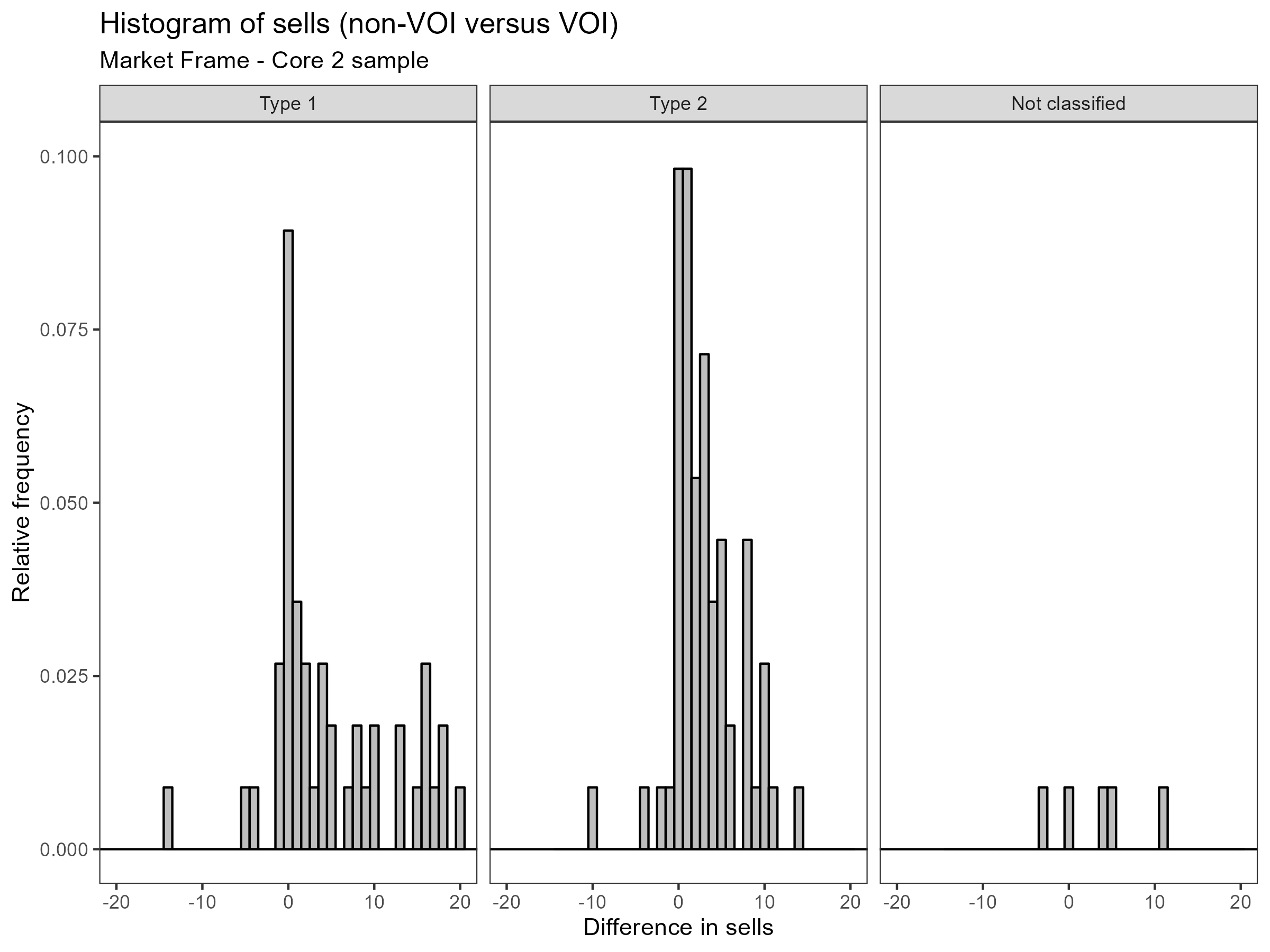} 
        \caption{Market frame}
        \label{fig:subfigb}
    \end{subfigure}
    \caption{Difference in selfish actions non-VOI versus VOI. Core 2 sample. Subjects classified as Type 1 if $\tau_1 > 0.95$ or Type 2 if $\tau_1 < 0.05$}
    \label{fig:diffsell_c2}
\end{figure}

\FloatBarrier

\clearpage

\section{Preference parameter estimation at the individual level} \label{app:individual}

\begin{figure}[htb!]
\centering
   \begin{subfigure}[b]{0.3\textwidth}
\hspace*{-2.8cm} 
   \centering
   \includegraphics[width=1.8\textwidth]{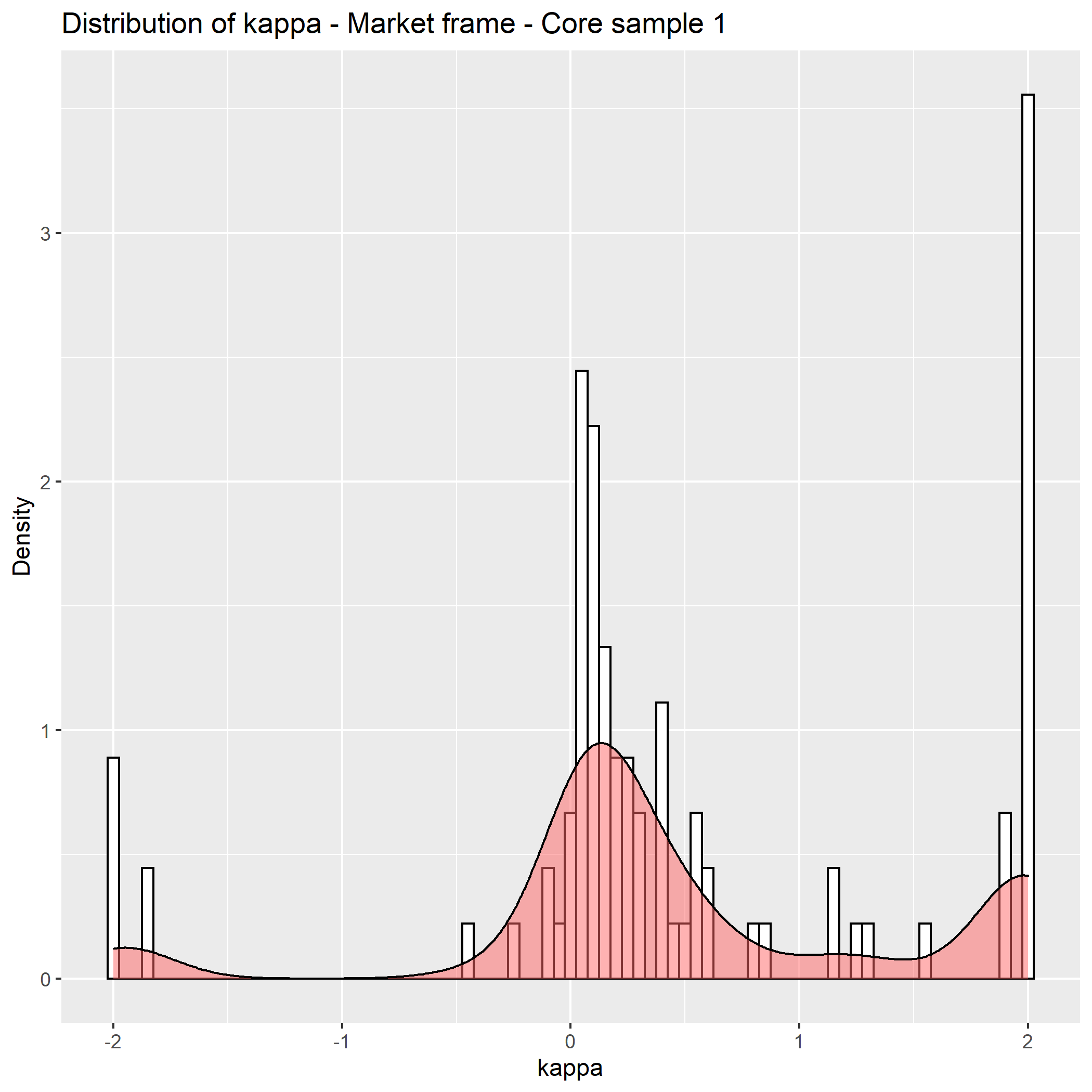}
   \caption{}
   \label{fig:kappadist_mkt_c1_2} 
\end{subfigure}

\begin{subfigure}[b]{0.3\textwidth}
\hspace*{-2.8cm} 
   \centering
   \includegraphics[width=1.8\textwidth]{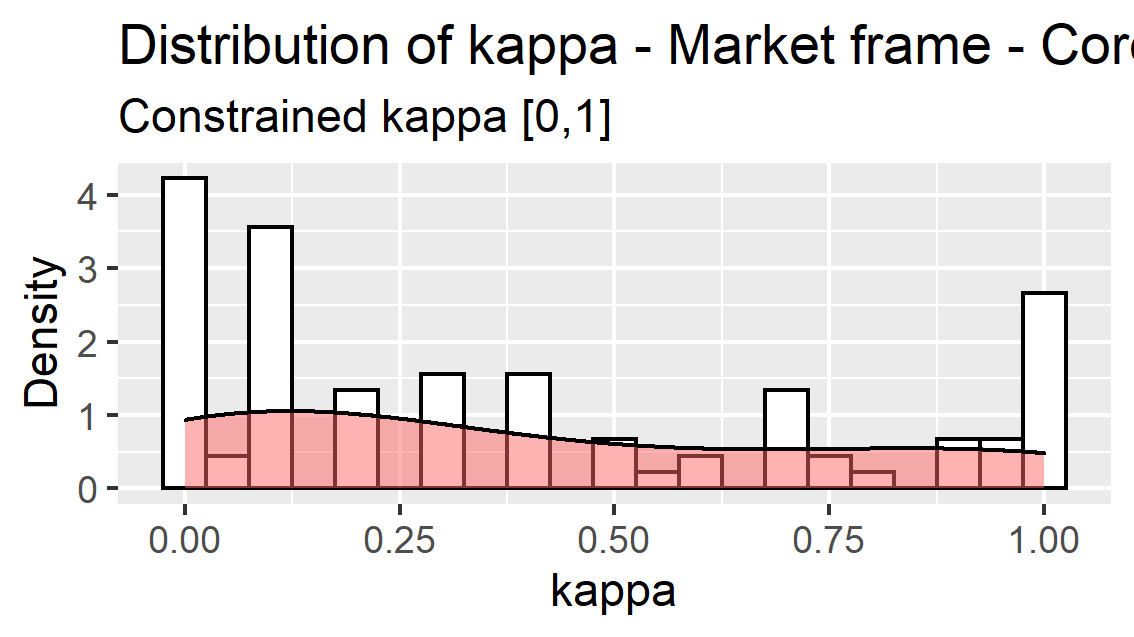}
   \caption{}
   \label{fig:kappadist_mkt_c1_constk}
\end{subfigure}
\caption{Distributions of individual estimates of $\kappa$ for the 100 subjects for whom the decision was different under VOI and non-VOI for at least one of the 20 payoffs, in the Market frame (treatment M). In panel (a), no restriction was imposed on the estimates. In panel (b) the estimates were restricted to the interval $[0,1]$. In panel (a) the estimates whose value is above 2 in absolute value are lumped together at -2 and 2.}
\label{kappadist_mkt_c1}
\end{figure}

\begin{figure}[htb!]
\centering
   \begin{subfigure}[b]{0.3\textwidth}
\hspace*{-2.8cm} 
   \centering
   \includegraphics[width=1.8\textwidth]{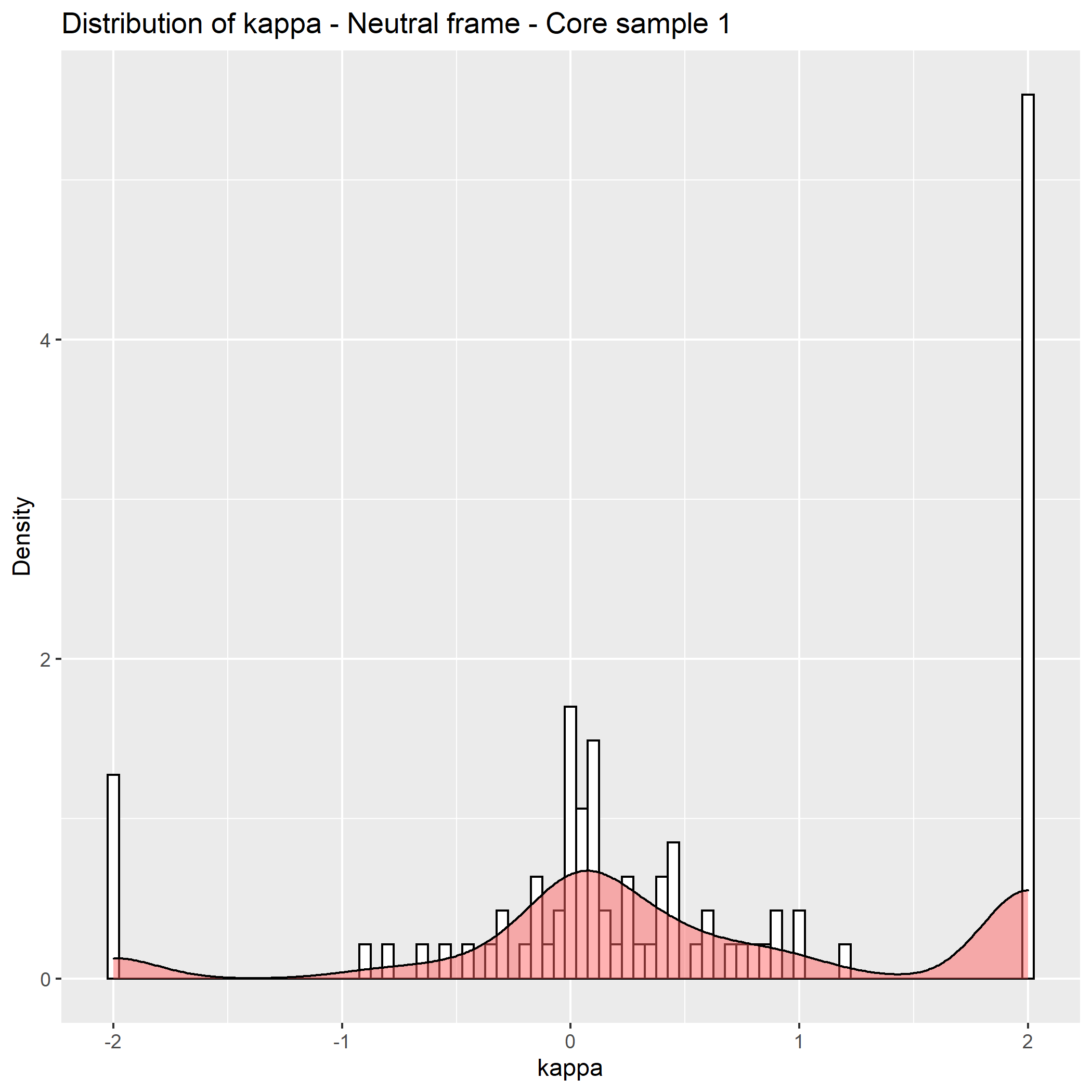}
   \caption{}
   \label{fig:kappadist_neut_c1_2} 
\end{subfigure}

\begin{subfigure}[b]{0.3\textwidth}
\hspace*{-2.8cm} 
   \centering
   \includegraphics[width=1.8\textwidth]{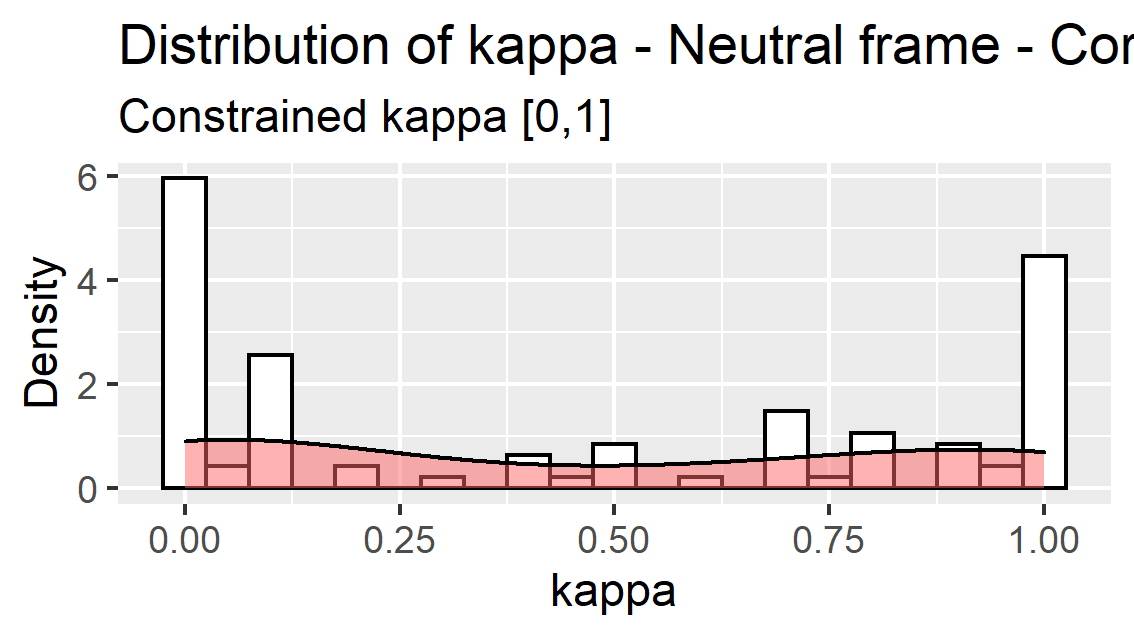}
   \caption{}
   \label{fig:kappadist_neut_c1_constk}
\end{subfigure}
\caption{Distributions of individual estimates of $\kappa$ for the 98 subjects for whom the decision was different under VOI and non-VOI for at least one of the 20 payoffs, in the neutral frame (treatment N). In panel (a), no restriction was imposed on the estimates. In panel (b) the estimates were restricted to the interval $[0,1]$. In panel (a) the estimates whose value is above 2 in absolute value are lumped together at -2 and 2. \label{kappadist_neut_c1}
}
\end{figure}

Figures \ref{kappadist_mkt_c1} and \ref{kappadist_neut_c1} show the distributions of the estimates of $\kappa$ at the individual level, for the Market and the Neutral frames, respectively. 

%%%%%%%%%%%%%%%%%%%%%%%%%%%%%%%%%%%%%%%%%%%%%%%%
%%%%%%%%%%%%%%%%%%%%%%%%%%%%%%%%%%%%%%%%%%%%%%%%

These figures show that for a large number of subjects the estimates are negative and/or exceed 2 in absolute value (this is especially true in the Neutral frame). This suggests that our experimental design is not ideal to structurally estimate individual preference parameters.
To see why, consider Figure \ref{fig:unaw_lemon1bis}, which shows the threshold values for $\beta$ and $\kappa$ (recall \eqref{betabar} and \eqref{kappabar}) for three of the payoffs used in the experiment: those labeled 1, 4, and 20 in Table \ref{tab:thresholds}. As indicated by the numbers in the blue balls, the solid lines correspond to payoff 1, the dashed ones to payoff 4, and the dotted ones to payoff 20. We discuss the challenge of identifying $(\beta_i,\kappa_i)$ for several different behavioral scenarios (a discussion which clearly generalizes to subjects taking decisions in the 20 payoff configurations of the experiment). For this discussion, recall that, for any given payoff, a subject $i$: (a) does not sell under either non-VOI or VOI if $(\beta_i,\kappa_i)$ is to the right of the vertical line; (b) sells under both non-VOI or VOI if $(\beta_i,\kappa_i)$ is below the downward-sloping  curve; and (c) makes a switch (i.e., sells under non-VOI but not under VOI) if $(\beta_i,\kappa_i)$ is between the vertical and the downward-sloping curve (recall Figure \ref{fig:unaw_lemon1}).

The first issue is the lack of bounds on $\beta$ and/or $\kappa$ inherent in individual estimations. Consider a subject who switches under payoff 1 and never sells under payoffs 4 and 20. Such behavior is consistent with any $(\beta,\kappa)$ in zone I (where the red ball labeled I appears). While the value of $\beta$ is thus bounded below by 0.33 and above by 0.6, the value of $\kappa$ is bounded below only. In a similar vein, a subject who switches under payoff 20 and sells under payoffs 1 and 4 is consistent with any $(\beta,\kappa)$ in zone II: for this subject, the value of $\beta$ is bounded above only, while the value of $\kappa$ is bounded below by the downward-sloping dashed and dotted lines. As a third example, a subject who switches under the three payoffs is consistent with any $(\beta,\kappa)$ in zone III. Again, identification of this subject's $(\beta,\kappa)$ is clearly impossible.

The second issue arises for subjects whose behavior is compatible with disjoint sets of values of $(\beta,\kappa)$. For example, consider a subject who switches under payoff 1 and 20, but not under payoff 4. Such a subject could have a $(\beta,\kappa)$ in zone I (consistent with the switch under payoff 1), in which case (s)he would have made a large mistake by switching under payoff 20; or (s)he could have a $(\beta,\kappa)$ in zone II (consistent with the switch under payoff 20), in which case (s)he would have made a large mistake by switching under payoff 1. 

By contrast, estimation of $(\beta,\kappa)$ at the aggregate level is possible. To see why, consider again Figure \ref{fig:unaw_lemon1bis}. Suppose that 20\% of the subjects switched under all the payoffs (zone III), 25\% switched under payoffs 1 and 4 and did not sell under payoff 20 (zone IV), 35\% switched under payoff 1 and did not sell under payoffs 4 and 20 (zone I), while the remaining 20\% switched only under payoff 20 and sold under payoffs 1 and 4 (zone II). If these decisions are interpreted as emanating from one single individual who sometimes makes mistakes, the estimated $(\beta,\kappa)$ of this hypothetical individual would likely be in zone V, or in zone IV close to zone V. The between-subject heterogeneity in behavior would thus help put boundaries on the estimates of $\beta$ and $\kappa$.

\usepgfplotslibrary{fillbetween}
\begin{figure}[ht] 	
  \captionsetup{width=0.7\linewidth}
\begin{center}
	\begin{tikzpicture}[scale=1.2]
	\begin{axis}[
	ticklabel style = {font=\footnotesize},
    legend style={font=\footnotesize},
	axis lines=center,
	grid = none,
	ymin=0,
	ymax=3.6,
    xtick={0.03,0.33,0.6},
    ytick={1.5},
	xmin = -0.6,
	xmax=1.0,
	xlabel={$\beta$},
	ylabel={$\kappa$}]

	\addplot[smooth,very thick, solid,name path=A] {1.5-2.5*x}; 
  	\addplot[smooth,very thick, solid,name path=F]coordinates {(0.6,0)(0.6,2.5)}; 
  	\addplot[smooth,very thick,dashed,name path=D] {0.5-1.5*x};
  	\addplot[smooth,very thick,dashed,name path=F]coordinates {(15/45,0)(15/45,2.5)};
  	\addplot[smooth,very thick, densely dotted,name path=F] {(0.5-16.5*x)/16};
  	\addplot[smooth,very thick,densely dotted,name path=F]coordinates {(5/165,0)(5/165,2.5)};
  	\node[ball color=blue,circle,text=white] at (0.6,2.9) {\footnotesize{1}};
    \node[ball color=blue,circle,text=white] at (15/45,2.9) {\footnotesize{4}};
  	\node[ball color=blue,circle,text=white] at (5/165,2.9) {\scriptsize{20}};
  	\node[ball color=red,circle,text=white] at (0.5,1.3) {\tiny{I}};
  	\node[ball color=red,circle,text=white] at (-0.4,0.73) {\tiny{II}};
  	\node[ball color=red,circle,text=white] at (-0.1,2.2) {\tiny{III}};
  	\node[ball color=red,circle,text=white] at (0.2,1.8) {\tiny{IV}};
  	\node[ball color=red,circle,text=white] at (0.2,0.6) {\tiny{V}};
  	% \node[ball color=red,circle,text=white] at (-0.05,0.8) {2};
  	% \node[ball color=red,circle,text=white] at (-0.015,1.95) {3};
 
	\end{axis}
	\end{tikzpicture}
	
\vspace{0.2cm}
\begin{minipage}{0.5\textwidth} % choose width suitably
\caption{\label{fig:unaw_lemon1bis} The numbers in the blue balls indicate the payoff to which the V-shaped lines correspond (see Table \ref{tab:thresholds}). The numbers in the red balls indicate zones with different switching behaviors between non-VOI and VOI.}
\end{minipage}
	\end{center}

    \end{figure}
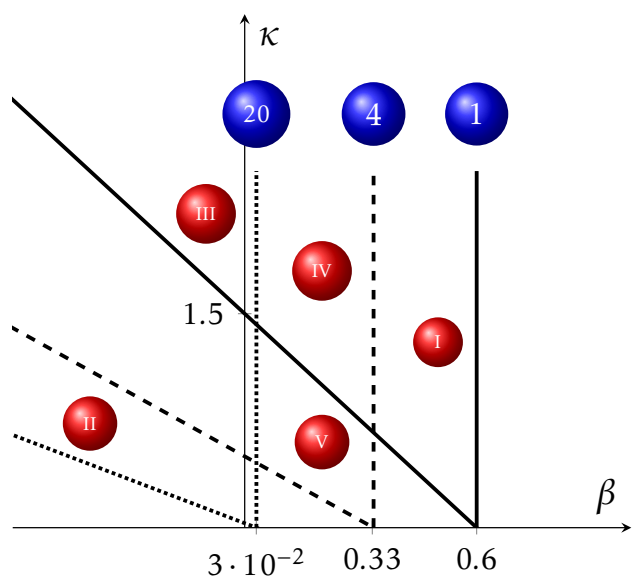

\clearpage

\section{Experimental instructions} \label{app:instructions}

\textbf{Screen 1 [common to all the treatments]:
} 

\bigskip 
\noindent 
Welcome to this experiment!

\bigskip 
\noindent 
Please read the following instructions carefully and from now on, do not communicate with any of the other participants. If you have any questions during the experiment, please raise your hand and wait until we come to you to answer your questions in private. 

\bigskip 
\noindent 
All the participants here today will be asked to take decisions. These decisions will generate points. At the end of the experiment points will be converted into money. Every point is worth 0.025 euros. Each participant also receives 4 euros for attending until the end and answering all questions.

\bigskip 
\noindent 
Your decisions during the experiment are anonymous. They will not be linked to your name in any way. Other participants can never trace your decisions back to you. Moreover, the amount of money you receive at the end will be handed over to you in an opaque envelope, and no other participant will see what is inside the envelope.

\bigskip 
\noindent 
During the experiment, your cell phone should be switched off and out of reach. And remember not to talk with the other participants. We would need to exclude you from the experiment (and the payment) if you breach these rules.

\bigskip 
\noindent 
The experiment consists of two parts, followed by a short questionnaire. At the beginning of each part, you will receive new instructions. Your decisions made in one part will never affect outcomes in the other part, so you can treat both parts as independent.
\newpage
\subsection{Remaining screens for the Market treatment}

\bigskip 
\noindent 
\textbf{Screen 2:}

\bigskip 
\noindent 
\textbf{Part I}

\bigskip 
\noindent 
In this part you will be asked to make choices in 20 different decision situations, which represent an interaction between a Seller and a Buyer.

\bigskip 
\noindent 
In each decision situation you will be paired with one of the other participants here today, each time with a different participant. 

\bigskip 
\noindent 
In each decision situation you are the Seller and the person you are paired with is the Buyer. You will be asked to choose one of two options, Sell and Not Sell, while the Buyer has no choice to make. Think of this as representing a situation in which the Buyer is willing to pay the price for a good that you own. 
Here is an example of a decision screen:

\begin{figure}[H]
\begin{center}
\includegraphics[scale=0.3]{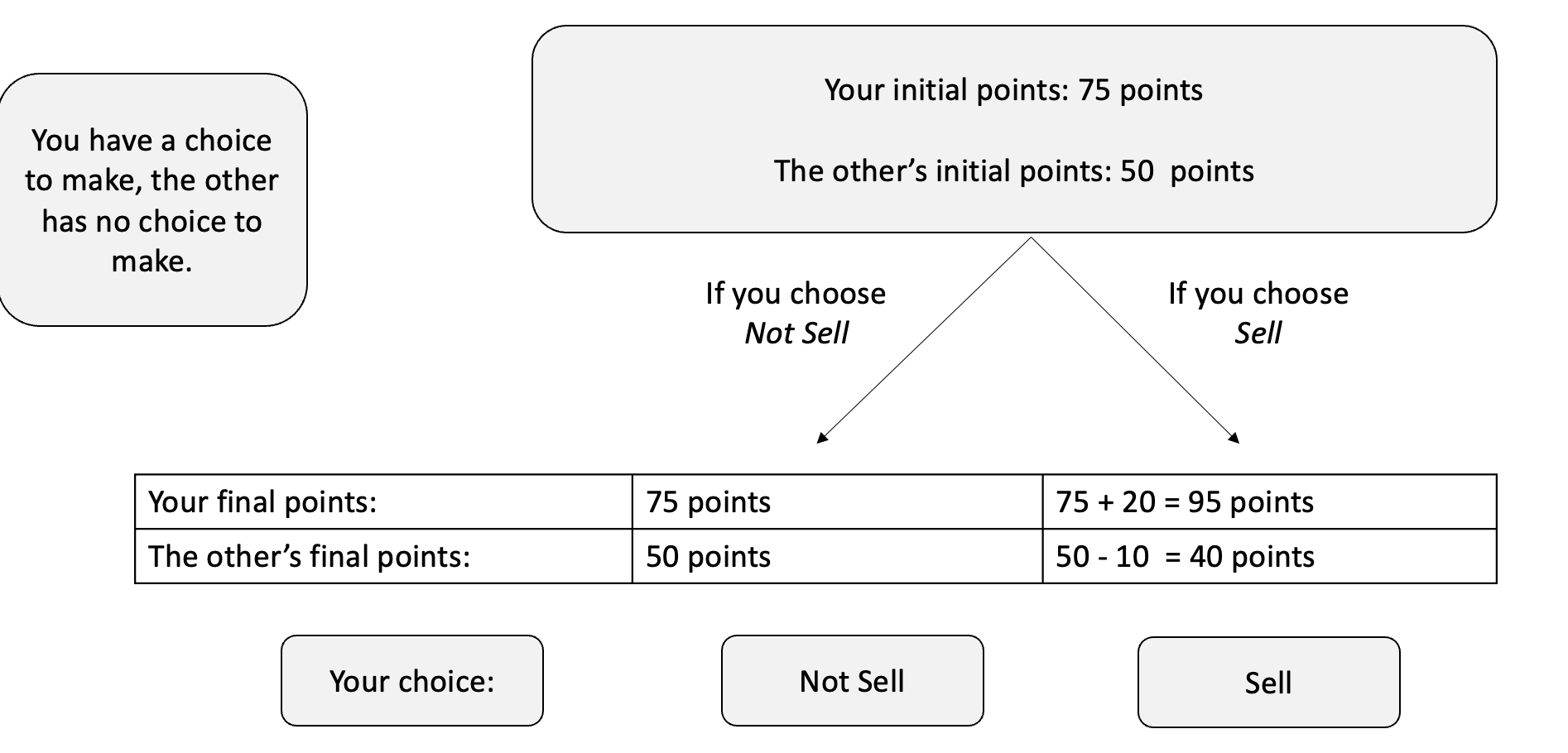}
\end{center}
\end{figure}

\bigskip 
\noindent 
The number of points that you get if you choose Not Sell can be interpreted as the value you attach to owning the good. The number of points that you get if you choose Sell can be interpreted as the amount of money you have if you Sell the good.

\bigskip 
\noindent 
The number of points that the Buyer gets if you choose Not Sell can be interpreted as the amount of money he/she has initially. The number of points that the Buyer gets if you choose Sell can be interpreted as the 
value he/she attaches to owning the good.

\bigskip 
\noindent 
As you can see on the decision screen above, the Buyer would be better off if you chose Not Sell, while you are better off if you Sell. Think of this as representing a situation in which the good that you sell has a defect which makes the Buyer enjoy owning the good less than you do.
At the end of the experiment one of these 20 decision situations will be randomly drawn. Your decision in this situation (and only this situation) will have an effect on the number of points you and the person you were paired with will get. This other participant will make no decision that affects your number of points, however. 

\bigskip 
\noindent 
Remember that:
\begin{itemize}
    \item all your decisions are anonymous, and no participant will ever learn with whom he/she was paired in any decision situation;
    \item all the decision situations are equally important, in the sense that they are all equally likely to count towards the amount of money you and the person you were paired with will receive at the end;
    \item each point will be converted to 0.025 euros at the end of the experiment.
\end{itemize}

\bigskip 
\noindent 
\textbf{Screen 3 [comprehension quiz]:
}

\begin{figure}[H]
\begin{center}
\includegraphics[scale=0.3]{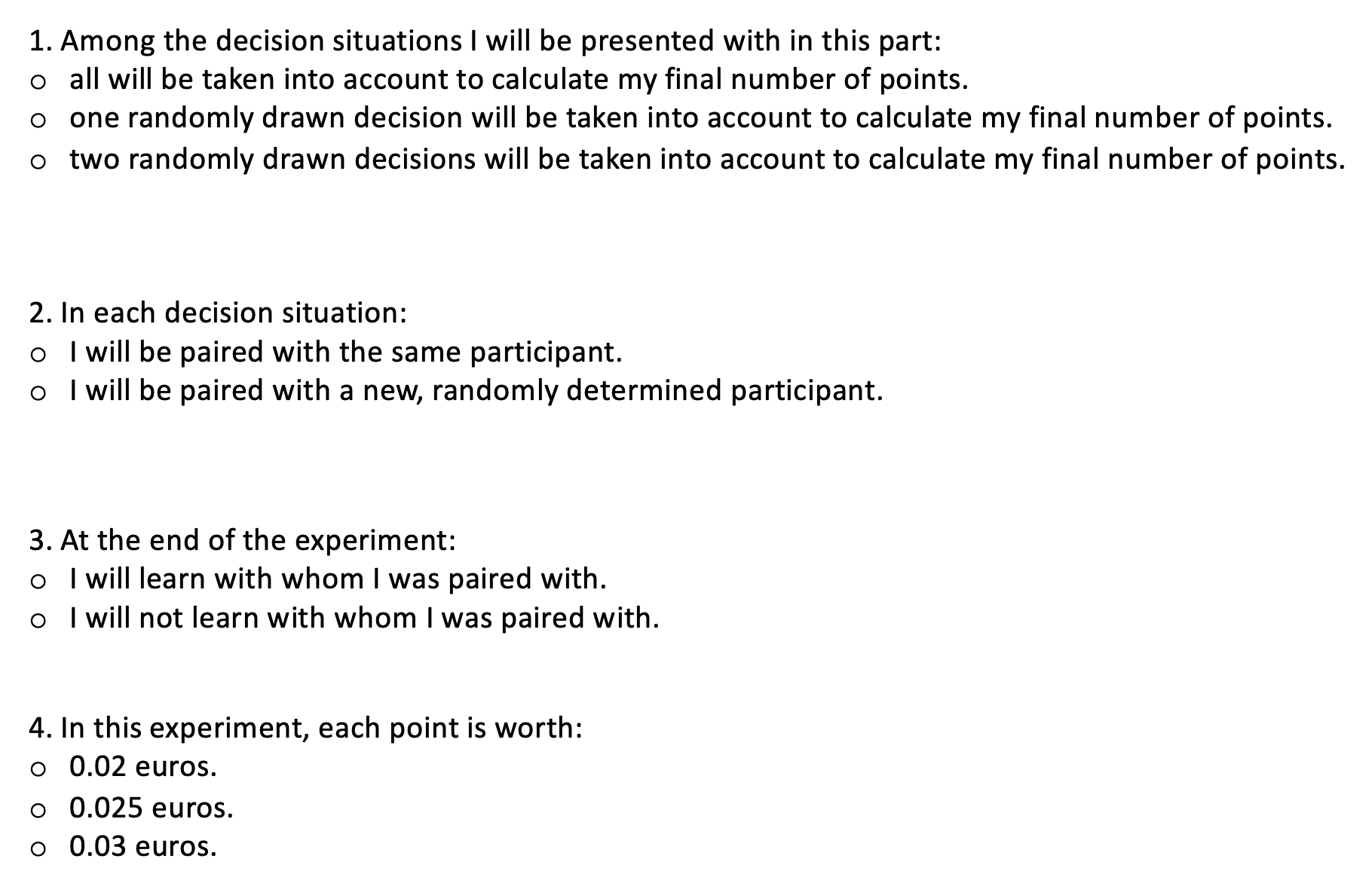}
\end{center}
\end{figure}

\bigskip 
\noindent 
\textbf{Screens 4-33:} //the first 20 decision situations//

\bigskip 
\noindent 
\textbf{Screen 34:
}

\bigskip 
\noindent 
\textbf{Part II
}

\bigskip 
\noindent 
In this part you will be asked to make choices in 20 different decision situations.

\bigskip 
\noindent 
In each decision situation you will be paired with one of the other participants here today, each time with a different participant. 

\bigskip 
\noindent 
In each decision situation there are two roles: the Seller role and the Buyer role. Either you or the person you are paired with will be assigned to the Seller role, while the other will be assigned to the Buyer role. The two different role assignments are equally likely. The person assigned to the Seller role chooses between two options, Sell and Not Sell, while the other person has no choice to make. However, because the role assignment will only be made at the end of the experiment, both you and the person you were paired with will be asked to state what you would do in Role A in each decision situation.
Here is an example of a decision screen:

\begin{figure}[H]
\begin{center}
\includegraphics[scale=0.3]{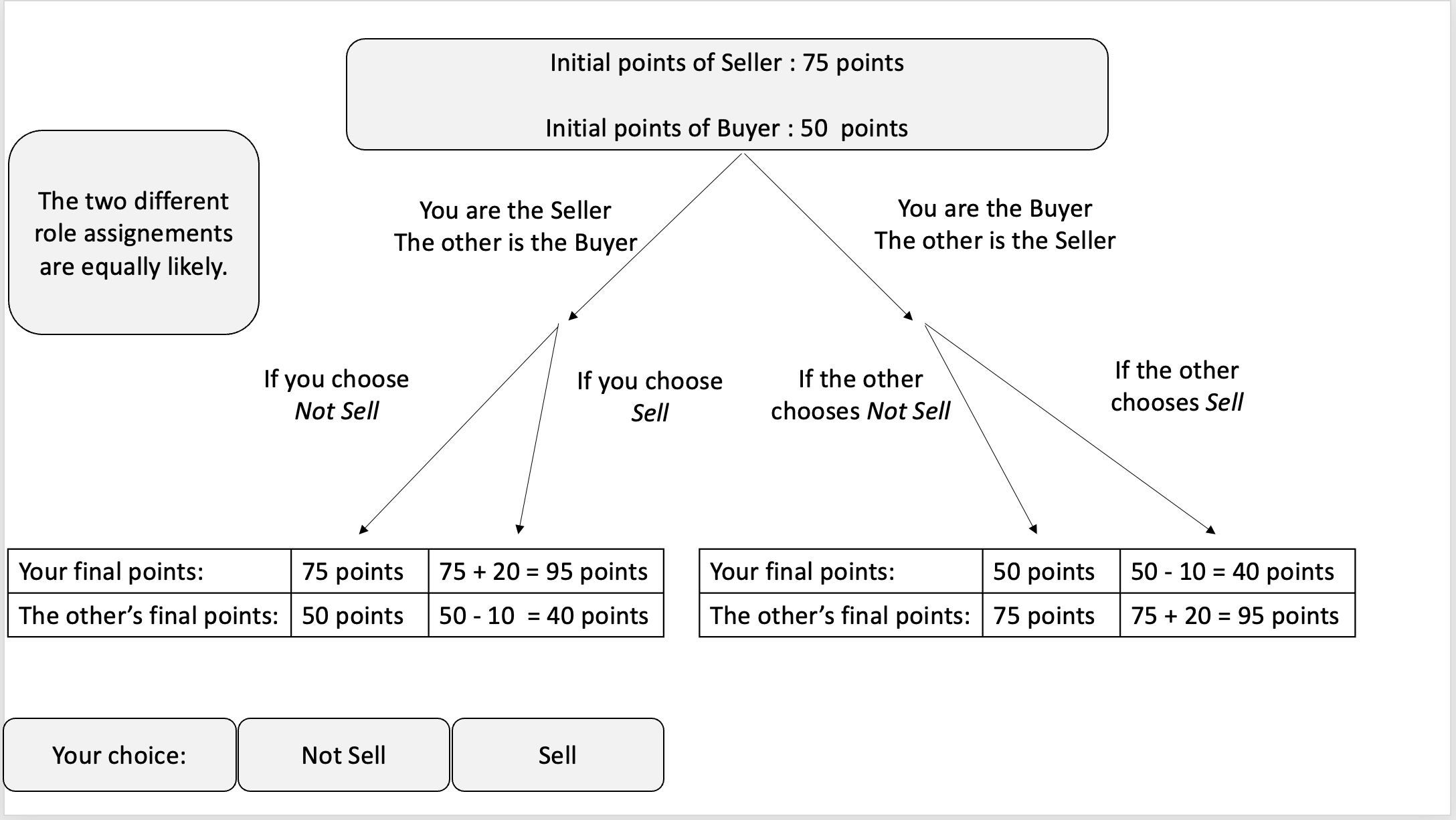}
\end{center}
\end{figure}

\bigskip 
\noindent 
The number of points that the Seller gets from choosing Not Sell can be interpreted as the value that the Seller attaches to owning the good. The number of points that the Seller gets from choosing Sell can be interpreted as the amount of money the Seller has if he/she Sells the good. The number of points that the Buyer gets if the Seller chooses Not Sell can be interpreted as the amount of money the Buyer has initially. The number of points that the Buyer gets if the Seller chooses Sell can be interpreted as the value the Buyer attaches to 
owning the good.

\bigskip 
\noindent 
As you can see on the decision screen above, the Buyer would be better off if the Seller chose Not Sell, while the Seller is better off if he/she chooses Sell. Think of this as representing a situation in which the good that the Seller sells has a defect which makes the Buyer enjoy owning the good less than the Seller does.

\bigskip 
\noindent 
At the end of the experiment one of these 20 decision situations will be randomly drawn. The role assignment for this decision situation will be randomly drawn. Recall that the two roles are equally likely. If you are assigned to the Seller role (this happens with probability 1/2), it is your decision in this situation that will have an effect on the number of points you and the person you were paired with will get. If you are assigned to the Buyer role (this happens with probability 1/2), it is the decision of the person you were paired with in this situation that will have an effect on the number of points you and the person you were paired with will get.

\bigskip 
\noindent 
Remember that:
\begin{itemize}
    \item all your decisions are anonymous, and no participant will ever learn with whom he/she was paired in any decision situation;
    \item all the decision situations are equally important, in the sense that they are all equally likely to count towards the amount of money you and the person you were paired with will receive at the end;
    \item each point will be converted to 0.025 euros at the end of the experiment.
\end{itemize}

\bigskip 
\noindent 
\textbf{Screen 35
[comprehension quiz]:}

\begin{figure}[H]
\begin{center}
\includegraphics[scale=0.3]{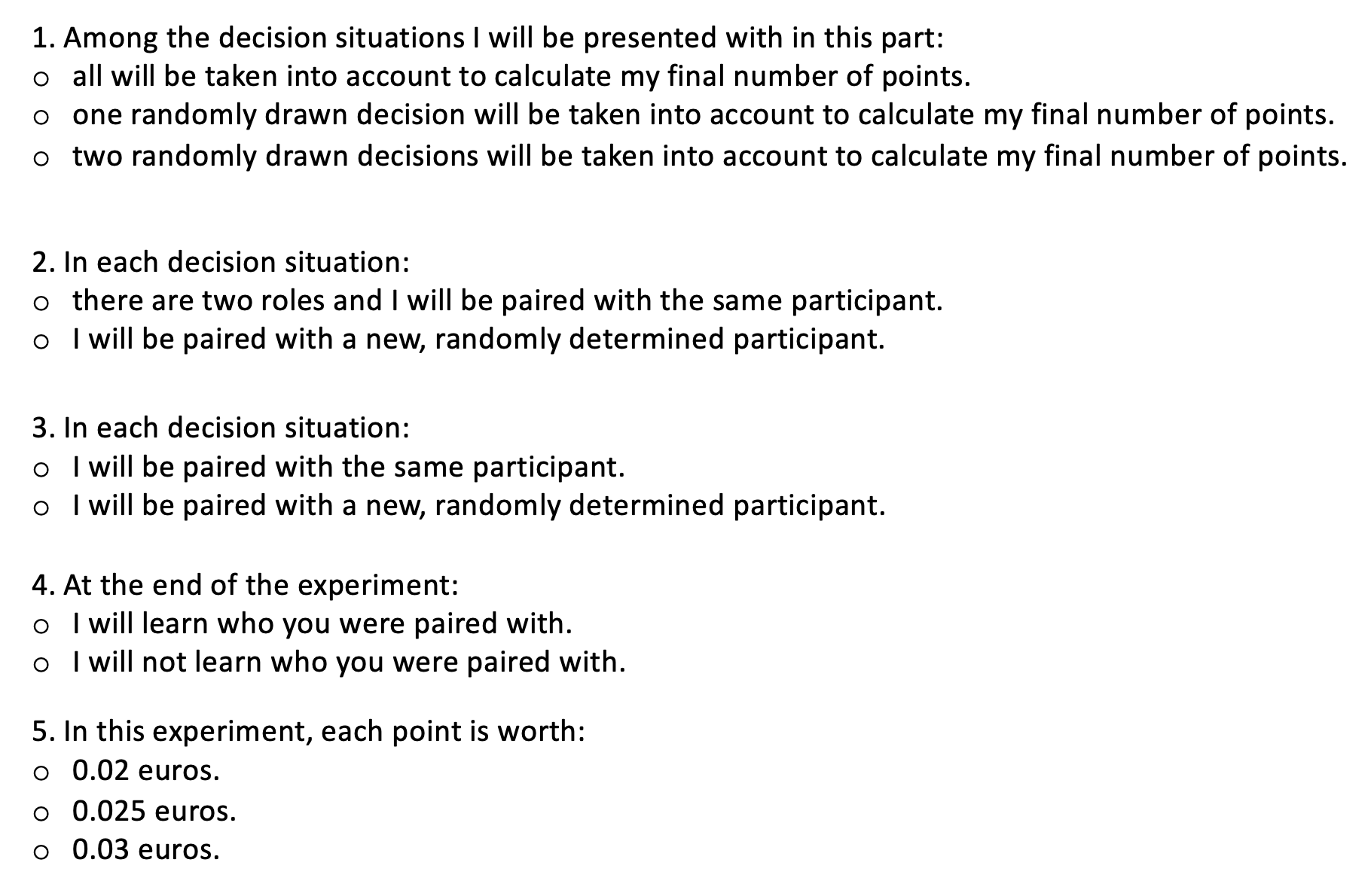}
\end{center}
\end{figure}

\bigskip 
\noindent 
\textbf{Screens 36-65: the 20 decision situations for Part II
}

\subsection{Remaining screens for the Neutral treatment}

\bigskip 
\noindent 
\textbf{Screen 2:}

\bigskip 
\noindent 
\textbf{Part I}

\bigskip 
\noindent 
In this part you will be asked to make choices in 20 different decision situations. 

\bigskip 
\noindent 
In each decision situation you will be paired with one of the other participants here today, each time with a different participant. 

\bigskip 
\noindent 
In each decision situation you will be asked to choose one of two options, X and Y, while the other participant has no choice to make.

\bigskip 
\noindent 
Here is an example of a decision screen:

\begin{figure}[H]
\begin{center}
\includegraphics[scale=0.3]{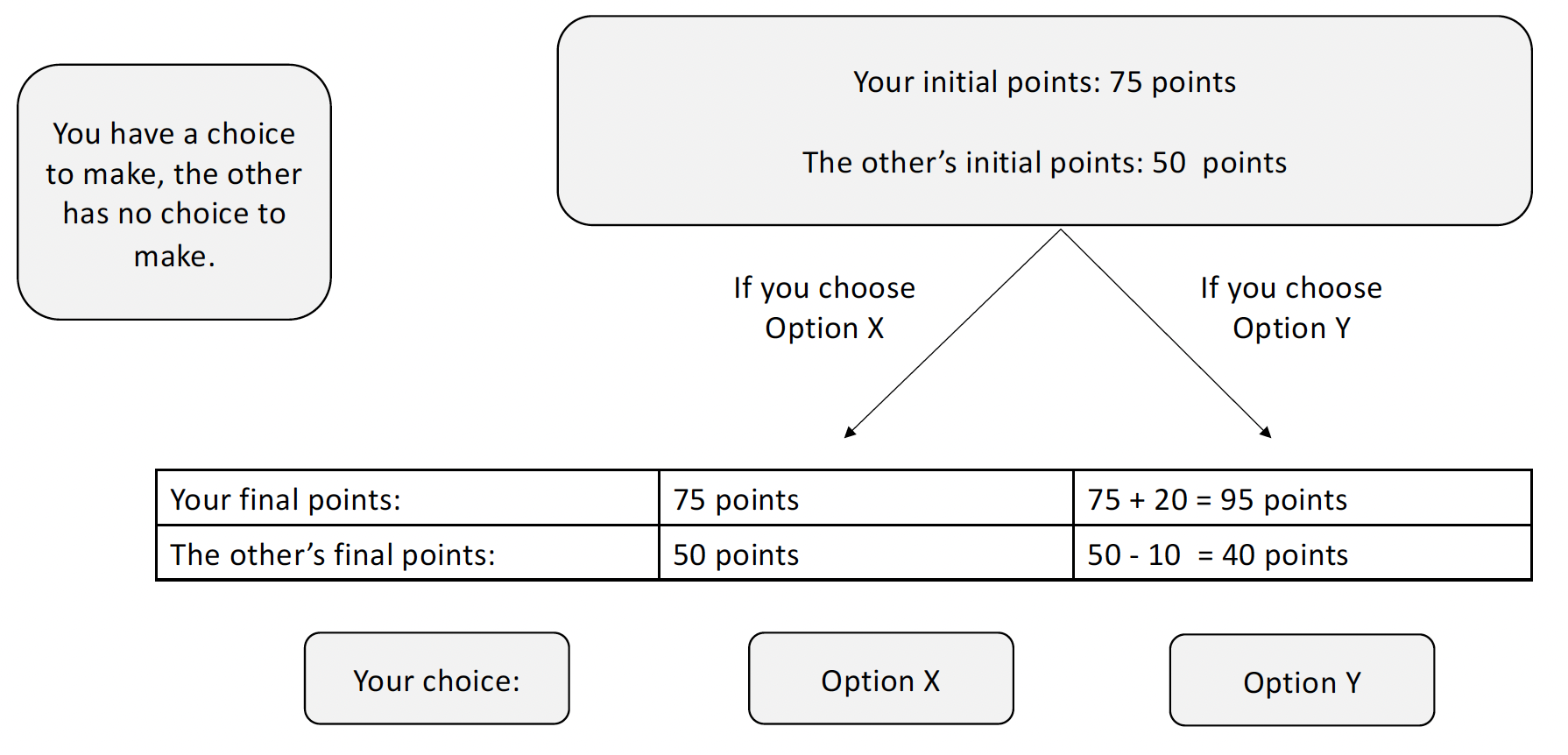}
\end{center}
\end{figure}

\bigskip 
\noindent 
At the end of the experiment one of these 20 decision situations will be randomly drawn. Your decision in this situation (and only this situation) will have an effect on the number of points you and the person you were paired with will get. This other participant will make no decision that affects your number of points, however. 

\bigskip 
\noindent 
Remember that:
\begin{itemize}
    \item all your decisions are anonymous, and no participant will ever learn with whom he/she was paired in any decision situation;
    \item all the decision situations are equally important, in the sense that they are all equally likely to count towards the amount of money you and the person you were paired with will receive at the end;
    \item each point will be converted to 0.025 euros at the end of the experiment.
\end{itemize}

\bigskip 
\noindent 
\textbf{Screen 3 [comprehension quiz]:}

\begin{figure}[H]
\begin{center}
\includegraphics[scale=0.3]{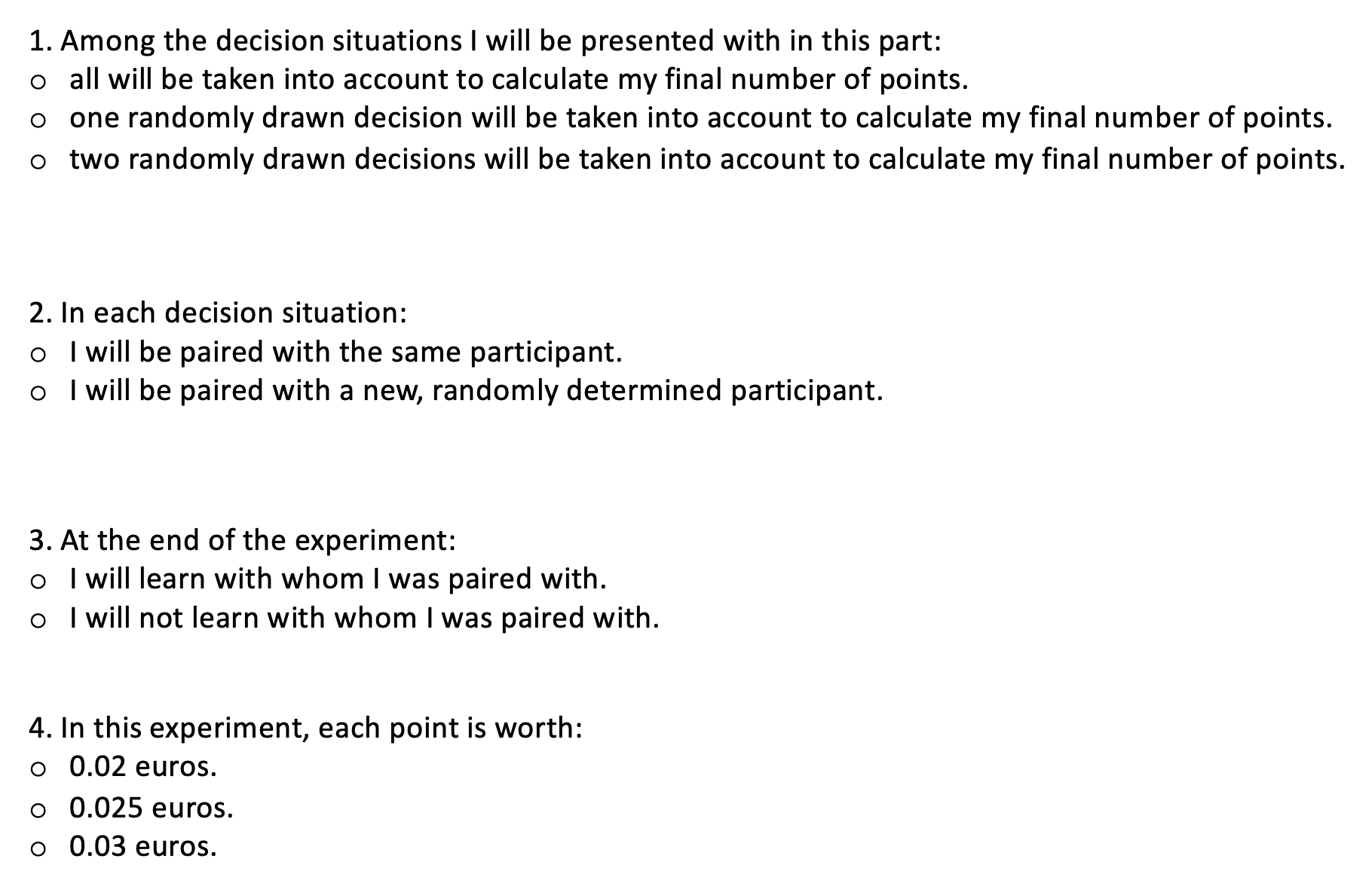}
\end{center}
\end{figure}

\bigskip 
\noindent 
\textbf{Screens 4-33: the 20 decision situations for Part I}

\bigskip 
\noindent 

\bigskip 
\noindent 
\textbf{Screen 34:
}

\bigskip 
\noindent 
\textbf{Part II
}

\bigskip 
\noindent 
In this part you will be asked to make choices in 20 different decision situations.

\bigskip 
\noindent 
In each decision situation you will be paired with one of the other participants here today, each time with a different participant. 

\bigskip 
\noindent 
In each decision situation there are two roles: Role A and Role B. Either you or the person you are paired with will be assigned to Role A, while the other will be assigned to Role B. The two different role assignments are equally likely. The person assigned to Role A chooses between two options, X and Y, while the other person has no choice to make. However, because the role assignment will only be made at the end of the experiment, both you and the person you were paired with will be asked to state what you would do in Role A in each decision situation.

\bigskip 
\noindent 
Here is an example of a decision screen:

\begin{figure}[H]
\begin{center}
\includegraphics[scale=0.3]{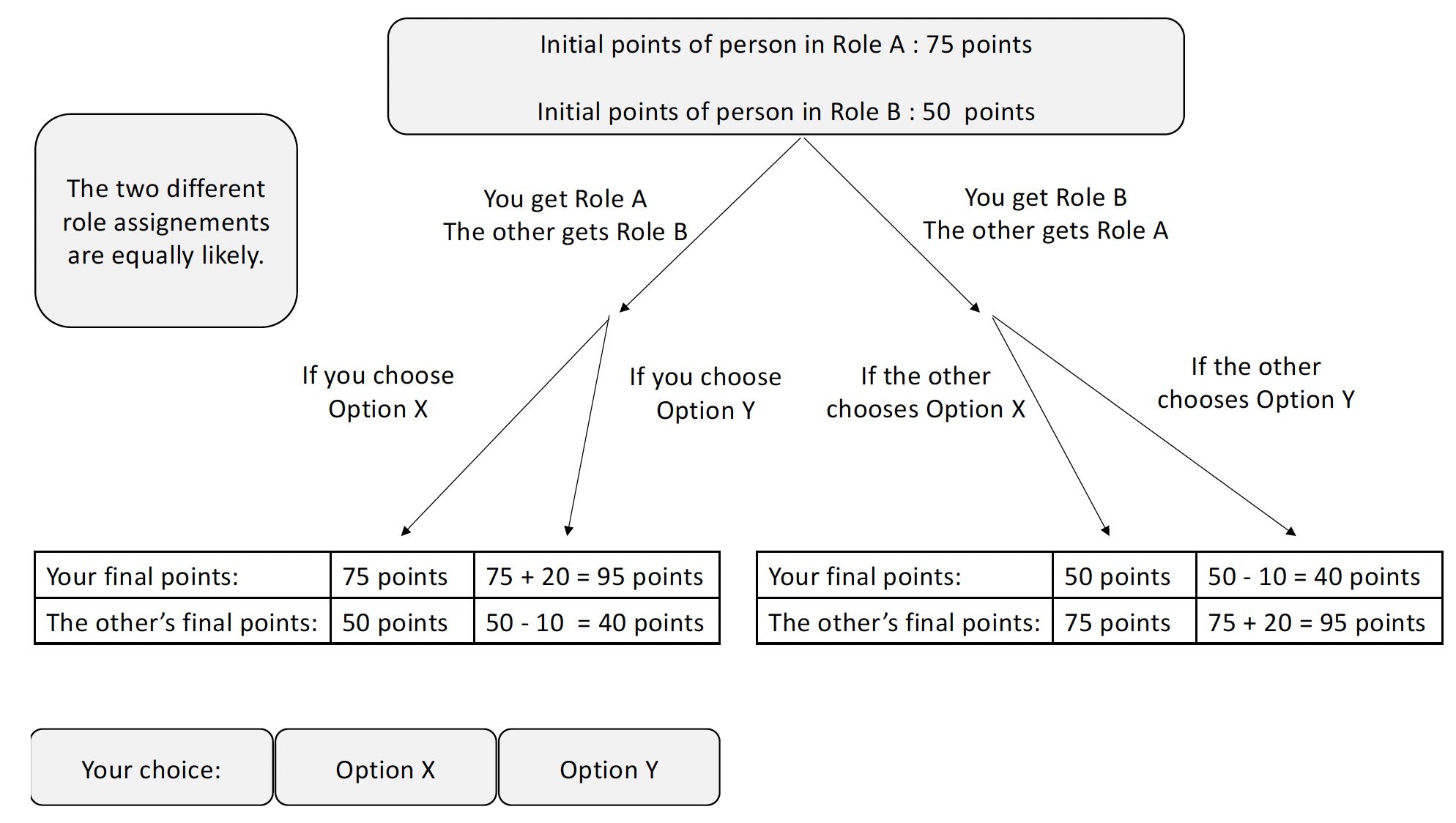}
\end{center}
\end{figure}

\bigskip 
\noindent 
At the end of the experiment one of these 20 decision situations will be randomly drawn. The role assignment for this decision situation will be randomly drawn. Recall that the two roles are equally likely. If you are assigned to Role A (this happens with probability 1/2), it is your decision in this situation that will have an effect on the number of points you and the person you were paired with will get. If you are assigned to Role B (this happens with probability 1/2), it is the decision of the person you were paired with in this situation that will have an effect on the number of points you and the person you were paired with will get.

\bigskip 
\noindent 
Remember that:
\begin{itemize}
    \item all your decisions are anonymous, and no participant will ever learn with whom he/she was paired in any decision situation;
    \item all the decision situations are equally important, in the sense that they are all equally likely to count towards the amount of money you and the person you were paired with will receive at the end;
    \item each point will be converted to 0.025 euros at the end of the experiment.
\end{itemize}

\bigskip 
\noindent 
\textbf{Screen 35 [comprehension quiz]:}

\begin{figure}[H]
\begin{center}
\includegraphics[scale=0.3]{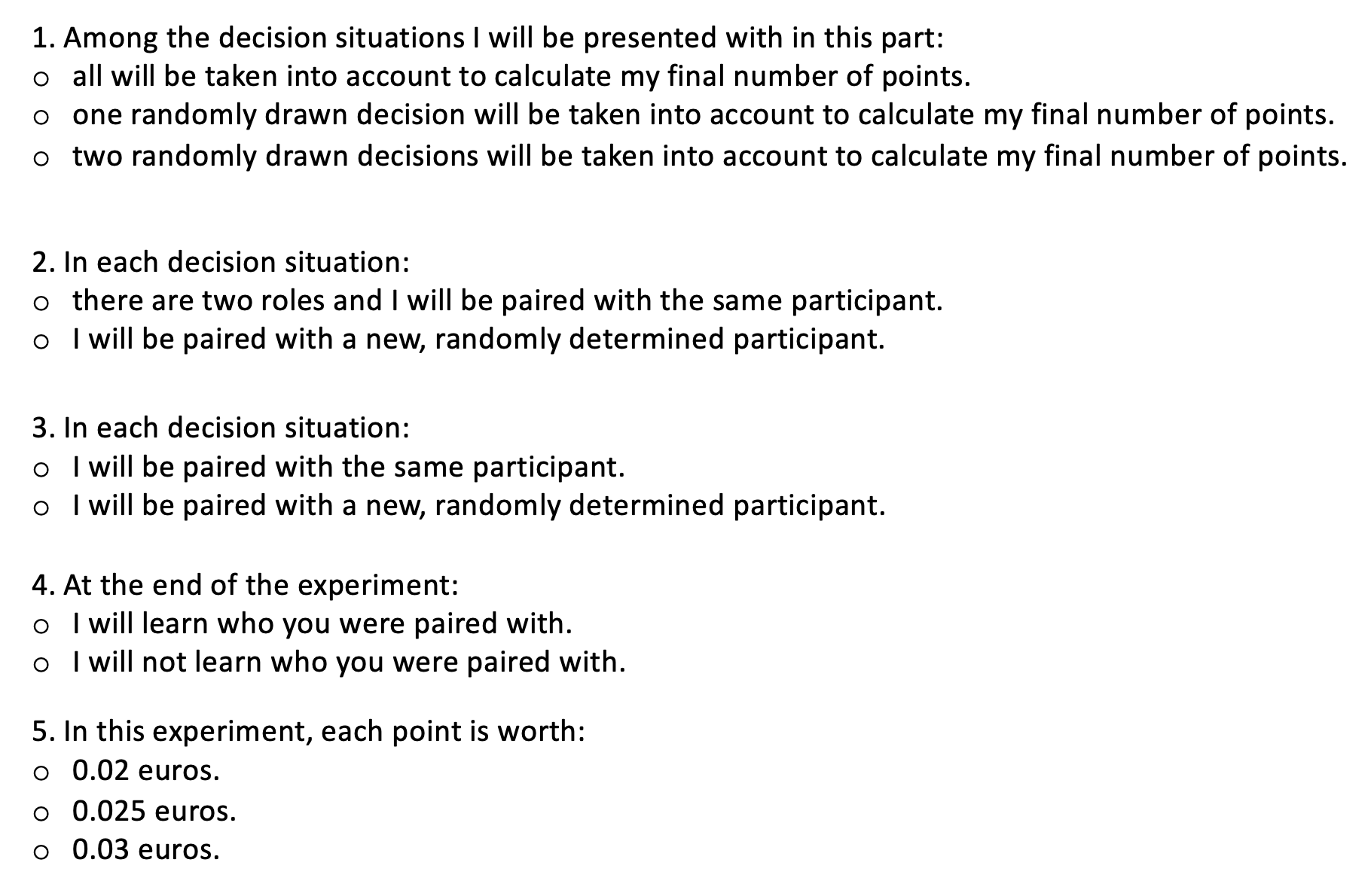}
\end{center}
\end{figure}

\bigskip 
\noindent 
\textbf{Screens 36-65: the 20 decision situations for Part II}

\subsection{Remaining screens for the two Mixed treatments}

\bigskip 
\noindent 
Each mixed treatment combines the Market VOI instructions (shown in Part II of subsection ``Remaining screens for the Market treatment'') and the Neutral VOI instructions (shown in Part II of subsection ``Remaining screens for the Neutral treatment'').

\subsection{Example of the two screens shown for one decision} \label{app:last}

\begin{figure}[H]
\begin{center}
\includegraphics[scale=0.4]{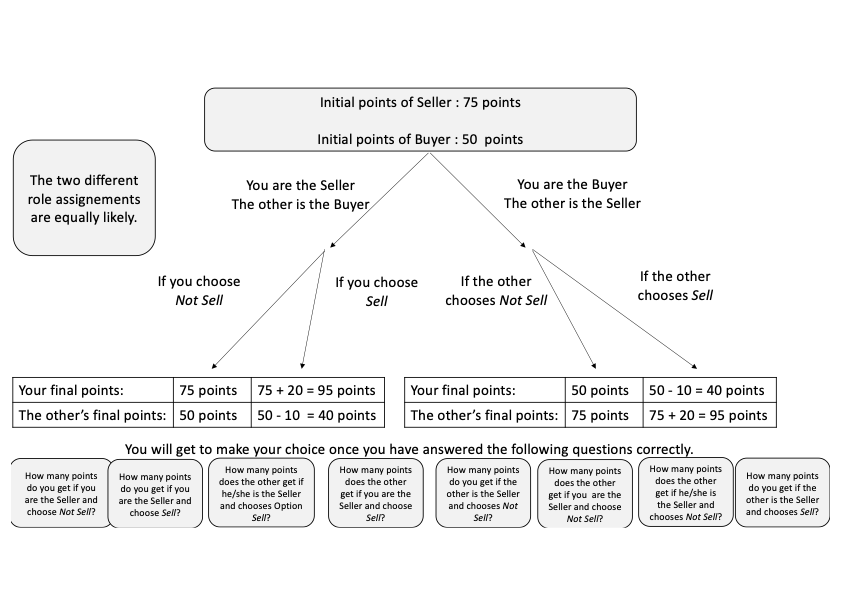}
\end{center}
\end{figure}

\begin{figure}[H]
\begin{center}
\includegraphics[scale=0.3]{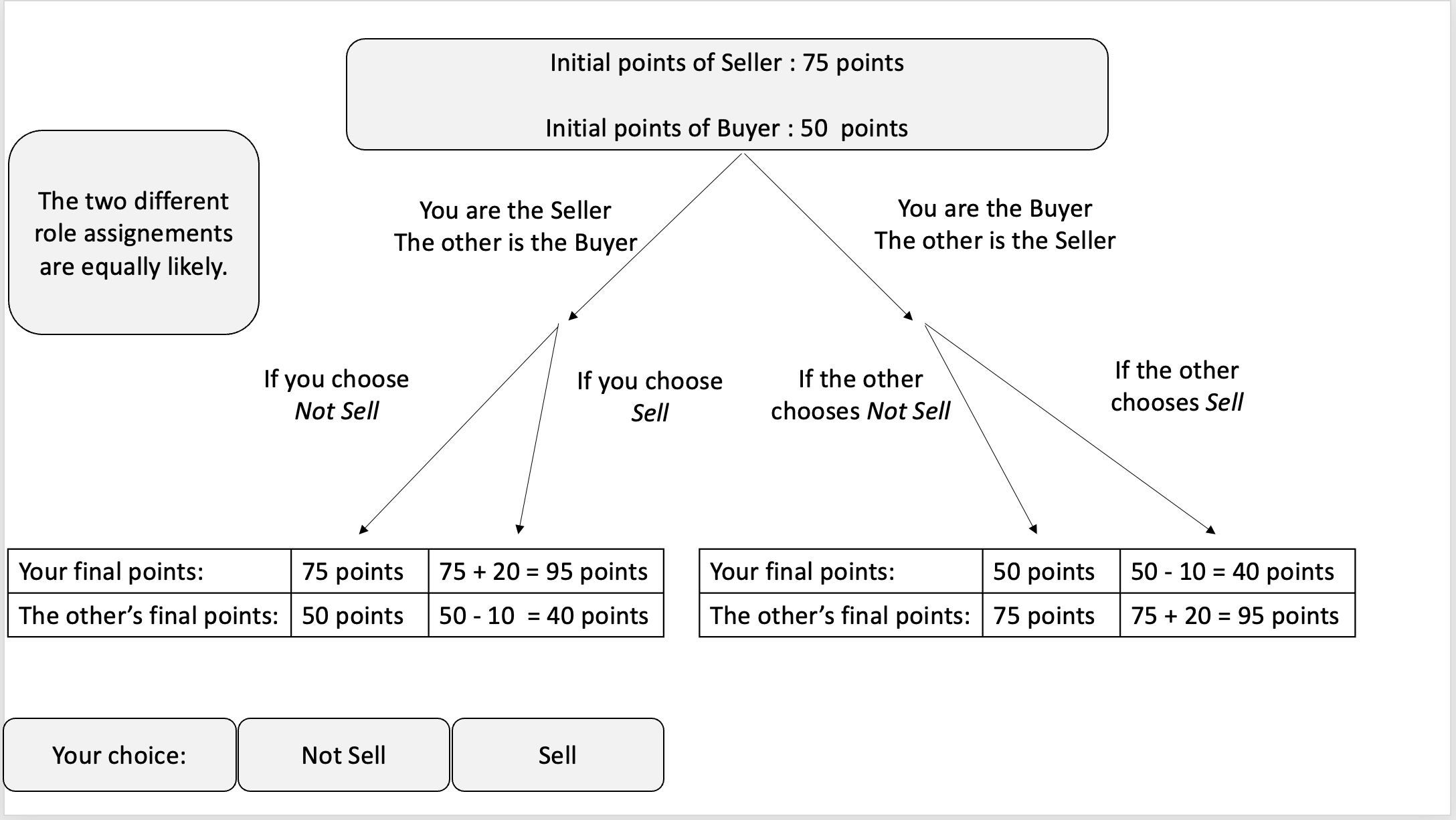}
\end{center}
\end{figure}

\end{appendices}

\end{document}